\documentclass[twocolumn]{aastex62}
\usepackage{xspace}
\graphicspath{{figures/}{./}}

\newcommand{\Grays}{$\gamma$-rays\xspace}
\newcommand{\GRays}{$\gamma$-Rays\xspace}
\newcommand{\GRay}{$\gamma$-Ray\xspace}
\newcommand{\gray}{$\gamma$-ray\xspace}

\newcommand{\Fermi}{\emph{Fermi}\xspace}
\newcommand{\FermiLAT}{\emph{Fermi}~LAT\xspace}
\newcommand{\fermiLAT}{\emph{Fermi}-LAT\xspace}

\received{February 4, 2019}
\revised{March 25, 2019}
\accepted{April 5, 2019}
\submitjournal{ApJ}

\shorttitle{Gamma-ray variability of bright FSRQs}
\shortauthors{Meyer et al.}

\begin{document}

\title{Characterizing the Gamma-Ray Variability of the Brightest Flat Spectrum Radio Quasars Observed with the \emph{Fermi} LAT}

\correspondingauthor{Manuel Meyer}
\email{mameyer@stanford.edu}

\author[0000-0002-0738-7581]{Manuel Meyer}
\affil{W. W. Hansen Experimental Physics Laboratory, Kavli Institute for
Particle Astrophysics and Cosmology, Department of Physics and SLAC National Accelerator Laboratory, Stanford University, Stanford, CA
94305, USA}

\author{Jeffrey D. Scargle}
\affil{Astrobiology and Space Science Division, NASA Ames Research Center, Moffett
Field, CA 94035-1000, USA}

\author{Roger D. Blandford}
\affil{W. W. Hansen Experimental Physics Laboratory, Kavli Institute for
Particle Astrophysics and Cosmology, Department of Physics and SLAC National Accelerator Laboratory, Stanford University, Stanford, CA
94305, USA}




\begin{abstract}

Almost 10~yr of $\gamma$-ray observations with the \emph{Fermi} Large Area Telescope (LAT) have revealed extreme  $\gamma$-ray outbursts from flat spectrum radio quasars (FSRQs), temporarily making these objects the brightest $\gamma$-ray emitters in the sky. 
Yet, the location and mechanisms of the $\gamma$-ray emission remain elusive. 
We characterize long-term $\gamma$-ray variability and the brightest $\gamma$-ray flares of six FSRQs.
Consecutively zooming in on the brightest flares, which we identify in an objective way through Bayesian blocks and a hill-climbing algorithm, 
we find variability on subhour time scales and as short as minutes for  two sources in our sample (3C~279, CTA~102) and weak evidence for variability at time scales less than the Fermi satellite's orbit of 95 minutes for PKS~1510-089 and 3C~454.3. 
This suggests extremely compact emission regions in the jet.
We do not find any signs for $\gamma$-ray absorption in the broad-line region (BLR), which indicates that $\gamma$-rays are produced at distances greater than hundreds of gravitational radii from the central black hole. 
This is further supported by a cross-correlation analysis between $\gamma$-ray and radio/millimeter light curves, which is consistent with $\gamma$-ray production at the same location as the millimeter core for 3C~273, CTA~102, and 3C~454.3.
The inferred locations of the $\gamma$-ray production zones are still consistent with the observed decay times of the brightest flares if the decay is caused by external Compton scattering with BLR photons. 
However, the minute-scale variability is challenging to explain in such scenarios.

\end{abstract}



\section{Introduction} \label{sec:intro}

More than half of the sources observed with the \Fermi Large Area Telescope (LAT) above 100\,MeV are active galaxies that produce particle outflows (jets) at almost the speed of light that are closely aligned to the line of sight \citep[see, e.g., the third \FermiLAT source catalog, i.e., the 3FGL;][]{3fgl}.
The broadband electromagnetic radiation observed from these so-called blazars spans decades in energy from radio frequencies up to very high \gray energies. 
It is often described with purely leptonic or a mixture of leptonic and hadronic emission models, involving both intrinsic and external radiation fields~\cite[e.g.,][and references therein]{Madejski:2016oqg}.
A common assumption is that the radiation is emitted by freshly accelerated   particles localized in ``plasmoids''  that move down the jet at relativistic speeds,
 leading to a strong doppler boost of the observed emission. 
Yet the origin, location, and even the very existence of such plasmoids are unknown.

Blazars display variability on timescales that can be as short as signal-to-noise limits allow and as long as the duration of the observations.
Flux doubling times as short as a few minutes have been observed at \gray energies in both BL Lac-type objects (BLL) and flat spectrum radio quasars (FSRQ) using  ground-based Cerenkov telescopes and the \FermiLAT~\cite[e.g.,][]{2007ApJ...669..862A,pks2155hess2007,pks1222magic2011,2013ApJ...762...92A,2014Sci...346.1080A,TheFermi-LAT:2016dss,2018ApJ...854L..26S}.
In these cases, causality arguments suggest extremely compact emission regions realized in, e.g., magnetic reconnection events, recollimation shocks, or magnetoluminescence~\cite[e.g.][]{Petropoulou:2016xat,Bodo:2017qqn,blandford:2017mag}.
In particular for FSRQs, the observation of \Grays beyond 10\,GeV suggests that these compact dissipation sites are located at distances of hundreds of Schwarzschild  radii from the central supermassive black hole. 
Otherwise, the \gray emission should be strongly attenuated through pair production on UV and optical photons that are emitted by the accretion disk and broad emission line clouds and scattered by the intercloud medium.
Meeting these constraints is challenging for standard emission scenarios, as extreme relativistic bulk motions of the plasma have to be invoked~\cite[e.g.,][]{2010MNRAS.405L..94T,2011A&A...534A..86T,TheFermi-LAT:2016dss}.  (For an alternative possibility see Sections~\ref{sec:blrabs}. and~\ref{sec:tcool})

After almost a decade of continuous all-sky observations, the \FermiLAT has accumulated a large sample of flares---\gray outbursts limited in time in which the source emission can increase typically by a factor of a few---from many FSRQs.
Our goal is to characterize flares and long-term behaviour of those FSRQs that have shown the brightest \gray flares over the course of the \Fermi mission. 
The most extreme flaring states enable us to perform a comprehensive search for \gray variability on time scales as short as minutes in order to investigate whether such short variability---and conversely compact emission sites---is a common phenomenon in FSRQ flares. 
Evidence for minute-scale variability has already been discovered in LAT observations of 3C~279~\citep{TheFermi-LAT:2016dss} and recently in CTA~102~\citep{2018ApJ...854L..26S}, but searches in other sources have been unsuccessful \citep{2017Galax...5..100N} 
or resulted in upper limits on the flux doubling time~\citep{2011A&A...530A..77F}.

The plethora of observed flares also enables us to perform a systematic study of the local temporal flare profiles, which could be diagnostic of particle injection, acceleration, and propagation.
\citet{2013MNRAS.430.1324N} investigated the brightest \gray flares in blazar in the first 4~yr of LAT data.
The author found that, on average, flares have a slight tendency toward rise times being shorter than decay times; however, no flare showed extreme asymmetry. 
\citet{2010ApJ...722..520A}, on the other hand, characterized the blazars in terms of their power spectral density (PSD) on longer timescales using 11 months of data.
The authors found that the distribution of the power-law slopes of the power spectra of bright blazars peaks around $-1.2$ or $-1.3$ with a scatter marginally larger than the observational uncertainty,  that is to say, intermediate between steep spectra (slope of $-2$, sometimes called Brownian noise) and less steep (slope of $-1$, sometimes called flicker noise).
Similar conclusions were reached by~\citet{2LAC} using 24 months of data. PSDs for both FSRQ and BLL could be well fitted with a power-law spectrum with an index of $1.15\pm0.10$.
These analyses can be significantly extended with almost a decade of continuous \fermiLAT observations.

Additionally, the high signal-to-noise spectra during flares enable us to search for spectral absorption features due to the interaction of \Grays with broad-line region (BLR) photons.
The detection of such features would locate the \gray emission region inside the BLR with important implications where particles dissipate their energy. 
Indeed, evidence for such absorption was reported in early  \fermiLAT observations~\citep{2010ApJ...717L.118P,2014ApJ...794....8S}, but a recent analysis of a large sample of over 100 FSRQs and more than 7\,yr of \fermiLAT observations could not confirm this result~\citep{2018MNRAS.477.4749C}.
Similar results were also reached by \citet{2019ApJ...874...47V}, who analyzed blazars detected above 50\,GeV~\citep{2016ApJS..222....5A}.
The absence of the absorption features can, in turn, be used to derive lower limits on the distance from the \gray emitting region to the central supermassive black hole. 

This paper is organized as follows. 
In Section~\ref{sec:data} we present the source selection and \FermiLAT data analysis.
In Section~\ref{sec:results-global}, we investigate the global-light curve properties before characterizing the temporal properties of the brightest flares in Section~\ref{sec:results-local}, which we identify by using an objective method.
We investigate the location of the \gray emitting region through searches for BLR absorption features in \gray spectra, a comparison between radiative cooling time scales and observed flare decay times, and a cross-correlation between long-term \gray and radio light curves in Section~\ref{sec:location}. 
Our findings and conclusions are summarized in Section~\ref{sec:conclusion}.

\section{Source Selection and Data Analysis}
\label{sec:data}

For our analysis, we select the FSRQs that show the brightest \gray flares as reported in the monitored source list\footnote{\url{https://fermi.gsfc.nasa.gov/ssc/data/access/lat/msl_lc/}}
with average daily fluxes $F \geqslant 10^{-5}\,\mathrm{cm}^{-2}\,\mathrm{s}^{-1}$ within $1\,\sigma$ statistical uncertainties above 100\,MeV.
This leaves us with six sources, listed in Table~\ref{tab:src-select}, together with their coordinates, redshift, and additional parameters taken from the literature, such as black hole mass and luminosity of the $\mathrm{H}\beta$ line, a measure of the BLR luminosity.  
All of the sources in this selection have at least one 
flare that is suitable
to search for intra-orbit variability and to derive high signal-to-noise spectra. 
All of the selected FSRQs are well-known \gray emitters, and individual flares from these objects have been studied in great detail~\citep[e.g.,][]{2010ApJ...714L..73A,2011ApJ...733...19T,2015ApJ...808L..48P,TheFermi-LAT:2016dss,2013ApJ...766L..11S,2015ApJ...809..164D,2018ApJ...854L..26S,2018A&A...617A..59K,2019ApJ...871...19Z,2011ApJ...733L..26A}. 
As noted in the Introduction, two of the sources (3C~279 and CTA~102) have already been shown to be variable on extremely short timescales~\citep{TheFermi-LAT:2016dss,2018ApJ...854L..26S}. 

Here we will introduce a novel objective method to identify a large set of flares in order to conduct a comprehensive search for the short variability in our source sample.  
Furthermore, PKS~B1222+216~\citep{2011ApJ...730L...8A}, 3C~279~\citep{2008Sci...320.1752M}, and PKS~1510-089~\citep{2011ApJ...730L...8A,2013A&A...554A.107H,2014A&A...569A..46A}  are among the seven FSRQs also detected above 100\,GeV with imaging air Cerenkov Telescopes. 
For 3C~454.3, the MAGIC and VERITAS telescopes only obtained upper limits on the flux during flaring states in 2007 \citep{2009A&A...498...83A} and 2010 \citep{2016AJ....151..142A}.

\begin{deluxetable*}{llccccccccc}
\tablewidth{0pt}
\tabletypesize{\scriptsize}
\tablecaption{ \label{tab:src-select}FSRQs selected for this study. }
\tablehead{
\colhead{Source name} &
\colhead{3FGL name} & 
\colhead{R.A.} & 
\colhead{Decl.} & 
\colhead{Redshift} & 
\colhead{$\log_{10}(M_\bullet / M_\odot)$\tablenotemark{a}} &
\colhead{$L_\mathrm{disk}$\tablenotemark{a}} & \colhead{$L(\mathrm{H}\beta)$\tablenotemark{a}} &
\colhead{$\delta_\mathrm{D}$\tablenotemark{b}} &
\colhead{$\Gamma_\mathrm{L}$\tablenotemark{b}} &
\colhead{$\theta_\mathrm{obs}$\tablenotemark{b}}\\
{} & {} & \colhead{[deg]} & \colhead{[deg]} & {} & {} & \colhead{$[10^{46} \mathrm{ergs}\,\mathrm{s}^{-1}]$} & \colhead{$[10^{43} \mathrm{ergs}\,\mathrm{s}^{-1}]$} &
{} & {} & \colhead{[deg]} 
}
\startdata
PKS~B1222+216 & 3FGL~J1224.9+2122 & 186.226  & 21.382 & 0.432 & 8.87\tablenotemark{c} & $1.61$ &  $2.79 \pm 0.56$\tablenotemark{d} & $7.4 \pm 2.1$ & $13.9\pm2.1$ & $5.6\pm1.0$\\
3C~273 &	3FGL~J1229.1+0202 & 187.266  & 2.051 & 0.158 & 	8.92 &	6.11 & 	15.40 & $4.3\pm1.3$ & $8.5\pm2.2$ & $6.4\pm2.4$ \\
3C~279 & 3FGL~J1256.1-0547 & 194.045  & $-5.786$ & 0.5362 	&	8.28 &	1.11 &	1.73 & $18.3\pm1.9$ & $13.3\pm0.6$ & $1.9\pm0.6$\\
PKS~1510-089 &	3FGL~J1512.8-0906 & 228.210  & $-9.106$ & 0.360 & 8.20 & 1.13 & 1.77 & $35.3\pm 4.6$ & $22.5\pm3.3$ & $1.2\pm0.3$\\
CTA~102 & 3FGL~J2232.5+1143 & 338.158  & 11.728 & 1.037 & 8.93\tablenotemark{c} & 4.00  &	8.93 $\pm$  6.00\tablenotemark{e} & $30.5\pm3.3$ & $21.7\pm1.3$ & $1.6\pm0.4$\\
3C~454.3 & 3FGL~J2254.0+1608 & 343.493  & 16.149 & 0.859 & 	8.83 &	7.19 & 	19.00 & $24.4\pm3.7$ & $13.8\pm1.4$ & $0.7\pm0.4$\\
\enddata
\tablenotetext{a}{Taken from \citet{2006ApJ...637..669L} if not noted otherwise.}
\tablenotetext{b}{Average jet values taken from \citet{2017ApJ...846...98J}.}
\tablenotetext{c}{From \citet{2014Natur.510..126Z}.}
\tablenotetext{d}{From \citet{2012RMxAA..48....9T}.}
\tablenotetext{e}{\citet{2012RMxAA..48....9T} gave the $L$(CIV) with $(255.7 \pm 17.2)\times10^{43}\mathrm{ergs}\,\mathrm{s}^{-1}$, and Eq.~7 from \citet{2006ApJ...637..669L} is used to convert this to $L$(H$\beta$).}
\tablecomments{The reported source positions are derived from 
our \gray analysis. $M_\bullet$ denotes the black hole mass, $L_\mathrm{disk}$ the accretion disk luminosity, $L(\mathrm{H}\beta)$ is the luminosity of the $\mathrm{H}\beta$ emission line, $\delta_\mathrm{D}$ denotes the relativistic Doppler boost factor, $\Gamma_\mathrm{L}$ is the bulk Lorentz factor, and $\theta_\mathrm{obs}$ is the angle between the jet axis and the line of sight. }

\end{deluxetable*}

\subsection{Data selection}

Our goal is to characterize both the long-term \gray behavior of the selected FSRQs as well as the brightest flares.
To this end, we select \Grays that were measured with the \FermiLAT between 2008 August 4 and 2018 January 30, yielding data over an interval of 114\,months, or almost 9.5\,yr.
The \FermiLAT is a pair conversion telescope designed to measure \Grays with energies from 20\,MeV to above 300\,GeV~\citep{2009ApJ...697.1071A}.

We follow the standard data selection recommendations\footnote{\url{https://fermi.gsfc.nasa.gov/ssc/data/analysis/documentation/Cicerone/Cicerone_Data_Exploration/Data_preparation.html}} and  restrict ourselves to \Grays in the energy range between 100\,MeV and 316\,GeV.\footnote{The upper energy bound  coincides with a bin edge of the instrumental response functions, which are logarithmically binned with 16 bins per decade.}
Below 100\,MeV, the effective area of the LAT quickly decreases, and the point spread function increases to above $\sim 6^\circ$\footnote{See, e.g., \url{http://www.slac.stanford.edu/exp/glast/groups/canda/lat_Performance.htm}} making a point-source analysis challenging. 
Since FSRQs usually have soft \gray spectra, we do not expect significant detection of these sources above our chosen maximum energy.

To mitigate contamination of \Grays originating from the Earth limb, we further limit the sample to events that have arrived at a zenith angle less than $90^\circ$, and we excise periods of bright \gray bursts and solar flares that have been detected with a test statistic $(\mathrm{TS}) > 100$.
The TS is defined as $\mathrm{TS} = -2\ln(\mathcal{L}_1 / \mathcal{L}_0)$, i.e., the log-likelihood ratio between the the maximized likelihoods $\mathcal{L}_1$ and $\mathcal{L}_0$ for the hypotheses with and without an additional source, respectively~\citep{mattox1996}.
We use the latest \texttt{Pass 8} instrumental response functions and Monte Carlo simulations~\citep{pass8} and select \Grays that pass the \texttt{P8R2 SOURCE} event selection. 
For each source, we analyze a $10^\circ \times 10^\circ$ region of interest (ROI) centered on the position of each source as provided in the 3FGL \citep[3FGL,][]{3fgl}.
We choose a spatial binning of $0.1^\circ\,\mathrm{pixel}^{-1}$ and eight energy bins per decade. 

\subsection{ROI optimization}
\label{sec:roi}

Our analysis proceeds iteratively, starting from the full time range and zooming in on bright flares and shorter time scales (see Section~\ref{sec:zoom}).
In a first step, we optimize the global \gray model of each ROI using the \textit{Fermi Science Tools} version 11-05-03\footnote{\url{http://fermi.gsfc.nasa.gov/ssc/data/analysis/software}} and \textsc{fermipy}, version 0.16.0+188\footnote{\url{http://fermipy.readthedocs.io}}~\citep{fermipy}.
The initial model consists of all \gray point sources within $15^\circ$ of the ROI center included in the 3FGL, as well as the standard templates for isotropic and Galactic diffuse emission.\footnote{For the Galactic diffuse emission, we use the file gll\_iem\_v06.fits and the file iso\_P8R2\_SOURCE\_V6\_v06.txt for the isotropic diffuse component; see: \url{ http://fermi.gsfc.nasa.gov/ssc/data/access/lat/BackgroundModels.html}}
After an initial optimization, we free the spectral normalization of all sources in the model. 
The spectral shape parameters, such as power-law indices, curvature, or exponential cut-off energies, are free to vary for sources within $5^\circ$ of the ROI center. 
We freeze all spectral parameters for sources with $\mathrm{TS} < 1$ or a predicted number of \Grays after the initial optimization less than $10^{-3}$ counts.
The normalizations of the diffuse backgrounds are left free during the fit,\footnote{
This is necessary because the background templates have been derived with a different data selection compared to the present analysis.
}
together with the spectral index of the Galactic diffuse background template.
After the fit has converged successfully, 
we relocalize the central \gray source and refit all spectral model parameters. The relocalized source positions are provided in Table~\ref{tab:src-select}.
After this step, we generate a $\mathrm{TS}$ map to search for additional point sources. For each pixel in the ROI, we add a putative point source with a power-law spectrum with index $\Gamma = 2$ and calculate its $\mathrm{TS}$. If  $\sqrt{\mathrm{TS}} \geqslant 5$. i.e.  a detection with a significance of just over $\sim 4\,\sigma$~\citep{3fgl}, we permanently add the source at the position of the highest $\mathrm{TS}$ value and reoptimize the spectral parameters for the whole ROI. This step is repeated until no further sources are found.

With the best-fit model for each ROI, we compute the \gray light curves for the FSRQs with an initial binning of 7~days. 
In each light-curve bin, we leave the spectral parameters free during the fit for sources within $3^\circ$ of the ROI center and, additionally, the normalizations of the Galactic and isotropic emission. If any of these sources have $\mathrm{TS} < 1$ or the number of predicted photons is less than $10^{-3}$, their parameters are fixed to the average values.

\subsection{Zooming In on Bright Flares Using an Objective Method to Identify Different Activity States}
\label{sec:zoom}

\begin{figure*}
    \centering
    \includegraphics[width = .9\linewidth]{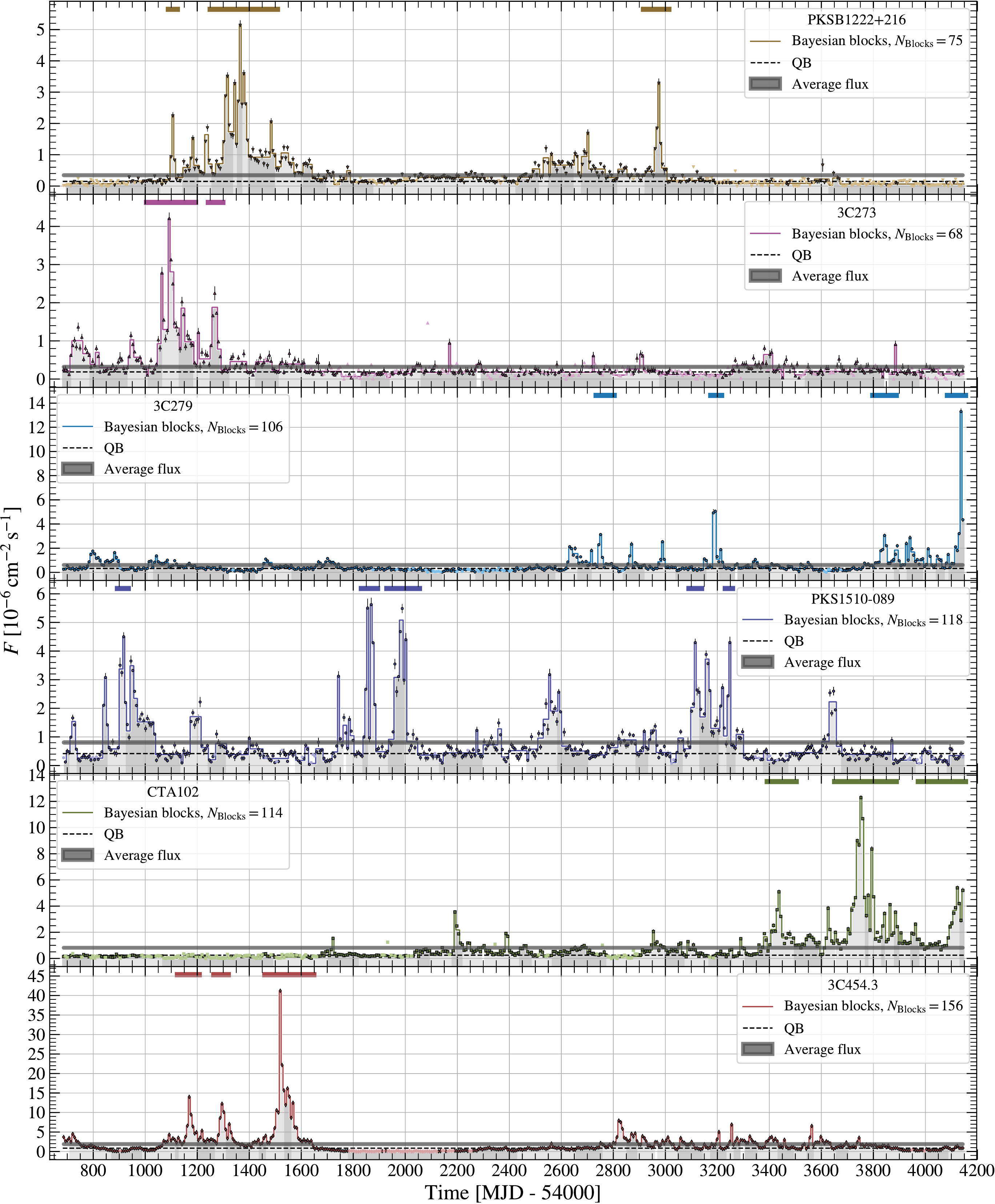}
    \caption{The \gray light curves with weekly binning for the considered FSRQs. Open symbols denote upper limits at the $2\,\sigma$ confidence level. The thin solid lines show the BBs and the gray shaded regions represent the identified HOP groups. The colored horizontal lines denote the time intervals identified as bright flares for which we derive light curves with finer binning.
    The gray shaded horizontal band denotes the average flux with $1\,\sigma$ statistical uncertainties, and the black dashed line shows the estimate of the QB introduced in Section~\ref{sec:qb}.}
    \label{fig:weekly}
\end{figure*}

The 9.5 yr light curves for all considered FSRQs are shown in Figure~\ref{fig:weekly}. If the source is detected with $\mathrm{TS} < 9$ within one time bin, or the flux in one bin is equal to or smaller than its statistical uncertainty $F_i \leqslant \sigma_i$, we show upper limits\footnote{We avoid 
problems with incorporating upper limits by simply ignoring them, because the BB algorithm does not require evenly spaced data. At these gaps, there
are just two possibilities:
(1) a single block spans the gap,
or 
(2) one block stops at the beginning of
the gap and another starts at the end.
The former indicates that the flux levels
before and after the gap are the same;
the latter indicates that they are not.
In neither case is there definitive 
information about the level in the
gap itself.}
at the $2\,\sigma$ confidence level instead.

The average source fluxes with their 1$\,\sigma$ statistical uncertainties, $\overline{F} \pm \sigma_{\overline{F}}$, derived from the likelihood maximization over the full 9.5\,years are shown as gray bands. 
These flux measurements and uncertainties determine the optimal step-function representations of the light curves 
using the Bayesian Block (BB) 
algorithm ~\citep[][]{2013ApJ...764..167S} maximizing the overall fitness function appropriate to \textit{point measurements}.
From all possible partitions of the data into blocks this algorithm finds the unique one maximizing the total fitness of the resulting step-function model.

These blocks provide an objective way to detect significant local variations in the light curve.
Several strong flares exceeding the average flux level are easily identified from this block representation. 

There is no generally accepted consensus on the best way to determine which data points belong to a flaring state and which characterize the quiescent level. \citet{2013MNRAS.430.1324N} suggested a simple definition that a flare is a continuous time interval associated with a flux peak in which the flux is larger than half the peak flux value. 
This definition is intuitive, however, and it is unclear how to treat overlapping flares and identify flux peaks in an objective way. 
Here we use a simple two-step procedure tailored to the block representation: (1) identify a block that is higher than both the previous and subsequent blocks as a \textit{peak}, and (2) proceed downward from the peak in both directions as long as the blocks are successively lower.
The two resulting monotonically decreasing sets of adjacent blocks are analogous to the \textit{watershed} concept of topological data analysis.
In fact, this approach was suggested by the 
HOP algorithm~\citep{1998ApJ...498..137E}, which is based on a bottom-up hill-climbing concept of great use in higher dimensions \citep[e.g.,][]{2011ApJ...727...48W}.\footnote{The name \textit{HOP} derives from the verb "to hop" (to each data element's highest neighbor) and is not an acronym \citep{1998ApJ...498..137E}.}

The time-series segments shown in 
Figure~\ref{fig:weekly} 
are the result of feeding
the block representations of the light curves to the HOP algorithm.
The combination of BB and the HOP algorithm provides an objective way to split a light curve into groups of quiescent and flaring episodes; we will refer to one connected flare episode as a \emph{HOP group} of consecutive BBs.

We iteratively zoom in on time ranges with bright \gray activity by identifying HOP groups where the peak BB fulfills the condition $F_{BB} \geqslant F_\mathrm{max} =  5\times\overline{F}$ and include adjacent blocks within the group with $F_{BB} \geqslant F_\mathrm{min} = \overline{F}$. 
This relatively conservative definition 
gives reliable group shape information 
at the cost of 
slightly underestimating the extent of the
groups and overestimates that of the quiescent 
intervals.
Furthermore, we prefer a criterion based on peak flux rather than, for instance, integrated flare luminosity.
This is because our approach promises the most straightforward way to find those time ranges with sufficient photon statistics to search for short-timescale flux variations and exponential cutoffs due to \gray absorption.
If multiple  adjacent HOP groups fullfill our criteria, they are combined into one time interval. 
The final time ranges are then extended by one time bin on either side.
For the identified time span, we reoptimize the spectral model of the ROI in the same way as described in Section~\ref{sec:roi} but without relocalizing the central FSRQ or adding new point sources. 
The results of the best-fit spectra for the different time ranges are provided in Appendix~\ref{sec:avg-spec}.
Subsequently, we calculate a light curve with finer binning and again select the time ranges of the highest \gray activity. We repeat this procedure twice, down to a binning equal to the good time intervals (GTIs), of the \Fermi satellite, which correspond to one passage of the source through the field of view of the satellite during one $\sim$95 minute orbit.
The choices of time binnings and values for $F_\mathrm{max}$ and $F_\mathrm{min}$ are summarized in Table~\ref{tab:zoom} together with the number of identified high \gray activity states (which might consist of several flares, as indicated by the HOP groups).

The values of the threshold fluxes $F_\mathrm{max}$ and $F_\mathrm{min}$ are somewhat arbitrary and are a compromise between including as many flares as possible and keeping the overall number of flares manageable. Note that the sole purpose of this exercise is to select the brightest flares for further analysis; consideration of a more
complete statistical sample is postponed to 
the future.

\begin{figure*}
    \centering
    \includegraphics[width = .99\linewidth]{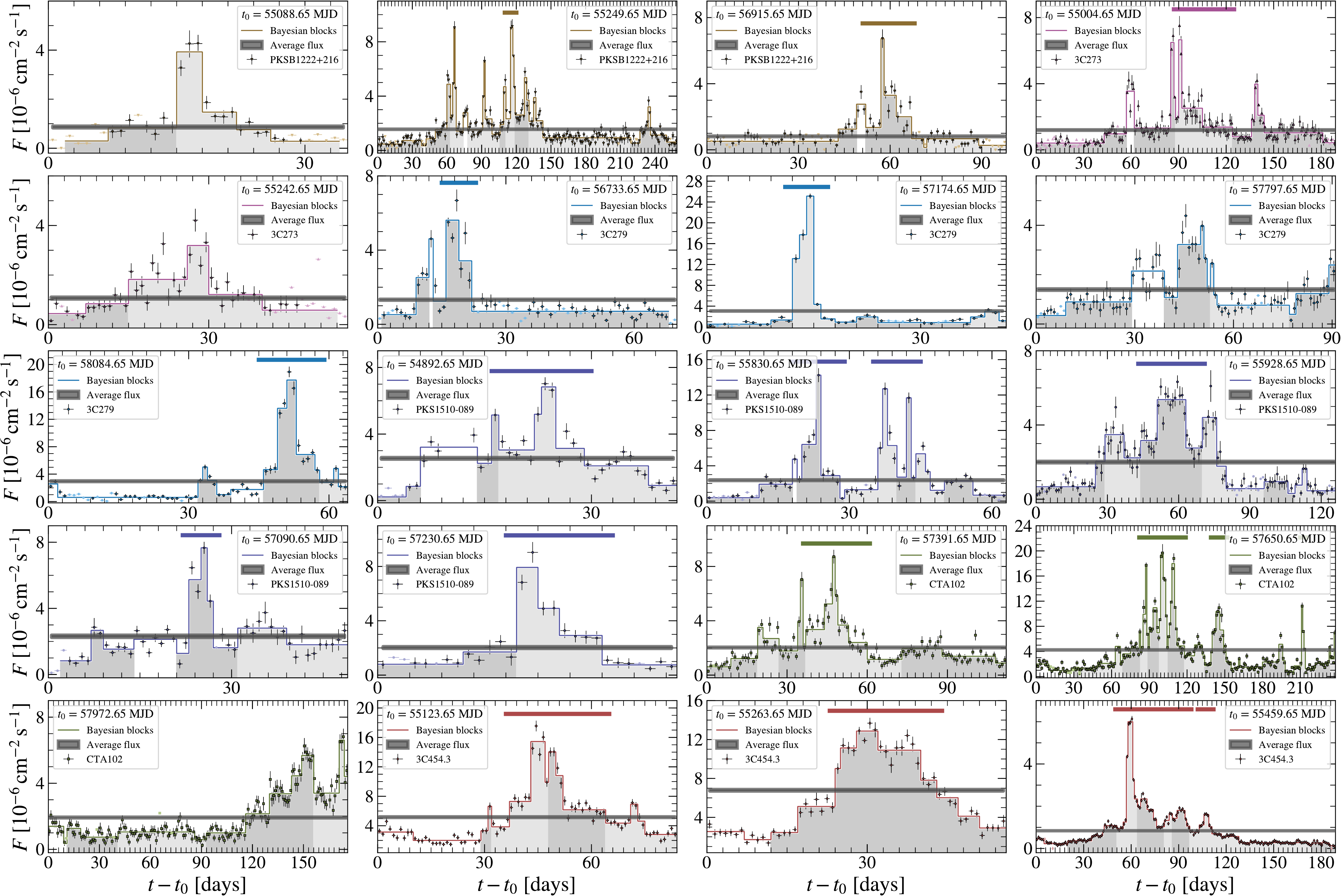}
    \caption{\label{fig:daily} Light curves with daily binning for the selected time ranges (horizontal lines in Figure~\ref{fig:weekly}). Symbols and lines are the same as in Figure~\ref{fig:weekly}. 
    As stated in Table~\ref{tab:zoom}, if all BB fluxes fail the criterion $F_{BB} \geqslant \mathrm{max}(10^{-5}\,\mathrm{cm}^{-2}\,\mathrm{s}^{-1}, 1.5 \bar{F})$, we include the HOP group with the maximum flux of the interval if that flux is above $5\times 10^{-6}\,\mathrm{cm}^{-2}\,\mathrm{s}^{-1}$. This is the case, e.g., for last two intervals of PKS~B1222+216 and the first interval of 3C~273 (last three panels of the top row).}
\end{figure*}

The
time ranges of the identified flares
for the weekly light curves are plotted as colored horizontal solid lines in Figure~\ref{fig:weekly}.
The intermediate daily light curves are provided in Figure~\ref{fig:daily}.
Figure~\ref{fig:gti} shows the GTI (equivalent to orbital) light curves with exponential fits of flare profiles that we will discuss in Section~\ref{sec:results-local}. 
The source exposure can vary significantly between two adjacent orbits, as the satellite rocks between the celestial north and south poles between orbits. 
This explains the large error bars on some of the time bins of the GTI light curves.

\begin{deluxetable}{cccc}
\tablewidth{1\linewidth}
\tablecaption{ \label{tab:zoom} Thresholds for BB fluxes in one HOP group to select time ranges of high \gray activity together with selected time binning and number of selected time ranges.
}
\tablehead{\colhead{Binning} & \colhead{$F_\mathrm{min}$} & \colhead{$F_\mathrm{max}$} & \colhead{$N_\mathrm{time~ranges}$}}
\startdata
7 days & $\overline{F}$ & $5\times\overline{F}$ & 20\\
1 day & $\overline{F}$ & $\mathrm{max}(10^{-5}\,\mathrm{cm}^{-2}\,\mathrm{s}^{-1}, 1.5 \times \overline{F})$\tablenotemark{a} & 21\\
GTI & $\overline{F}$ & $2\times\overline{F},~\mathrm{TS} \geqslant 150$\tablenotemark{b} & 7\\
\enddata
\tablenotetext{a}{We choose here the absolute flux (rather than the flux relative to the average) as a threshold in order to be consistent with our initial source selection. However, because of the high average flux of 3C~454.3, we also include the max argument. If $F_\mathrm{max} = 1.5 \times \overline{F}$ we set $F_\mathrm{min} = 10^{-5}\,\mathrm{cm}^{-2}\,\mathrm{s}^{-1}$. }
\tablenotetext{b}{Motivated by the high $\mathrm{TS}$ found for the flare of 3C~279~\citep{TheFermi-LAT:2016dss}, we also demand that at least one GTI of each HOP group be detected with $\mathrm{TS} \geqslant 150$ in order to ensure enough statistics to search for variability on time scales of minutes.}
\tablecomments{The criteria are applied to all sources. Furthermore, if no interval fulfills the $F_\mathrm{BB} \geqslant F_\mathrm{max}$ criterion for the weekly or daily light curves, we include the HOP group of the maximum flux if that flux exceeds $5\times10^{-6}\,\mathrm{cm}^{-2}\,\mathrm{s}^{-1}$, i.e., we change $F_\mathrm{max}$ to the maximum value $F_\mathrm{BB}$.}
\end{deluxetable}

In a last step, we derive light curves on sub-GTI time scales. 
The time bin size is calculated from the adaptive binning method of \citet{lott2012}, where we choose bins of constant flux uncertainty of $\sim20\,\%$. 
In this step, we use the spacecraft information in time steps of 1\,s instead of 30\,s. Additionally, we compute the effective area in five bins of the azimuthal spacecraft coordinates because, on such short timescales, the exposure dependence on the azimuth should not be averaged over.\footnote{See \url{https://fermi.gsfc.nasa.gov/ssc/data/analysis/scitools/help/gtltcube.txt}}
We discuss these light curves in Section~\ref{sec:sub-gti}.

Light curves and fit results for the different time intervals and binnings are provided online.\footnote{
See \url{https://www-glast.stanford.edu/pub_data/1605/} and \url{https://zenodo.org/record/2598791}.
}

\section{Results for Global Light-curve Properties}
\label{sec:results-global}
We first present results derived from the weekly \gray light curves spanning the full 9.5\,yr time range, which we refer to as \emph{global light-curve properties}, before deriving results from the local light curves on GTI and sub-GTI time scales in Section~\ref{sec:results-local}.

\begin{figure}
    \includegraphics[width = .99\linewidth]{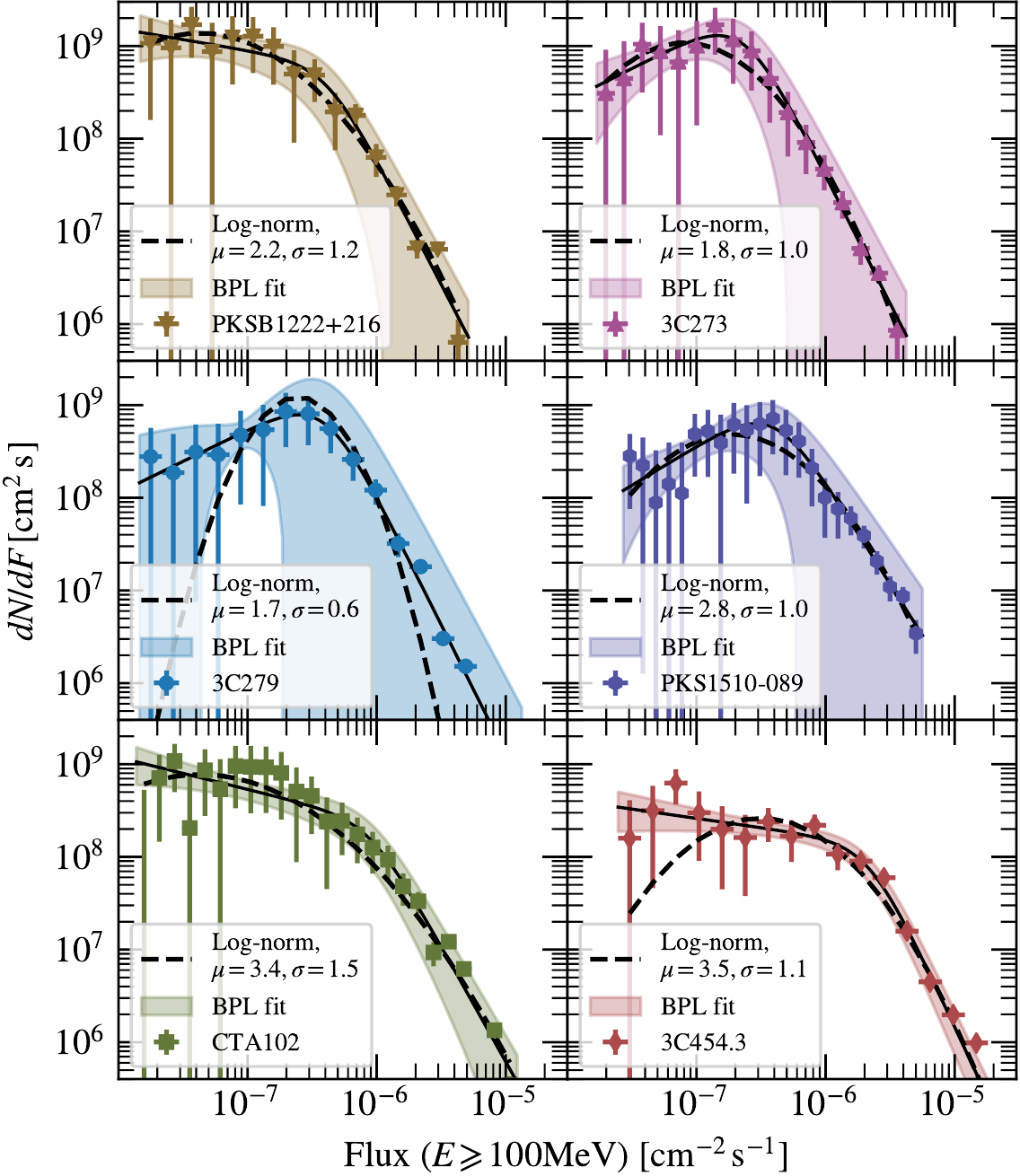}
    \caption{\label{fig:fluxpdf} Distribution of the fluxes of the weekly 9.5~yr \gray light curves. The BPL fit is shown as a black solid line with $1\,\sigma$ uncertainties (shaded bands). The black dashed line indicates a fit with a lognormal distribution with location and scale parameters $\mu$ and $\sigma$ as indicated in the legends.}
\end{figure}

From the weekly light curves in Figure~\ref{fig:weekly}, it is evident that the FSRQs show strong flares that exceed the average flux by a factor of a few, while the quiescent level is relatively stable. 
Such behavior is typical for FSRQs, and we 
further quantify it by calculating the flux distribution, $dN/dF$, of the weekly fluxes for bins with $\mathrm{TS} \geqslant 9$ and $F_i \geqslant \sigma_i$. 
The results are shown in Figure~\ref{fig:fluxpdf}. The flux bins are chosen according to the algorithm of~\citet{knuth2006}, and the error bars are calculated under the assumption that the observed weekly fluxes, $F_i$, $i = 1,\ldots,n$, are Gaussian-distributed numbers with a standard deviation equal to the measurement uncertainty $\sigma_i$.\footnote{
With this assumption, the uncertainty of finding $x$ entries in the $j$th flux bin of width $\Delta F_j = F_{\mathrm{hi},j} - F_{\mathrm{lo},j}$ is given by the sum of the Bernoulli probabilities $p_{ij}$, $\sum_{i = 1}^n p_{ij}(1-p_{ij})$, where $p_{ij} =  \left[\mathrm{erf}\left((F_{\mathrm{hi},j} - F_i) / \sqrt{2\sigma_i^2}\right) - \mathrm{erf}\left((F_{\mathrm{lo},j} - F_i) / \sqrt{2\sigma_i^2}\right)\right]/2$, and $\mathrm{erf}$ is the error function.
}
We fit the flux distribution with a smoothly broken power law (BPL) of the form 
\begin{equation}
    \frac{dN}{dF} = N_0 \left( \frac{F}{F_0}\right)^{\alpha_\mathrm{low}}
        \left[ 1 + \left(\frac{F}{F_\mathrm{br}}\right)^s \right]^{\frac{\alpha_\mathrm{high} - \alpha_\mathrm{low}}{s}},
        \label{eq:dndf}
\end{equation}
with the smoothing factor $s$ fixed to 3. 
The results of a $\chi^2$ minimization are summarized in Table~\ref{tab:global}.
Generally, below the break flux $F_\mathrm{br}$, the flux distribution is flat, $\alpha_\mathrm{low}\sim 0$.
Above $F_\mathrm{br}$, 
which lies between $\sim2\times10^{-7}$ and $\sim2\times10^{-6}\,\mathrm{cm}^{-2}\mathrm{s}^{-1}$, $dN/dF$ declines  with power-law indices $\alpha_\mathrm{high} \lesssim -2.2$, making the brightest flares rare events.
The power-law distribution of the occurrence of flares is a natural prediction of self-organized criticality, commonly observed in solar flares and also in blazars~\citep[see, e.g.,][and references therein]{2016SSRv..198...47A}.    
Furthermore, it is clear that the flux distribution is very different from Gaussian behavior but
compatible with a lognormal distribution (black dashed lines in Figure~\ref{fig:fluxpdf}). 
Lognormal flux distributions are commonly observed at \gray energies for blazars~\citep[e.g.,][]{2010A&A...524A..48T,2015ApJ...810...14A,2018arXiv180504675S}
and can be interpreted as evidence for a connection between the  
modulations in the accretion rate and the jet activity~\citep{2009A&A...503..797G}.

\subsection{Determination of the QB Level}
\label{sec:qb}
As mentioned in the Introduction, 
there is no rigorous way to distinguish the 
flux in flares from the flux 
in a {\em quiescent background} (QB).
It is not even guaranteed that
there is a corresponding astrophysical distinction.
Nevertheless, some progress can be made,
especially given the assumption that the QB is
truly constant.

The minimum flux observed over time 
might be taken as an estimate of the true QB,
but it is a rather a crude lower limit, biased downward 
because of the scatter due to observational errors.
On the other hand, long overlapping tails of flares 
could contribute an approximately constant flux level
yielding an upward bias to some QB estimates.
For the present, we assume that the overlap of flare tails
is negligible and propose a statistical
procedure addressing the observational error bias.

This algorithm is based on an approximate 
separation of the distribution function of the observed
fluxes into two components,
(a) a low end and (b) a high end, dominated by
the QB and the flares, respectively.
While not relying on (b) being devoid of any QB flux,
we do assume that (a) contains almost entirely QB flux.
In the picture outlined above, this amounts to the
assumption that the flares are isolated (no overlap) 
and the intervening intervals are essentially pure QB.
The algorithm implements this separation
using only flux values with no regard to 
their time sequence.  It is based on finding the
flux interval that maximizes the symmetry of 
the resulting distribution.

In the following pseudo-code, the term ``distribution''
refers to the distribution of low values (a) in a putative
flux interval defining the QB.
\begin{enumerate}
\item Sort the $N$ flux values (with $\mathrm{TS} \geqslant 9)$ in increasing order:
$F_{1} \leqslant F_{2} \leqslant \dots F_{N-1}   \leqslant F_{N} $
\item Define a search grid of flux values $\phi_{m}$, $m = 1,\ldots,M$, 
to serve as candidates for the peak of the distribution. We take these 
values on an evenly spaced grid between the minimum and mean flux $\bar{F}$:
$\phi_m \in [F_1 + \epsilon (\bar{F} - F_1); \bar{F}]$.
By definition, the QB is very unlikely to be less than the
minimum flux or exceed the mean.
The factor $\epsilon$ is a small number to  avoid edge effects 
near the generally sparse lower end of the flux distribution.
\item For each $m$, define two intervals  of equal length straddling  $\phi_m$: 
    \begin{enumerate}
    \item LOW: from $F_{1}$ to $\phi_m$; and
    \item HIGH: from $\phi_m$ to $\phi_m + (\phi_m - F_{1})$.
    \end{enumerate}

\item Construct fine grids of flux values spanning these two intervals.
\item Compute the cumulative distributions (CDFs) of the
corresponding flux values.
\item Reverse the  CDF for the HIGH interval .
\item Normalize both CDFs to unity at the peak $\phi_m$
\item Compute the ratio of the posterior probabilities that the CDFs come from different distributions or the same distribution,  using the algorithm of \citet{Wolpert1996}.
\item Find the value of $\phi_m$ that minimizes this ratio, which measures the asymmetry of the total CDF of HIGH and LOW.
\item Report the median flux, $F_\mathrm{QB}$, in this
optimally symmetric distribution as the QB estimate.
\end{enumerate}
\noindent
In the limit of moderately finely gridded bins 
used in step 5, the CDF estimate is effectively
unbinned (little or no dependence
on the binning).

We show our estimate for the QB level in Figure~\ref{fig:weekly} as black dashed lines and report the values in Table~\ref{tab:global}.
In general, these $F_\mathrm{QB}$ values match 
the visual flux baselines extremely well. 
In the case of 3C~454.3, it is slightly higher than the minimum flux level observed for this source around MJD 55,800-56,200. The reason is our applied $\mathrm{TS}$ cut and a contamination of $F_\mathrm{QB}$ from the tails of the flares. 
We have tested the latter point with simulations drawing random numbers from a uniform distribution and Cauchy distributions to emulate flares. 
For the Cauchy distributions, we randomize the height, width, and position. 
Applying our algorithm to these simulated histograms, we find that $F_\mathrm{QB}$ slightly overestimates the true uniform background. Therefore, we conclude that $F_\mathrm{QB}$ can be seen as a firm upper limit on a truly constant QB level. 
We also note that using the peak or mean of the distribution in step 10, instead of the median,  only changes the results marginally. 
We have furthermore tested different metrics instead of the algorithm of \citet{Wolpert1996}, namely, the Kolmogorov-Smirnov (K-S) test and the least squares between the CDFs. 
We again find comparable results. However, the K-S test estimates underpredict the true QB level in simulations, while the least squares give estimates that are too high  (higher than the Wolpert estimate).
We also note that the $F_\mathrm{QB}$ values are either consistent with or lower than the break flux of the BPL fit, $F_\mathrm{br}$.
This is expected, as $F_\mathrm{QB}$ marks the median of the QB while $F_\mathrm{br}$ probably indicates the transition from the QB to the flaring states. 

\subsection{Determination of the PSD}
We further characterize the global \gray light curves in terms of their PSD,
which usually can be described with simple power laws in frequency,  $\mathrm{PSD} \propto 1 / \nu^\beta$.
An analysis by \citet{2010ApJ...722..520A} of the first 11~months of LAT data from 106 blazars revealed that 
these objects have $\beta$ values between 1 and 2, the intermediate regime between flicker noise ($\beta = 1$) and Brownian motion ($\beta = 2)$. 
In addition to the noise behavior, 
we will use the derived PSDs in Section~\ref{sec:gammaradio} to simulate \gray light curves in order to calculate the significance of a correlation between radio and \gray emission. 

The best-fit PSDs are estimated from the periodograms and simulated light curves following the method described in detail in \citet{2014MNRAS.445..437M} and \citet{2013MNRAS.433..907E} and summarized briefly below.
The observed periodograms $P(\nu)$ as a function of frequency (inverse time) $\nu$ are calculated from the absolute square of the Fourier transformation of the light curve (Eq.~3 in \citealt{2014MNRAS.445..437M}). 
We include all data points detected with $\mathrm{TS} \geqslant 9$ and perform a linear interpolation between gaps in the light curve to guarantee an even sampling. 
Since we are using weekly binned light curves and bright FSRQs, the gaps are small and at most six consecutive data points long (42\,days) in the case of PKS~B1222+216. The number of nondetected bins is less than $\sim13\,\%$ for all sources.
In contrast to~\citet{2014MNRAS.445..437M}, we do not apply a window function (see the discussion below). 

We compare the observed periodogram with simulated light curves, which we produce with the method of \citet{1995A&A...300..707T} for power-law PSDs with values $0 \leqslant \beta \leqslant 3$ in steps of $\Delta\beta = 0.05$. For each $\beta$ value, 100 light curves are generated, each one a 100 times longer than the actual observation to account for possible red-noise leakage. Splitting the simulated light curves (without overlap) leaves us with $10^4$ realizations. 
The light curves are initially produced with a regular time binning equal to 0.7\,days and  rebinned into 7~day light curves through averaging. 
The same observational gaps and interpolation as in the observed light curves are applied. 
The periodograms are then calculated for the simulated light curve in the same way as for the observed one.

To fix the normalization of the PSD model, \citet{2014MNRAS.445..437M} suggested variance matching; i.e., they rescaled the simulated flux data points with a factor $A^{-1}$, where $A^2 = \sigma_\mathrm{sim}^2 / (\sigma_\mathrm{obs}^2 - \bar{\sigma_i^2})$, with $\sigma_\mathrm{sim}^2$ ($\sigma_\mathrm{data}^2$) the variance of the simulated (observed) light curve and $\bar{\sigma_i^2}$ the variance of the observational noise.
For the \gray light curves, we choose to follow \citet{2013MNRAS.433..907E} instead and iteratively match the probability distribution of the simulated fluxes to the observed ones, given by the $dN/dF$ distributions shown in Figure~\ref{fig:fluxpdf}. 
The reason is that the algorithm of \citet{1995A&A...300..707T}, which is used by \citet{2014MNRAS.445..437M}, produces light curves with Gaussian-distributed fluxes, which is clearly not the case at \gray energies.\footnote{Furthermore, the variance matching relies on Parseval's theorem, from which it follows that the light-curve variance is equal to the integrated PSD. However, Parseval's theorem is only valid for square-integrable functions, i.e. $\beta \geqslant 2$, and thus not strictly applicable for smaller values of $\beta$ commonly observed at \gray energies.}
In a final step, we add uncertainties to the light curves by randomly drawing with replacements from the observed uncertainties $\sigma_i$ and adding a Gaussian random number $\mathcal{N}(0,\sigma_i)$ to the simulated flux values.

\begin{figure*}
    \centering
    \includegraphics[width = 0.4\linewidth]{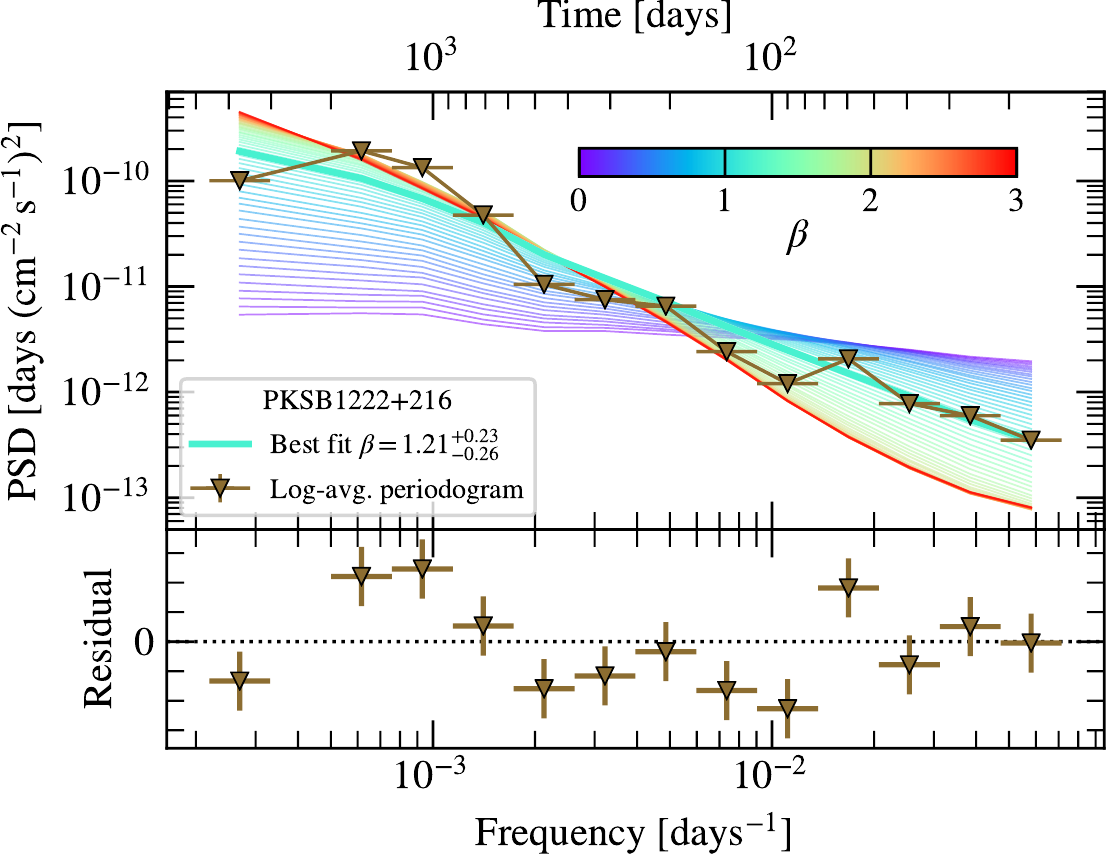}
    \includegraphics[width = 0.4\linewidth]{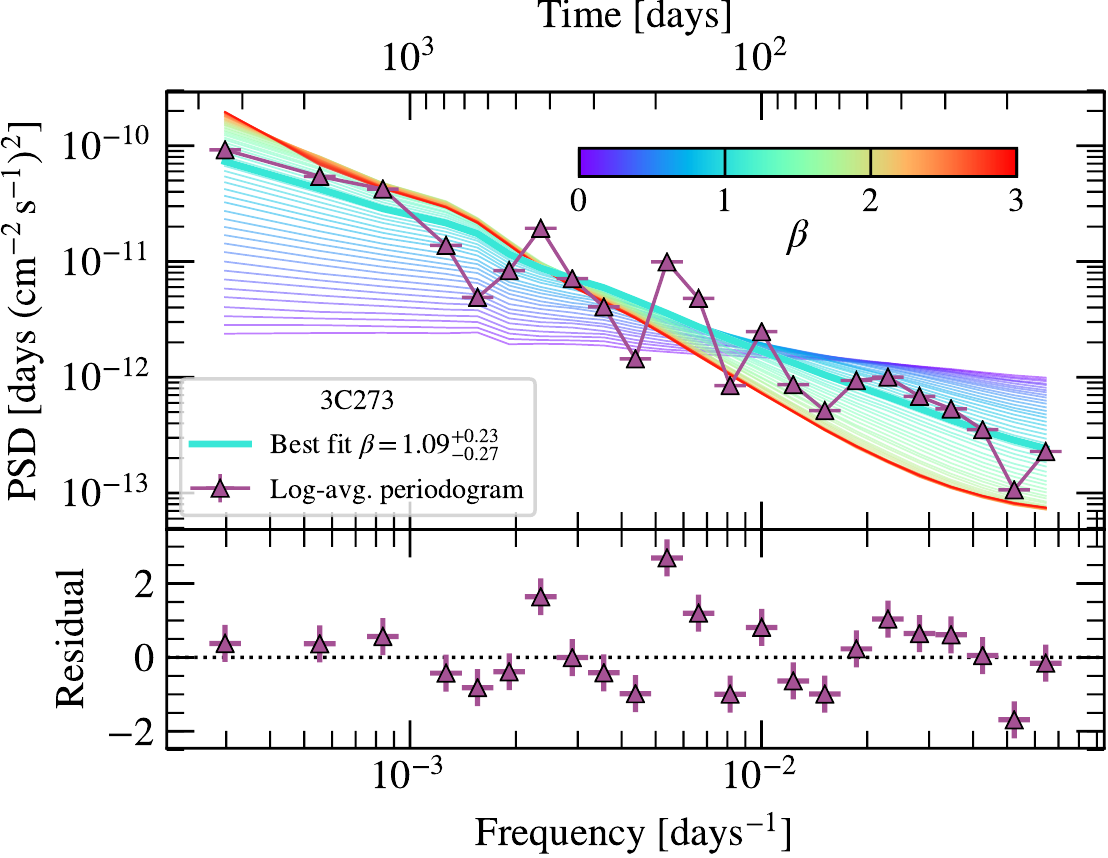}
    \includegraphics[width = 0.4\linewidth]{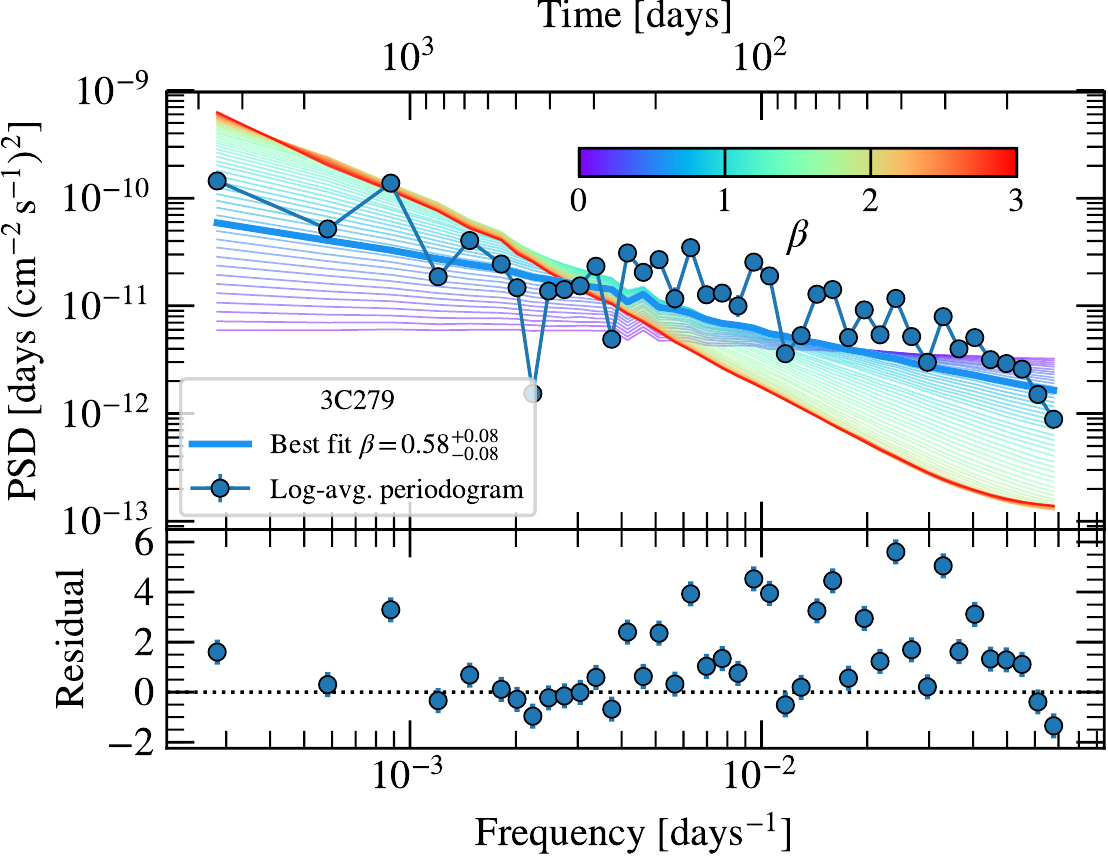}
    \includegraphics[width = 0.4\linewidth]{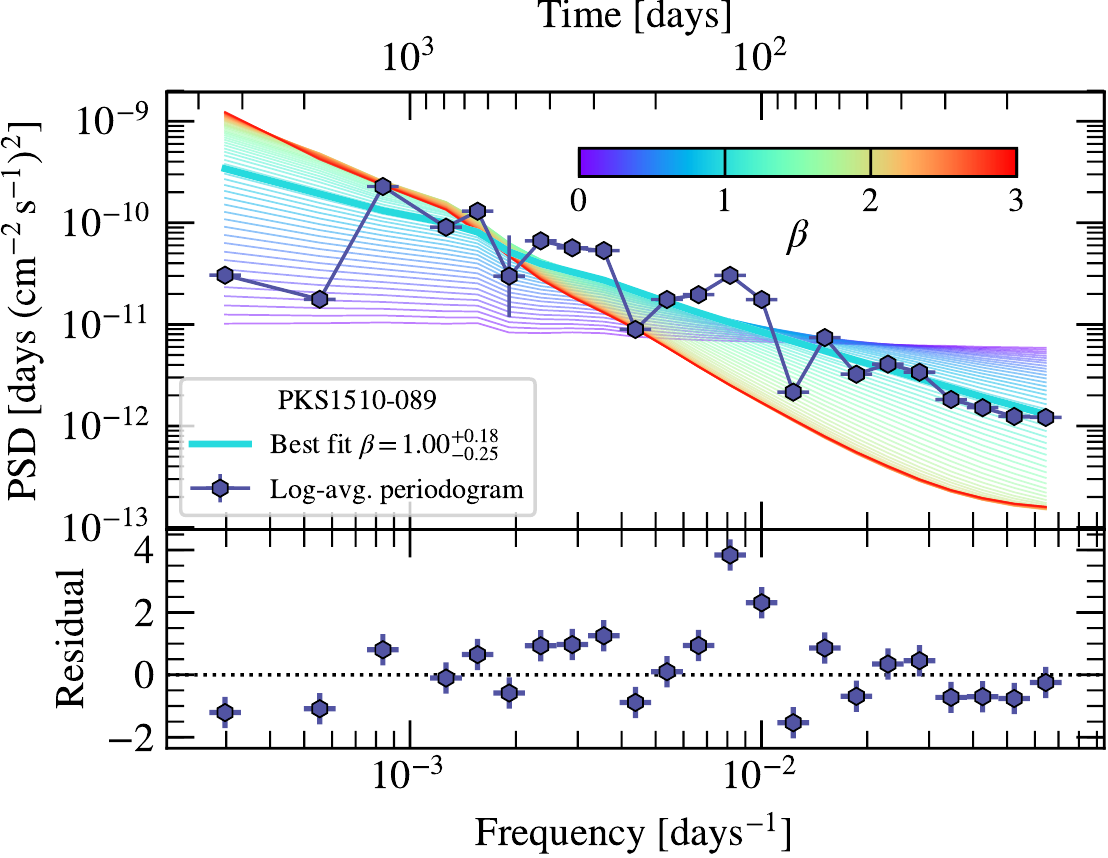}
    \includegraphics[width = 0.4\linewidth]{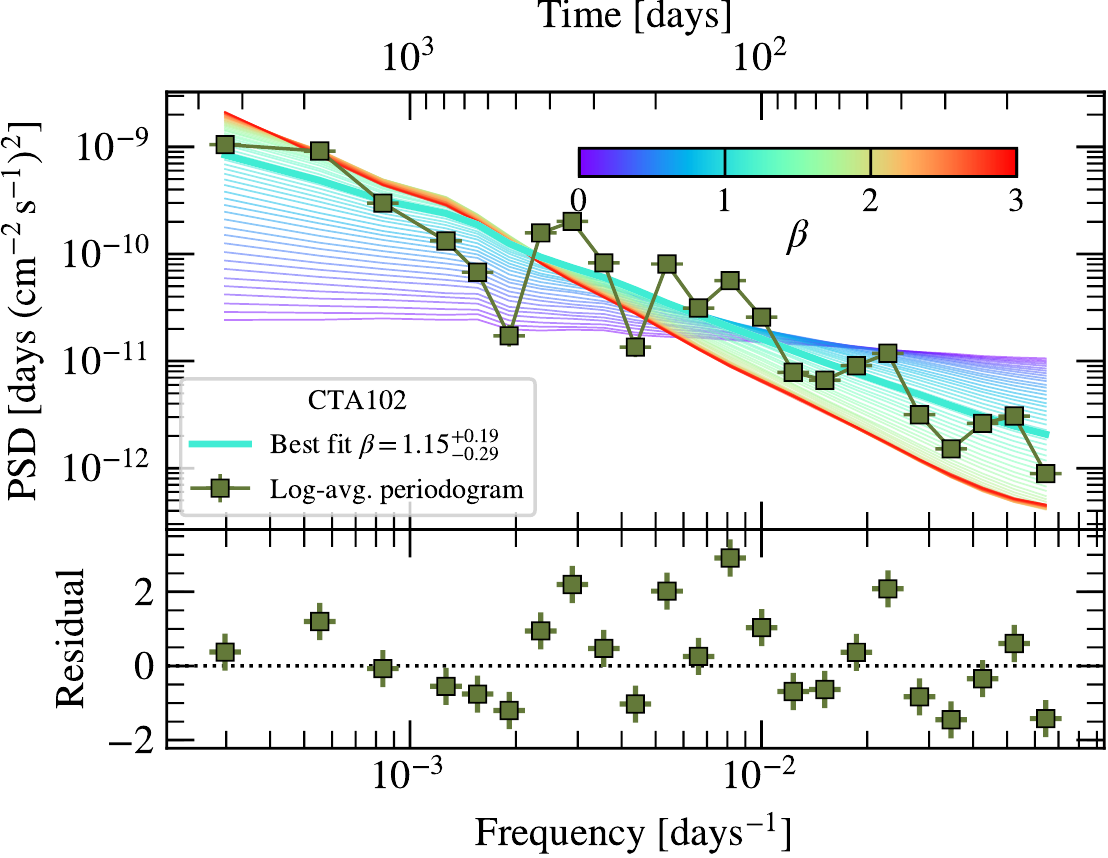}
    \includegraphics[width = 0.4\linewidth]{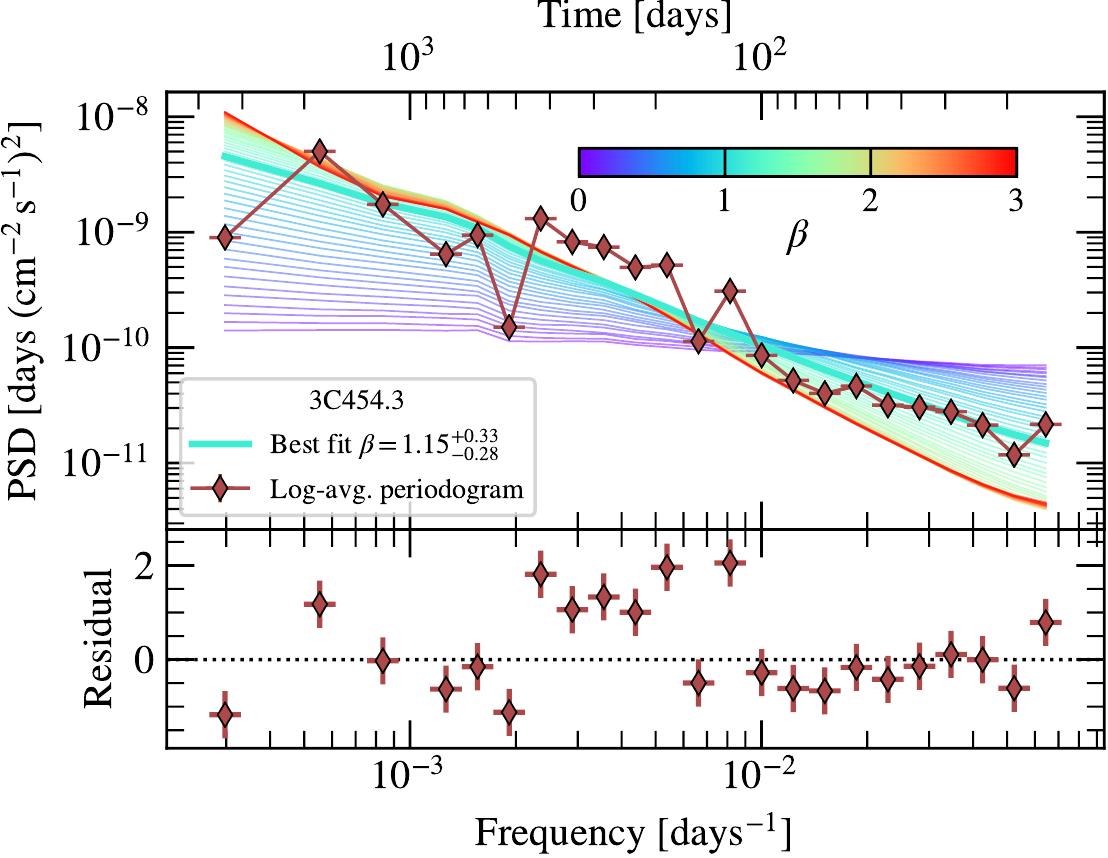}
    \caption{Periodograms of the observed (symbols) and simulated light curves (colored lines). The simulated periodograms follow power-law PSDs between $\beta = 0$ (purple line) and 3 (red line) in steps of $\Delta\beta = 0.05$. The bottom panels show the residuals with respect to the best fit, which is indicated in the legend and as a thick solid line in the top panels.}
    \label{fig:periodograms}
\end{figure*}

The peridograms of the observed and simulated light curves, ${P}_\mathrm{obs}$ and ${P}_\mathrm{sim}$, are averaged in
31 logarithmic bins\footnote{
The minimum and maximum frequencies are chosen such that the broadest possible frequency range is covered. We have  explicitly checked that changing the number of bins does not significantly change the results.
\citep{1993MNRAS.261..612P} between $\nu_\mathrm{min} = 1 / T$ (where $T$ is the full duration of the light curve with $N$ measurements) and $\nu_\mathrm{max} = N  / (2T)$ (the Nyquist frequency)
}
and compared by means of a $\chi^2$ test~\citep{2014MNRAS.445..437M},
\begin{equation}
    \chi^2(\beta) = \sum_{\nu_\mathrm{min}}^{\nu_\mathrm{max}}\frac{(P_\mathrm{obs}(\nu) - \overline{P}_\mathrm{sim}(\nu,\beta))^2}{\Delta\overline{P}_\mathrm{sim}(\nu,\beta)^2}.\label{eq:chi2psd}
\end{equation}
Here $\overline{P}_\mathrm{sim}(\nu)$ and $\Delta\overline{P}_\mathrm{sim}(\nu)^2$ are the mean and variance of the periodograms obtained from the simulated light curves.
The averaged periodograms of the simulated light curves and the observed ones are shown in Figure~\ref{fig:periodograms} and the best-fit average periodogram is shown as a thick solid line. 
The quality of the the best-fit value $\hat\beta$ with a corresponding minimum $\chi^2$ value $\hat\chi^2\equiv\chi^2(\hat\beta)$, is evaluated from the light curves simulated with $\beta = \hat\beta$ in the following way. We form the distributions of simulated $\chi^2$ values, $\chi^2_\mathrm{sim}$, by replacing $P_\mathrm{obs}(\nu)$ with $P_\mathrm{sim}(\nu,\beta)$ in Eq.~(\ref{eq:chi2psd}),
\begin{equation}
    \chi^2_\mathrm{sim}(\beta,\beta') = \sum_{\nu_\mathrm{min}}^{\nu_\mathrm{max}}\frac{(P_\mathrm{sim}(\nu,\beta) - \overline{P}_\mathrm{sim}(\nu,\beta'))^2}{\Delta\overline{P}_\mathrm{sim}(\nu,\beta')^2},\label{eq:chi2psd_sim}
\end{equation}
and calculate the $p$-value, $p_\beta$, as the  fraction of simulations that result in $\chi^2_\mathrm{sim}(\hat\beta,\hat\beta) > \hat\chi^2$.\footnote{From the $10^4$ simulations, we effectively derive a histogram of the $10^4$ $\chi^2_\mathrm{sim}(\hat\beta,\hat\beta)$ values from which we then determine $p_\beta$.}
The confidence interval for $\hat\beta$ is derived by determining the $\Delta\chi^2_\mathrm{sim}(\hat{\beta},\beta)$ value from simulations such that 95\,\% of the time, the simulated (true) $\beta$ value is contained within $\Delta\chi^2$.\footnote{Put differently, the $10^4$ simulated light curves provide $10^4$ $\chi^2$ curves as functions of $\beta$ from which we calculate the value of $\Delta\chi^2$ such that the true (simulation-input) $\beta$ value is contained within that interval in 95\,\% of the time.}
The same $\Delta\chi^2$ value is then applied to the observed $\chi^2$ curve.
The results of our PSD analysis are summarized in Table~\ref{tab:global} where we also report the value of $\beta$ obtained from a linear regression in log-log space. 
In general, the periodograms are well fit by our method, as indicated by the $p_\beta$-values and observed in Figure~\ref{fig:periodograms}. 
The only exception is 3C~279 where only two of the $10^4$ simulated light curves result in a $\chi^2_\mathrm{sim}(\hat\beta,\hat\beta) > \hat\chi^2$.
The steep $\chi^2$ curve for this source also explains the small error bars on the reconstructed value of $\beta$.
The reason might be
the variation of the underlying PSD with time, as found by \citet{TheFermi-LAT:2016dss}. Another possibility is
the specific 7 day binning we have chosen here.

Our results are compatible with the PSD slopes found by \citet{2013ApJ...773..177N} at the 1\,$\sigma$-2\,$\sigma$ level but less so for the PSD power laws obtained by \citet{2014ApJ...786..143S}. 
The reason for the discrepancy with the latter analysis might be due to the different binning schemes and time ranges (\citealt{2014ApJ...786..143S} used an adaptive binning for 4~yr of LAT data instead of a constant binning and 9.5~yr used here).

\begin{deluxetable*}{l|ccc|c|ccc}
\tablewidth{0pt}
\tablecaption{ \label{tab:global}Global \gray light-curve properties.}
\tablehead{\colhead{Source name} & 
\colhead{$\alpha_\mathrm{low}$} &
\colhead{$\alpha_\mathrm{high}$} & 
\colhead{$F_\mathrm{br} [10^{-6}\mathrm{cm}^{-2}\mathrm{s}^{-1}]$} &
\colhead{$F_\mathrm{QB} [10^{-7}\mathrm{cm}^{-2}\mathrm{s}^{-1}]$} &
\colhead{$\beta_\mathrm{slope}$} &
\colhead{$\hat\beta$} &
\colhead{$p_\beta$}
}
\startdata
PKS~B1222+216 & $-0.24^{+0.41}_{-0.27}$ & $-2.70^{+0.33}_{-0.43}$ & $0.42^{+0.28}_{-0.15}$ & 1.49 & 1.23 & $1.12^{+0.21}_{-0.26}$ & 0.423 \\
3C~273 & $0.70^{+0.40}_{-0.54}$ & $-2.77^{+0.24}_{-0.27}$ & $0.23^{+0.07}_{-0.07}$  & 1.87 & 1.14 & $1.09^{+0.24}_{-0.27}$ & 0.330 \\
3C~279 & $0.68^{+0.27}_{-0.40}$ & $-2.80^{+0.21}_{-0.23}$ & $0.40^{+0.10}_{-0.10}$ & 3.23 & 0.67 & $0.61^{+0.10}_{-0.07}$ & $2\times10^{-4}$ \\
PKS~1510-089 & $0.84^{+0.72}_{-0.48}$ & $-2.21^{+0.23}_{-0.26}$ & $0.40^{+0.16}_{0.12}$ & 4.19 & 0.88 & $1.00^{+0.18}_{-0.25}$ & 0.129 \\
CTA~102 & $-0.35^{+0.22}_{-0.18}$ & $-2.50^{+0.16}_{-0.19}$ & $0.90^{+0.38}_{-0.26}$ & 2.49 & 1.21 & $1.18^{+0.16}_{-0.32}$ & 0.138 \\
3C~454.3 & $-0.20^{+0.17}_{-0.13}$ & $-3.00^{+0.13}_{-0.14}$ & $2.19^{+0.39}_{-0.32}$ & 8.59 & 1.05 & $1.15^{+0.32}_{-0.28}$ & 0.274 \\
\enddata
{
\tablecomments{Columns 2-4 indicate the best-fit values for the BPL fit (Equation.~(\ref{eq:dndf})) to the $dN/dF$ distributions, column 5 reports our estimate of the QB flux, and columns 6-8 show the best-fit results for the PSD. Here $\beta_\mathrm{slope}$ gives the result for a linear regression of the periodograms, and $\hat\beta$ is the best-fit value of the $\chi^2$ minimization with corresponding $p_\beta$-value. The interval around $\hat\beta$ is at 95\,\% confidence.}
}
\end{deluxetable*}

\section{Results for local light-curve properties}
\label{sec:results-local}

\begin{figure*}
    \centering
    \includegraphics[width = .99\linewidth]{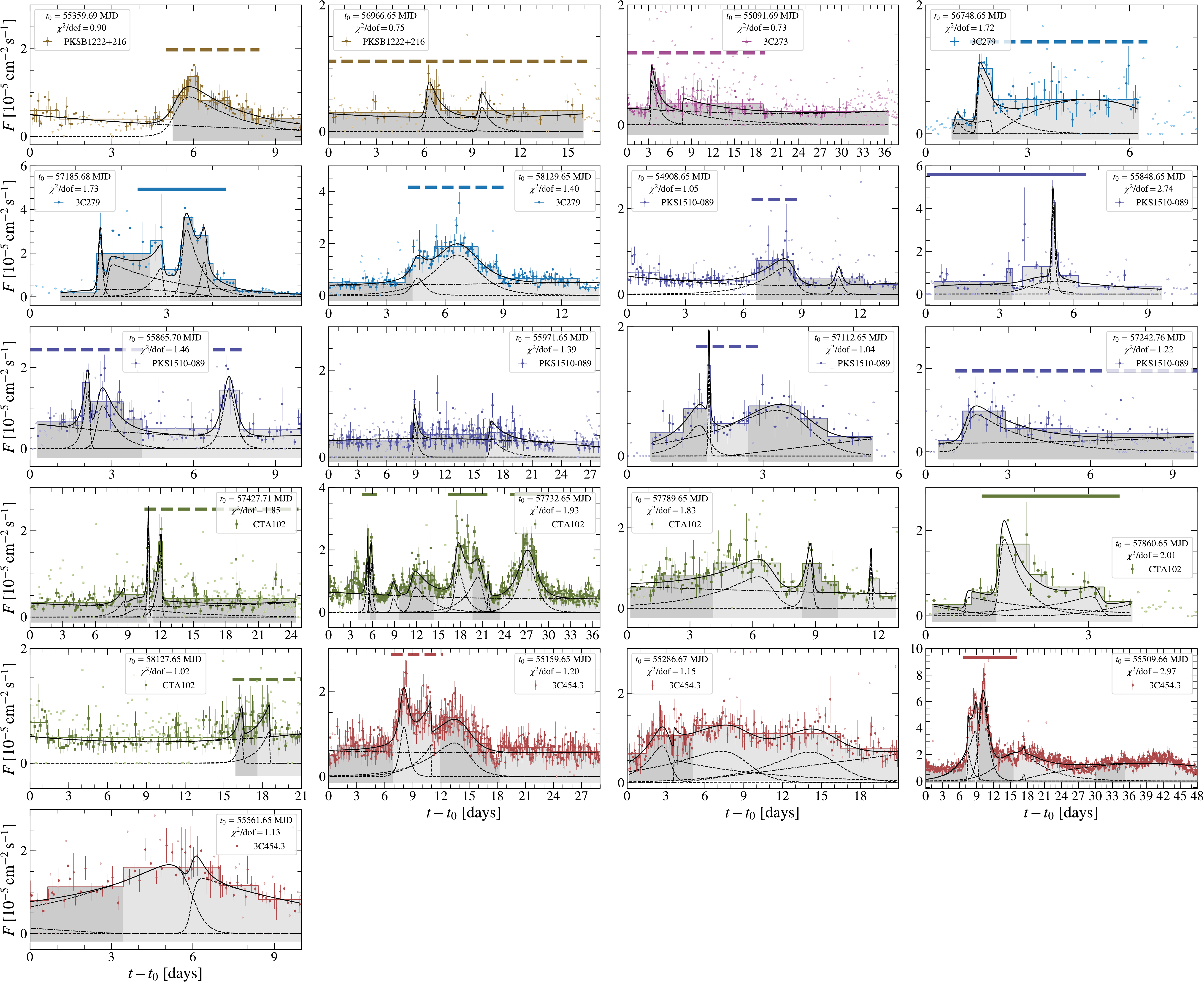}
    \caption{ Light curves with one bin per GTI for the selected time ranges (solid horizontal lines in Figure~\ref{fig:daily}). Solid and dashed black lines show fits to the light curve with exponential flare profiles discussed in Section~\ref{sec:results-local}. Other symbols and lines are the same as in Figure~\ref{fig:weekly} and ~\ref{fig:daily}.
    The sub-GTI light curves are derived for the time intervals indicated as solid horizontal lines, where at least one orbital bin shows $\mathrm{TS}\geqslant 150$ (see Section~\ref{sec:sub-gti}). 
    Average spectra to search for a cutoff due to \gray absorption in the BLR are derived from the time intervals indicated as either solid or dashed horizontal lines (see Section~\ref{sec:blrabs}).
    }
    \label{fig:gti}
\end{figure*}

We proceed with deriving local properties of the \gray flares, focusing first on the light curves with one bin per GTI that are shown in Figure~\ref{fig:gti}.
The average best-fit spectral parameters for the entire time spans of the daily, orbital, and suborbital light curves (the time ranges for the suborbital light curves are indicated as solid horizontal bars in Figure~\ref{fig:gti}) are summarized in Table~\ref{tab:avg-spec} in Appendix~\ref{sec:avg-spec}. 

\subsection{Flare profiles and asymmetry}
\label{sec:flare-profile}
We again use BBs and HOP groups to identify different states in 
the orbital light curves (see Figure~\ref{fig:gti}). 
To assess the time profile of the flares,
we fit each HOP group $i$ with a sum of exponential profiles, $F_{\mathrm{flare},i}$,
using a $\chi^2$ minimization, and
\begin{eqnarray}
    F_{\mathrm{flare},i}(t) &=& 
    \sum\limits_{j = 1}^{N_i} F_{0,ij}\nonumber\\
    &\times&\left[\exp\left(\frac{t - t_{0,ij}}{\tau_{\mathrm{rise},ij}}\right) + \exp
    \left(\frac{t_{0,ij} - t}{\tau_{\mathrm{decay},ij}}\right)\right]^{-1}\!\!\!,
    \label{eq:flareHOP}
\end{eqnarray}
where $t_{0,ij}$ are the times when the flare flux is equal to $F_{0,ij} / 2$, and $\tau_{\mathrm{rise},ij}$ and $\tau_{\mathrm{decay},ij}$ are the flare rise and decay times, respectively.
All light-curve points are included that fulfill $\mathrm{TS}\geqslant9$ and $F_i \geqslant 3\sigma_i/2 $.
The number of flare profiles per HOP group, $N_i$, is either one or two and determined during the fit using the Bayesian information criterion (BIC), defined as $\mathrm{BIC} = n_\mathrm{par}\ln(n) + \chi^2$, where $n_\mathrm{par}$ is the number of fit parameters ($n_\mathrm{par} = 4$ for $N_i = 1$), and $n$ is the number of data points within one HOP group $i$. 
Two flare profiles are selected if the difference between the two BIC values is $\Delta\mathrm{BIC} = \mathrm{BIC}(N_i = 2) - \mathrm{BIC}(N_i = 1) < 0$.
The reason for allowing $N_i > 1$ is that the flare profile in Eq.~\ref{eq:flareHOP} does not capture the long-lasting plateaus of a flare (see, e.g., all flares of 3C~279 or the last panel with the flare of 3C~454.3 starting at 55,551.65\,MJD in Figure~\ref{fig:gti}).

After each HOP group is fitted individually and $N_i$ is determined, 
we refit the entire light curve, which consists of $N_\mathrm{HOP}$ groups, with the function 
\begin{equation}
    F_\mathrm{flare}(t) = \sum\limits_{i = 1}^{N_\mathrm{HOP}}F_{\mathrm{flare},i}(t) + F_\mathrm{bkg}(t),
\end{equation}
where $F_\mathrm{bkg}(t)$ is an order 2 polynomial to describe a slow  varying background.
The fit results are shown as black solid lines in Figure~\ref{fig:gti}.
In general, the $\chi^2$ values divided by the degrees of freedom (dof) are between 1 and 2 (see the legends in Figure~\ref{fig:gti}). Given the large values of dof, the fit qualities are generally poor. This is not unexpected, as we only allow up to two flare profiles per HOP group and no arbitrary functions. Already with this choice, there are probably some spurious flares identified, see, e.g., the second flare profile in the first PKS~1510-089 flare (starting at 54,908.65\,MJD). 
Nevertheless, the overall light-curve evolution is well captured, which allows us to describe the local flare properties from the ensembles of flare profiles. 

\begin{figure}
    \centering
    \includegraphics[width = .9\linewidth]{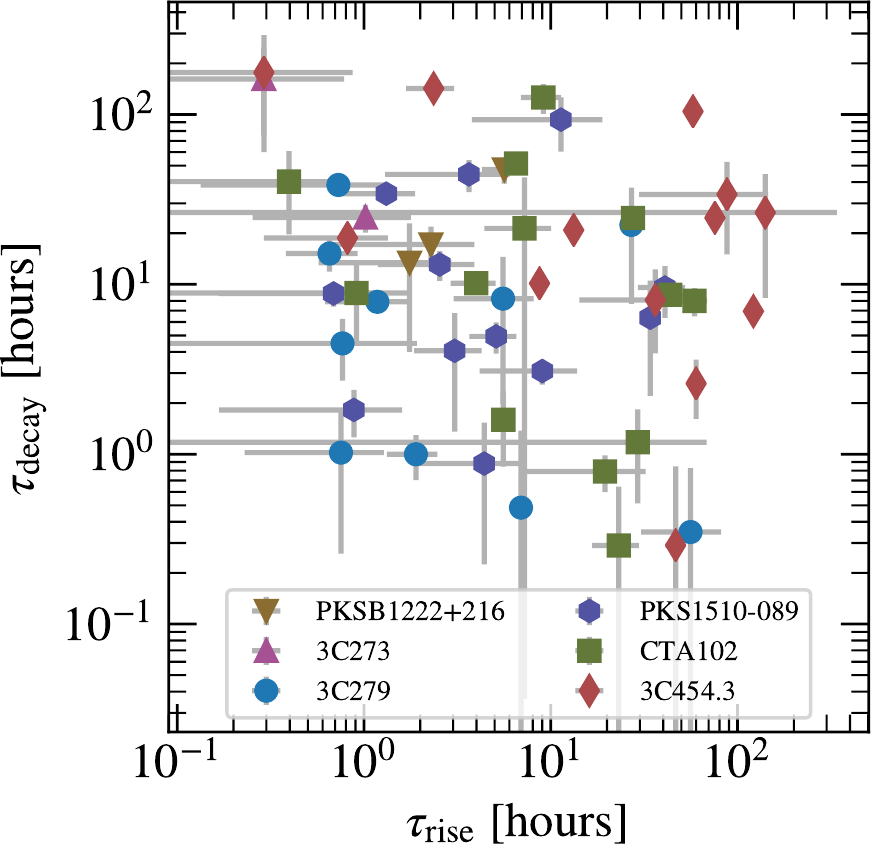}
    \caption{
    Distribution of rise and decay times for the flare profiles fitted to the data. 
    }
    \label{fig:times-hist}
\end{figure}

We show the distribution of rise and decay times in  Figure~\ref{fig:times-hist} for flares with a time-integrated flux $>10^{-7}\,\mathrm{cm}^{-2}$. 
Remarkably, all sources show values of $\tau_\mathrm{rise}$ and $\tau_\mathrm{decay}$ that are close or below the horizon-crossing time scale of the central supermassive black hole,
$t_g = r_g / c$, where $r_g = 2 G M_\bullet / c^2 $ is the gravitational radius, $G$ is the gravitational constant, $M_\bullet$ is the black hole mass (taken from Table~\ref{tab:src-select}), and $c$ is the speed of light.
This results in values between $\sim0.5$ and $\sim2.5$~hr using the black hole masses listed in Table~\ref{tab:src-select}.

\begin{figure*}
    \centering
    \includegraphics[width = .8\linewidth]{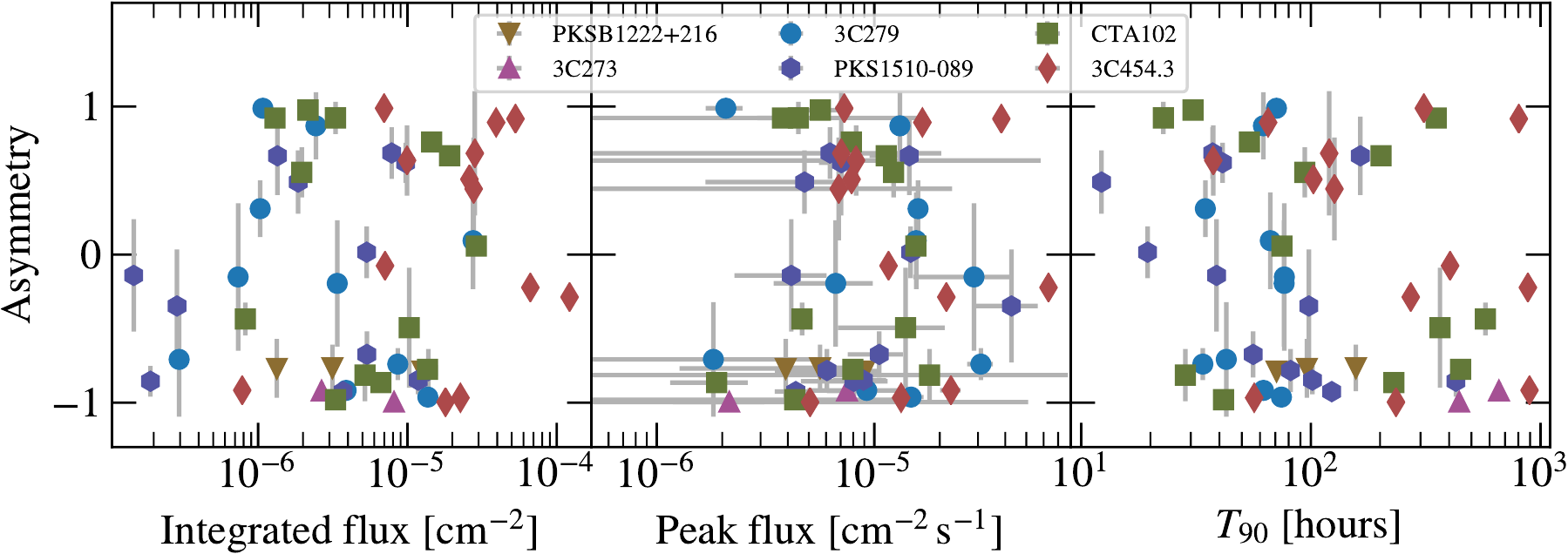}
    \caption{Flare asymmetry versus integrated flux (left), peak flux (middle), and flare duration $T_{90}$ (right) for the fitted flare profiles shown in Figure~\ref{fig:gti}.}
    \label{fig:asym}
\end{figure*}

From the rise and decay times, we can calculate the flare asymmetry as
\begin{equation}
    A = \frac{\tau_\mathrm{rise}-\tau_\mathrm{decay}}
    {\tau_\mathrm{rise}+\tau_\mathrm{decay}},
\end{equation}
so that $A < 0$ for fast-rise exponential-decay (FRED) type flares, 
as expected from an injection of energetic particles on time scales faster than the subsequent cooling through radiative processes such as inverse Compton (IC) scattering or synchrotron emission.
The asymmetry is shown versus integrated flux, peak flux, and flare duration $T_{90}$, defined as the time around the flare peak that contains 90\,\% of the integrated flux, in Figure~\ref{fig:asym}.
The peak flux for each flare of each HOP group is derived from the maximum of  Eq.~\ref{eq:flareHOP} with respect to time (suppressing indices $ij$),

\begin{equation}
    F_{\mathrm{peak}} = \frac{F_{0} \tau_\mathrm{rise}}{\tau_\mathrm{rise} + \tau_\mathrm{decay}}\left(\frac{\tau_\mathrm{decay}}{\tau_\mathrm{rise}}\right)^{\frac{\tau_\mathrm{decay}}{\tau_\mathrm{rise} + \tau_\mathrm{decay}}}.
\end{equation}
The error bars on the peak flux and asymmetry are derived from standard Gaussian error propagation from the fit uncertainties.

The median of the asymmetry is found to be $-0.195$, i.e., FRED-type flares are more common than the opposite. In general, the flares show a versatile behaviour and no clear trends are seen from Figure~\ref{fig:asym}. This is also reflected in the fact that we do not find any significant correlation between the asymmetry, integrated and peak flux, rise and decay times, as well as flare duration using Kendall's $\tau$.

We also investigate whether subsequent flares in each panel of Figure~\ref{fig:gti} show a trend with time in peak flux, asymmetry, or duration. 
For consecutive flares, we calculate the difference between, e.g., the peak fluxes, and calculate the $p$-value of a binomial distribution assuming an equal probability of finding negative and positive differences.
For 32 values of differences the $p$-values for the peak flux, asymmetry, as well as for the flare duration are close to 0.1 (14, 13, and 13 positive values for 32 trials, respectively) indicating no particular evolution of these quantities with time. 
As a systematic check, we have repeated the entire profile fit for time reversed versions of the light curves. In general we find good agreement between the fitted profiles. Correcting for the sign reversal, the median of the asymmetry changes from -0.195 to -0.198, suggesting that the systematic error is of the order of 2\,\%. 

\begin{figure*}
    \centering
    \includegraphics[width = .9 \linewidth]{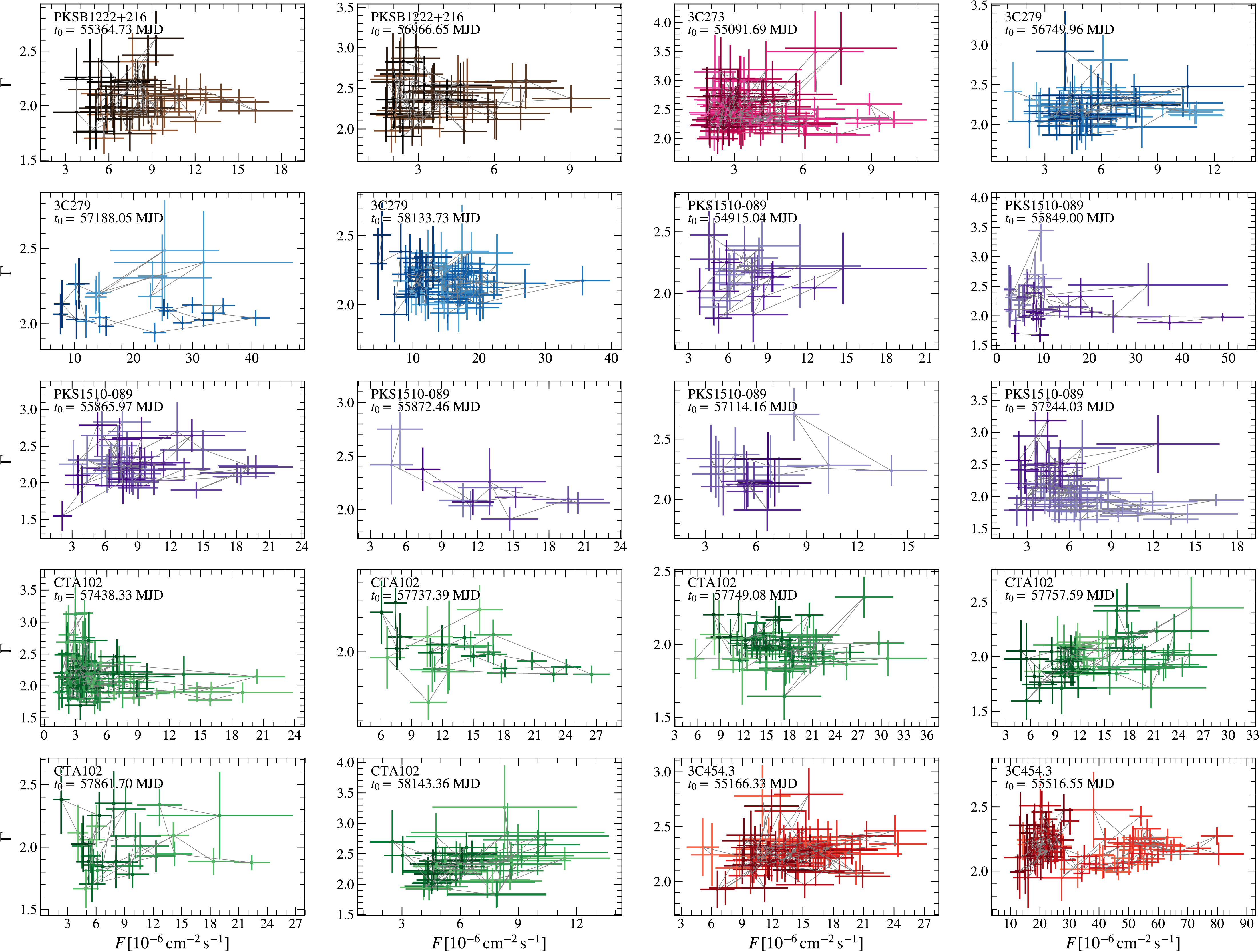}
    \caption{Spectral evolution of the flux $F$ and power-law index $\Gamma$ during the brightest flares on orbital time scales. Light colors refer to earlier times and dark colors to later times.}
    \label{fig:specvar}
\end{figure*}

We also find complex behavior of the spectral evolution during the flares. Evidence for a  ``harder-when-brighter'' trend is found for some sources and flares, which is however not significant. 
We therefore cannot draw any firm conclusions from the spectral evolution, which we show 
in Figure~\ref{fig:specvar}. 

\subsection{Sub-GTI light curves}
\label{sec:sub-gti}
We search for sub-orbital variability in a subset of orbital light curves, namely in those where at least one orbital bin is detected with $\mathrm{TS} \geqslant 150$. 
In this way we ensure high photon statistics and reduce the number of trials when searching for minute-scale variability (for comparison, the orbital light curve bin for which \citet{TheFermi-LAT:2016dss} measured minute scale variability in 3C~279 is detected in our analysis with $\mathrm{TS} \sim 400$). 
The selected time regions are indicated with solid horizontal lines regions in Figure~\ref{fig:gti}, whereas the dashed lines show the time intervals selected with the criteria in Table~\ref{tab:zoom} that do not pass the additional $\mathrm{TS}$ cut.

\begin{figure*}
    \centering
    \includegraphics[width = .9\linewidth]{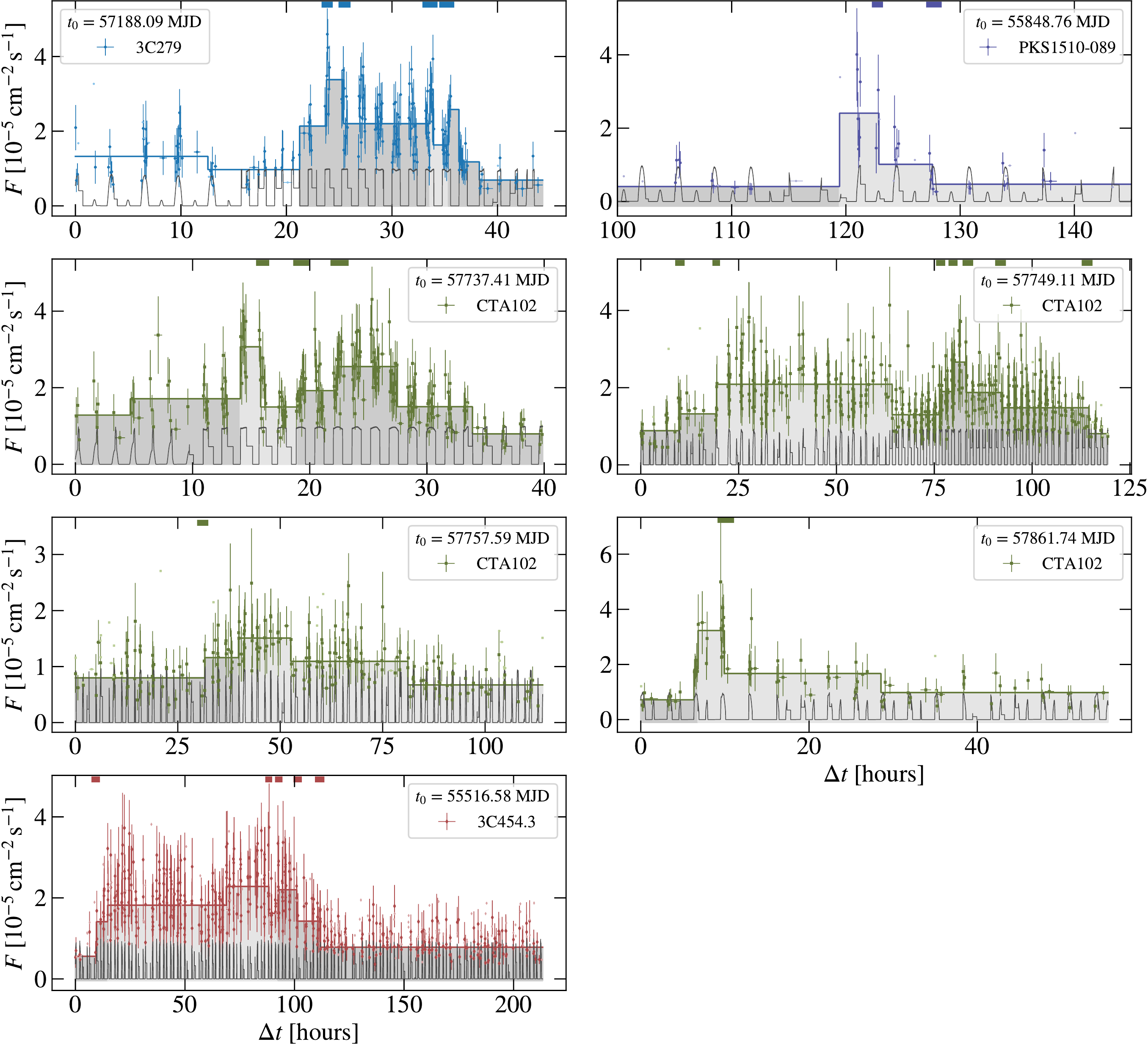}
    \caption{Sub-GTI light curves for the HOP groups of the orbital light curves that contain one bin with $\mathrm{TS} \geqslant 150$. The solid horizontal lines indicate GTIs that encompass a significant flux change found by the BB algorithm. 
    The thin dark gray lines at the bottom of each panel
    show the relative exposure at 1\,GeV. The ToO campaigns for 3C~279 (upper left panel) and CTA~102 (first column, second row) are clearly visible, as the exposure is constant over several GTIs. }
    \label{fig:lc_minutes}
\end{figure*}

\begin{figure*}
\centering
\includegraphics[width = .95\linewidth]{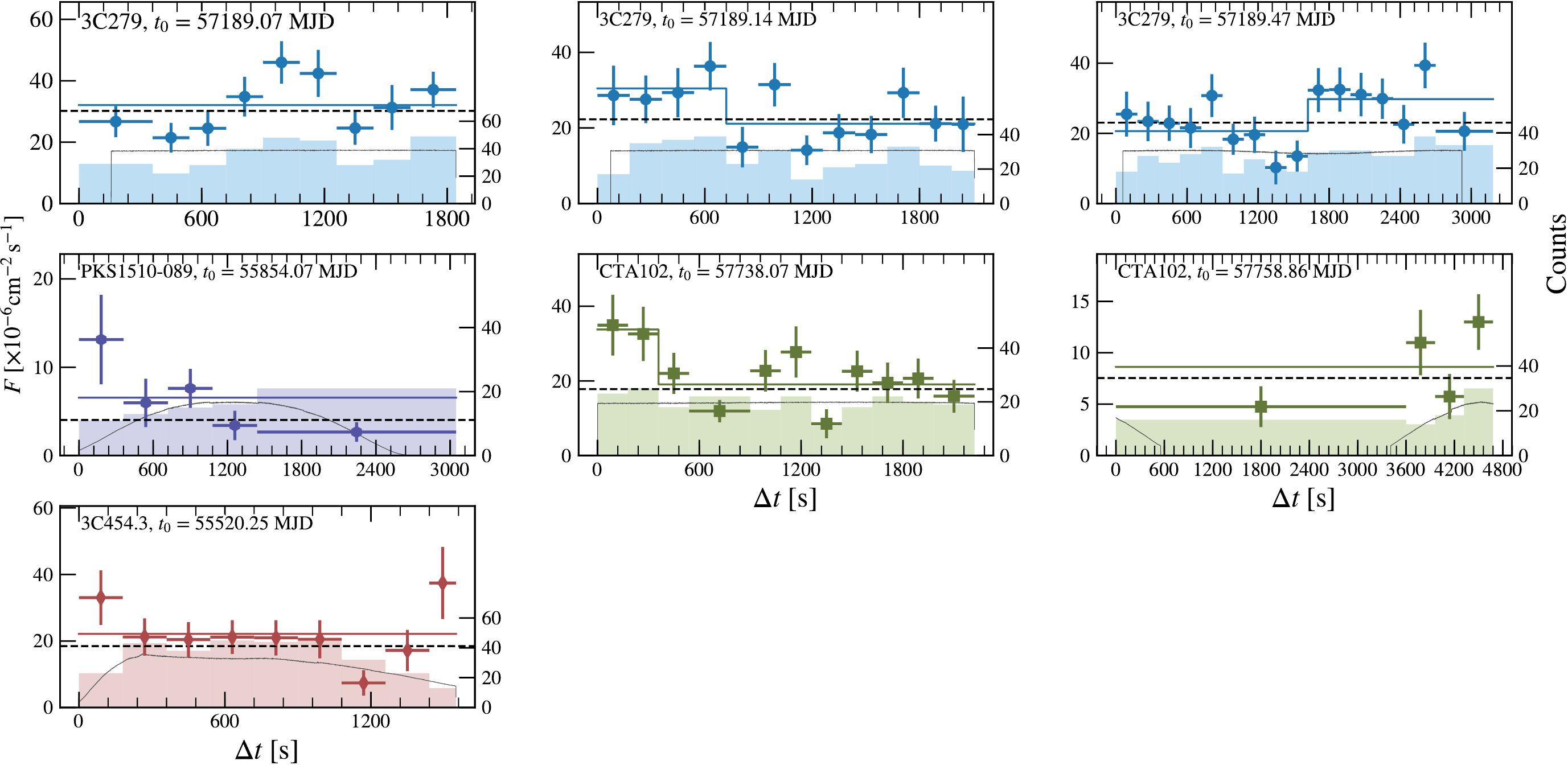}
\caption{Light curves of single GTIs for which the BBs for the full time interval indicate a flux change within the GTI (solid horizontal lines in Figure~\ref{fig:lc_minutes}) and for which a fit with a constant flux results in a fit significance of less than 0.1. The colored thin lines indicate BBs if only the data within the GTI is taken into account. The black dashed line is the best-fit value of the constant flux. The colored histograms show the number of counts in each bin and the thin dark grey curves indicate the relative source exposure as a function of time. }
\label{fig:singel-gtis}
\end{figure*}

The resulting light curves, binned such that the uncertainty in each bin is of the order of $\sim20\,\%$ \citep[using the adaptive binning introduced by][]{lott2012}, are shown in Figure~\ref{fig:lc_minutes}. 
In order to make an objective selection of GTIs that we test against the hypothesis of a constant flux, we consider only those where the BBs indicate a significant flux change within the GTI.
Naively, one could expect that a BB change within one GTI would correspond to a significance of 95\,\% for a nonconstant flux, since this is the threshold we have selected in the BB algorithm~\citep{2013ApJ...764..167S}. However, the BB algorithm also takes data before and after the particular GTI into account and only provides qualitative evidence for minute-scale variability. 
Therefore, we test each bin selected in this way against the hypothesis of constant flux using a simple $\chi^2$ test. 
The best-fit constant flux is given by $\hat{F} = (\sum (F_i / \sigma_i^2))(\sum F_i^{-2})^{-1}$. 
For the GTIs where the constant fit results in a pretrial $p$-value of less than 0.1, we show the sub-GTI light curves in Figure~\ref{fig:singel-gtis} and report the pre- and posttrial fit probabilities in Table~\ref{tab:minute}.
We count each tested GTI as one trial. 
We also provide the minimum values for the variability times for pairs of fluxes $F_i$ and $F_j$ measured at $t_i$ and $t_j$, respectively, given by~\citet{1999ApJ...527..719Z},
\begin{equation}
t_{\mathrm{var},ij} = \frac{F_i + F_j}{2}\left|\frac{t_i - t_j}{F_i - F_j}\right|.
\end{equation}
The pretrial $p$-values for rejecting the constant flux hypothesis are all around $2\,\sigma$. 
The trial correction leaves only one GTI for 3C~279 and two GTIs for CTA~102 close to or above the $2\,\sigma$ evidence for a variable flux. 
However, inspecting the light curve of the second GTI of CTA~102 (starting at MJD\,57,758.86, see Figure~\ref{fig:singel-gtis}), the high $\chi^2$ value might be the result of our chosen binning; the first bin actually spans a long time range for which the exposure is mostly zero.  
For the other two GTIs, the suggested variability timescales are between 3 and 4\,minutes.

In comparison to previous results for 3C~279 and CTA~102, our method results in lower-significance detections of minute-scale variability.
Furthermore, \citet{2018ApJ...854L..26S} found evidence for minute-scale variability in a different orbit compared to our results.
This is due to trial correction but probably also due to differences in the analysis.
For example, we use a finer binning for the exposure in the azimuthal direction to take this dependence for such short observations into account.
The change in exposure is, however, below 10\,\%.
More importantly, we use a different binning within one GTI, which can also change the significance. 
Taking the $\chi^2$ test and the BBs together, we conclude that we find evidence that minute-scale variability is a common phenomenon during bright FSRQ flares. 

\begin{deluxetable}{cccccc}
\tablewidth{0pt}
\tabletypesize{\scriptsize}
\tablecaption{ \label{tab:minute}Results from sub-GTI light curves on minutes-scale variability.}
\tablehead{\colhead{$t_0$} & \colhead{$\Delta t$}  & \colhead{$\chi^2 / \mathrm{d.o.f.}$} & \colhead{$p$-value} & \colhead{$p$-value}& \colhead{$\mathrm{min}(t_\mathrm{var})$}  \\ 
{} [MJD] & [mins] & {} & (Pretrial) & (Postrial) & [mins]}
\startdata
\hline
\multicolumn{6}{c}{3C~279}\\
\hline
57,189.07 & 30.72 & 1.93 & 0.051 (1.95$\sigma$) & 0.188 (1.32$\sigma$) & $5.6\pm2.8$ \\
57,189.14 & 35.13 & 1.68 & 0.071 (1.81$\sigma$) & 0.254 (1.14$\sigma$) & $3.6\pm1.4$ \\
57,189.47 & 53.08 & 1.94 & 0.015 (2.42$\sigma$) & 0.060 (1.88$\sigma$) & $3.7\pm1.4$ \\
\hline
\multicolumn{6}{c}{PKS~1510-089}\\
\hline
55,854.07 & 50.80 & 2.01 & 0.091 (1.69$\sigma$) & 0.173 (1.36$\sigma$) & $7.9\pm5.0$ \\
\hline
\multicolumn{6}{c}{CTA~102}\\
\hline
57,738.07 & 37.04 & 2.30 & 0.011 (2.55$\sigma$) & 0.032 (2.14$\sigma$) & $2.8\pm1.0$ \\
57,758.86 & 78.00 & 2.62 & 0.049 (1.97$\sigma$) & 0.049 (1.97$\sigma$) & $7.8\pm3.7$ \\
\hline
\multicolumn{6}{c}{3C~454.3}\\
\hline
55,520.25 & 25.83 & 1.96 & 0.048 (1.98$\sigma$) & 0.216 (1.24$\sigma$) & $3.2\pm1.6$ \\
\enddata
{
\tablecomments{The number of trials is counted for each flare individually and given by the number of horizontal solid lines in each panel of Figure~\ref{fig:lc_minutes}.}
}
\end{deluxetable}

\section{Location of the $\gamma$-ray emitting region}
\label{sec:location}

As discussed in Sec.~\ref{sec:intro}, the location of the gamma-ray emission region(s) in blazar jets is a matter of considerable debate.
From the rich data set of the six FSRQs studied here, we attempt to constrain the position of the emitting region using three independent approaches. 
First, in Section~\ref{sec:blrabs}, we search for absorption signatures in the LAT spectra caused by pair production of \Grays with photons of external photon fields. We use spectra during the brightest flares identified in the orbital light curves in Figure~\ref{fig:gti} (dashed and solid horizontal bars).
The derived constraints on the position are used to calculate energy dependent cooling times in Section~\ref{sec:tcool}, which will be compared against our results for the decay times for the whole energy range (see Section~\ref{sec:results-local}) and for energy dependent light curves.
As shown by~\citet{2012ApJ...758L..15D}, if the flux decay is dominated by radiative cooling in external radiation fields, the energy dependence of the decay times can be used to distinguish inverse-Compton cooling in the radiation fields of the BLR or the dust torus.
This provides additional information about the position of the \gray emission region.
In Section~\ref{sec:gammaradio}, we search for time lags between \gray and radio emission. 
In the scenario where the non-thermal emission is triggered by, e.g., shocks propagating downstream through the jet, a time lag can be translated into the spatial separation between the radio and \gray emitting regions~\citep{2014MNRAS.445..428M}. 
With information about the location of the radio core, the position of the \gray emitting region can be constrained~\citep[e.g.,][]{2014MNRAS.441.1899F}. 

\subsection{Results from spectral fits to \gray data}
\label{sec:blrabs}
\subsubsection{\gray attenuation}
The attenuation due to pair production on a radiation field of soft photons should manifest itself as a cut-off feature in the \gray spectrum. 
The cut-off energy depends on the distance of the \gray emitting region to the central black hole, $r$, and the photon density of the considered photon field.
For FSRQs, photon densities of external radiation fields of the accretion disk, the BLR, and the extended dust torus usually dominate those of internal synchrotron emission~\citep[see,e.g.,][]{2012ApJ...758L..15D}.
The most relevant external photon field for the \gray energies which can be probed with \FermiLAT is the BLR. 
Pair production on photons from the dust torus or the accretion disk only becomes important at energies beyond 1\,TeV, even when the \Grays are produced close to the central black hole~\citep{finke2016}.

\begin{figure*}
    \centering
    \begin{tabular}{ccc}
    \includegraphics[width=0.32\linewidth]{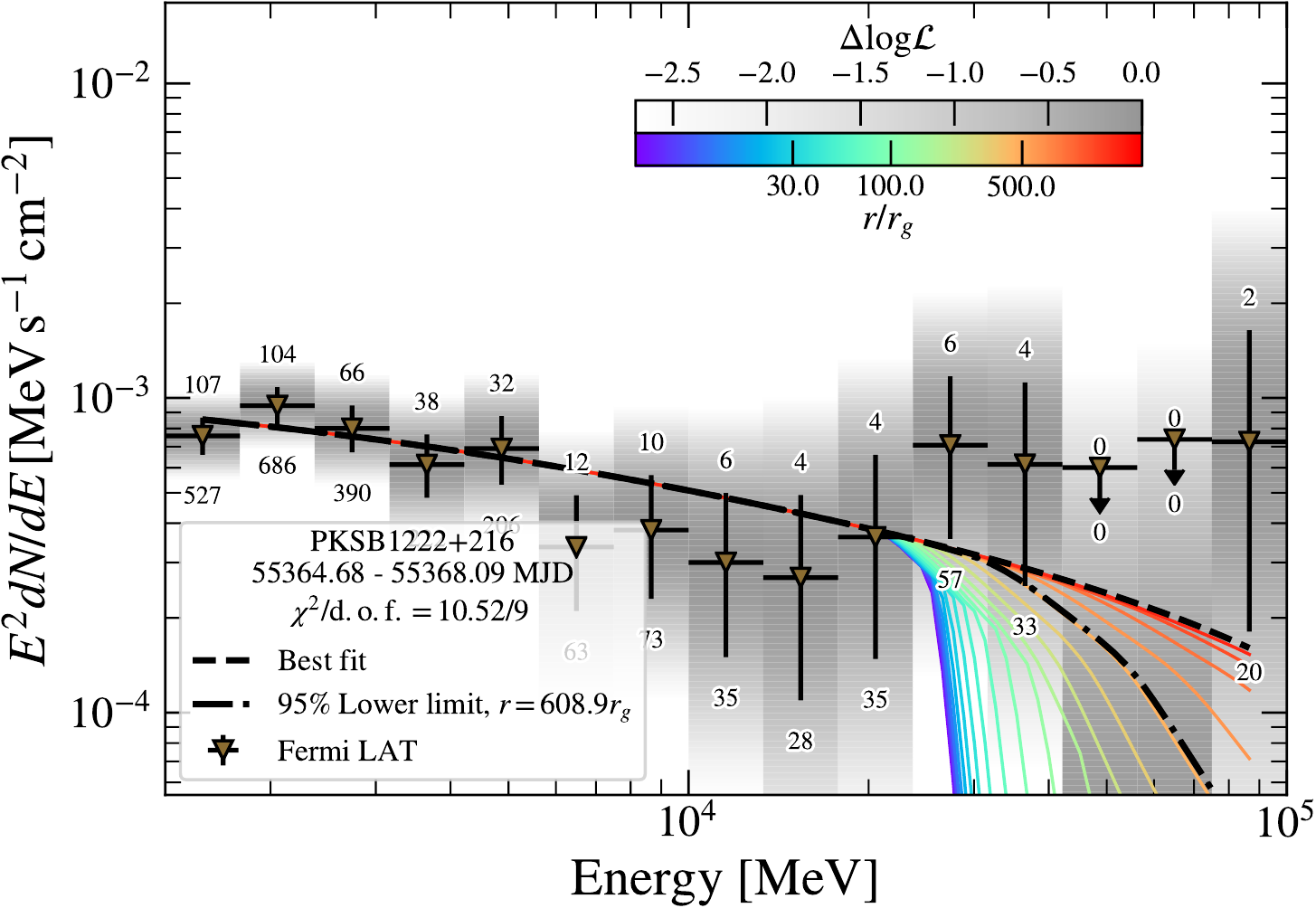} &
    \includegraphics[width=0.32\linewidth]{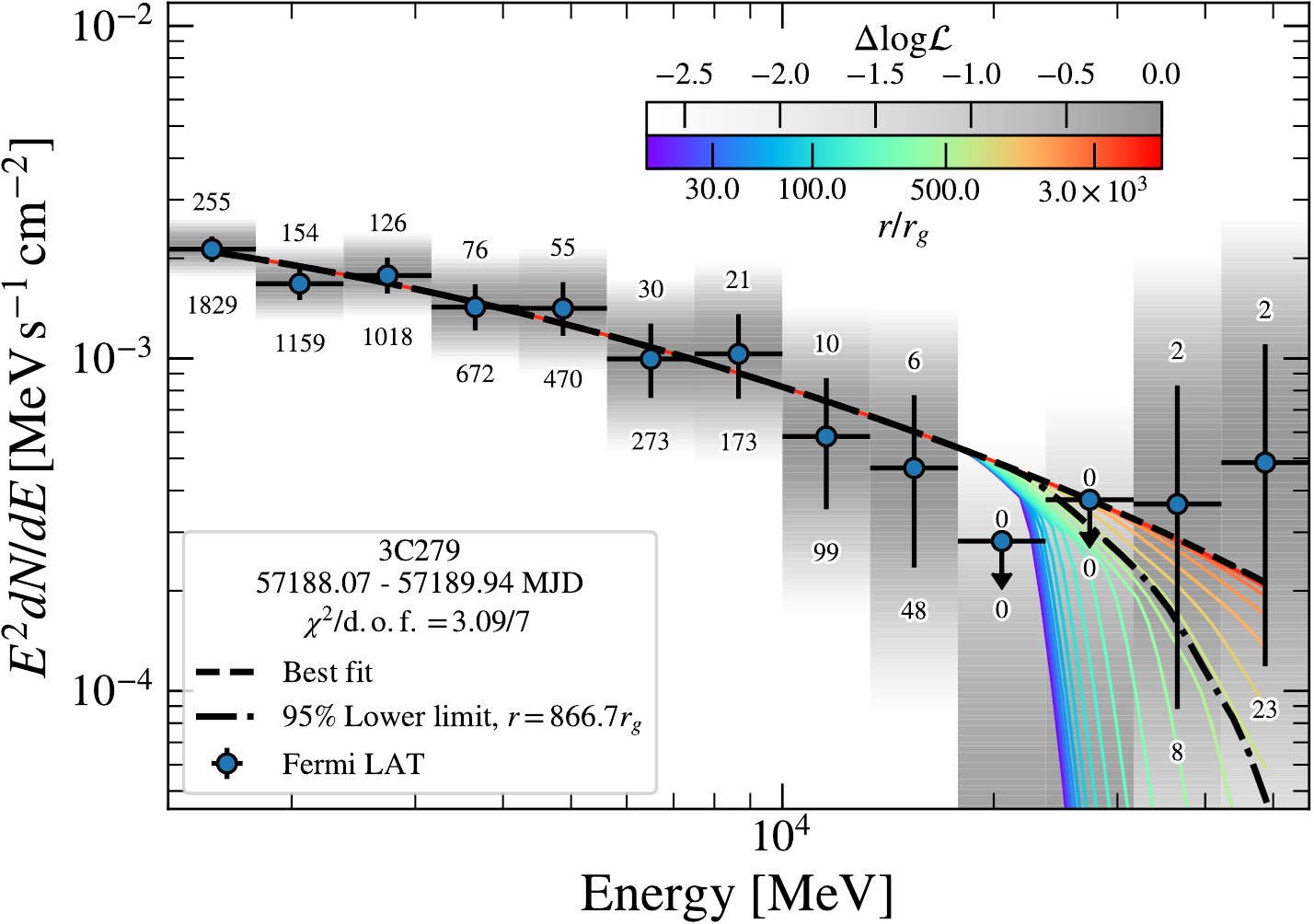} & 
    \includegraphics[width=0.32\linewidth]{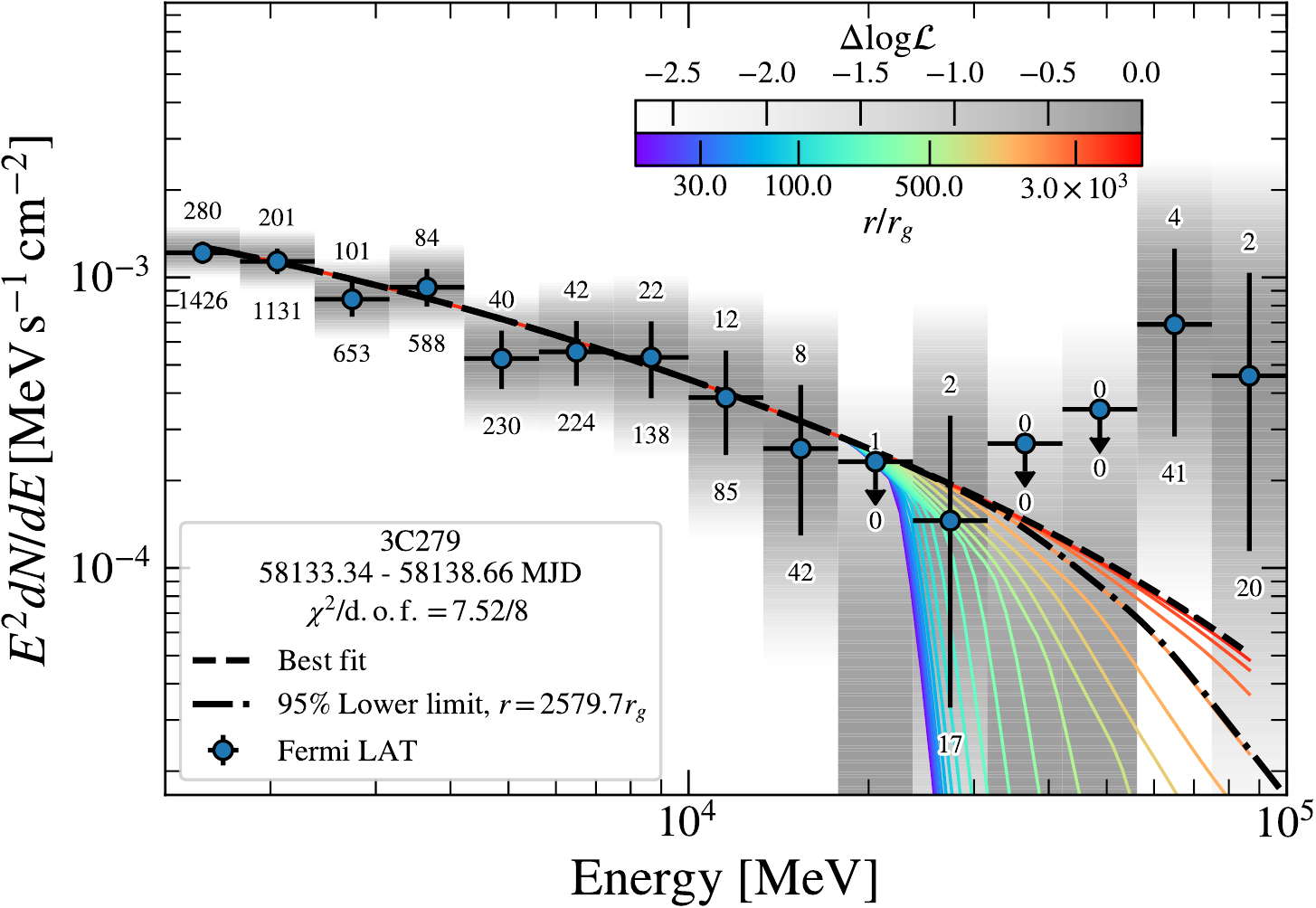}\\
    \includegraphics[width=0.32\linewidth]{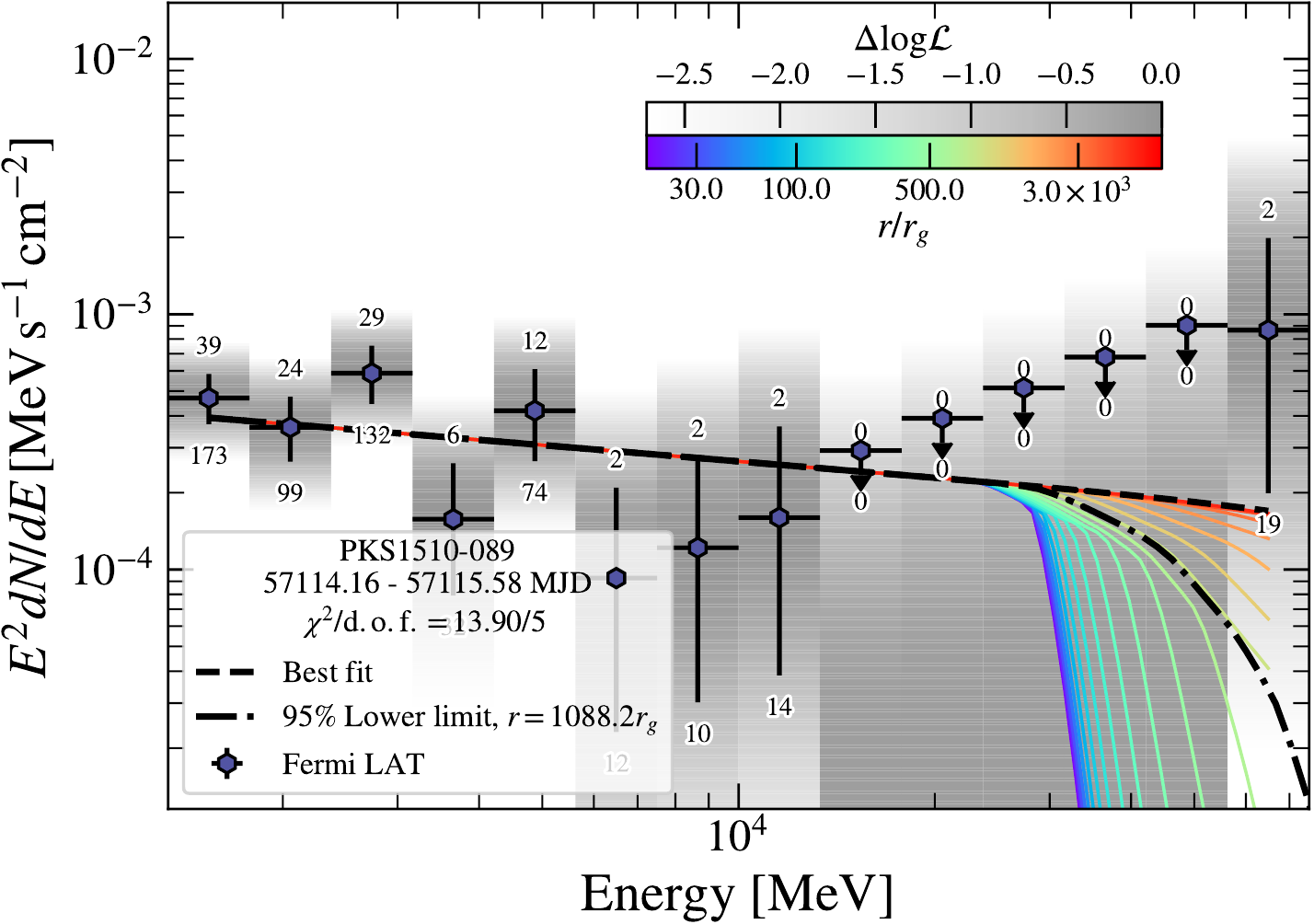} &
    \includegraphics[width=0.32\linewidth]{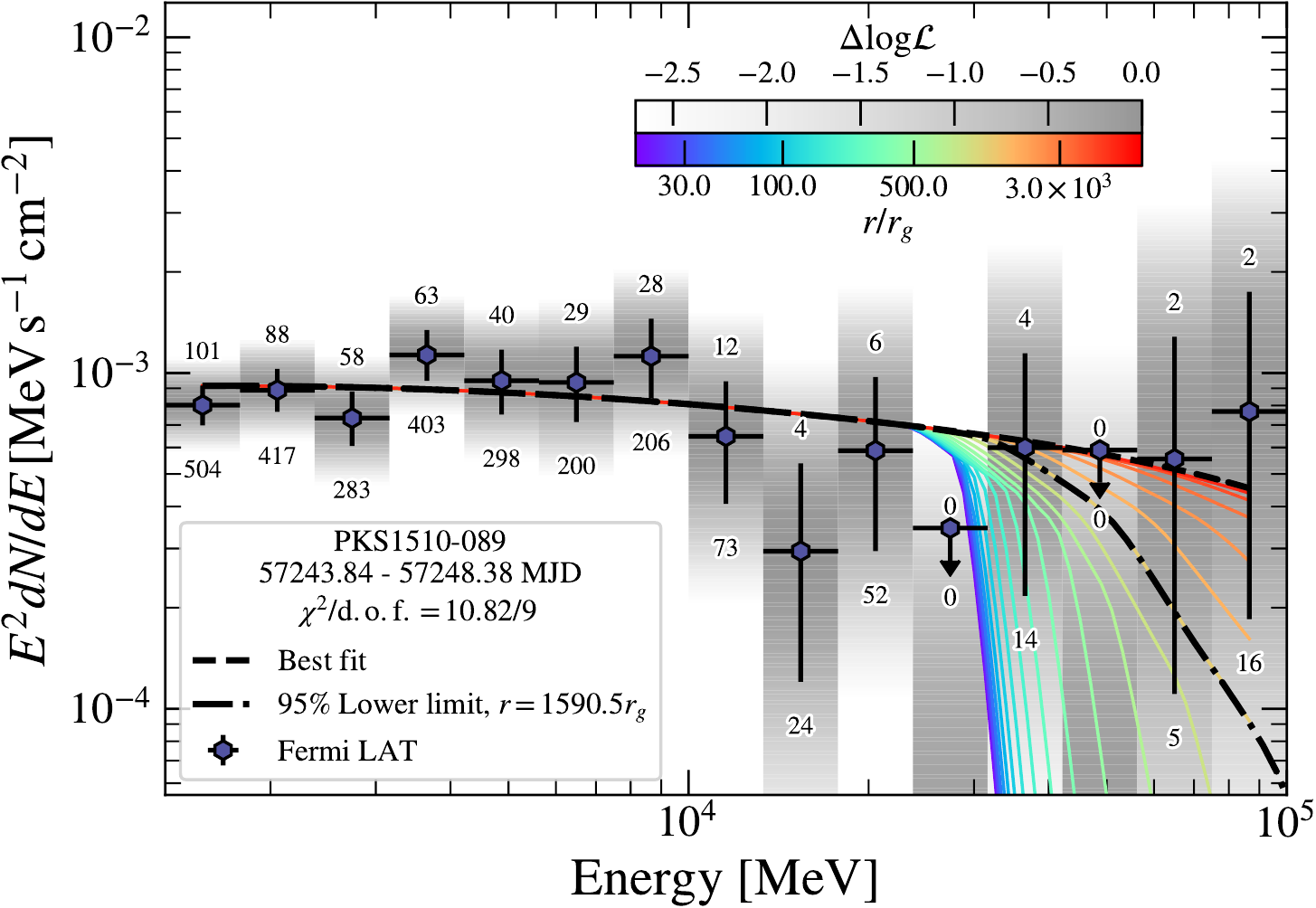} & 
    \includegraphics[width=0.32\linewidth]{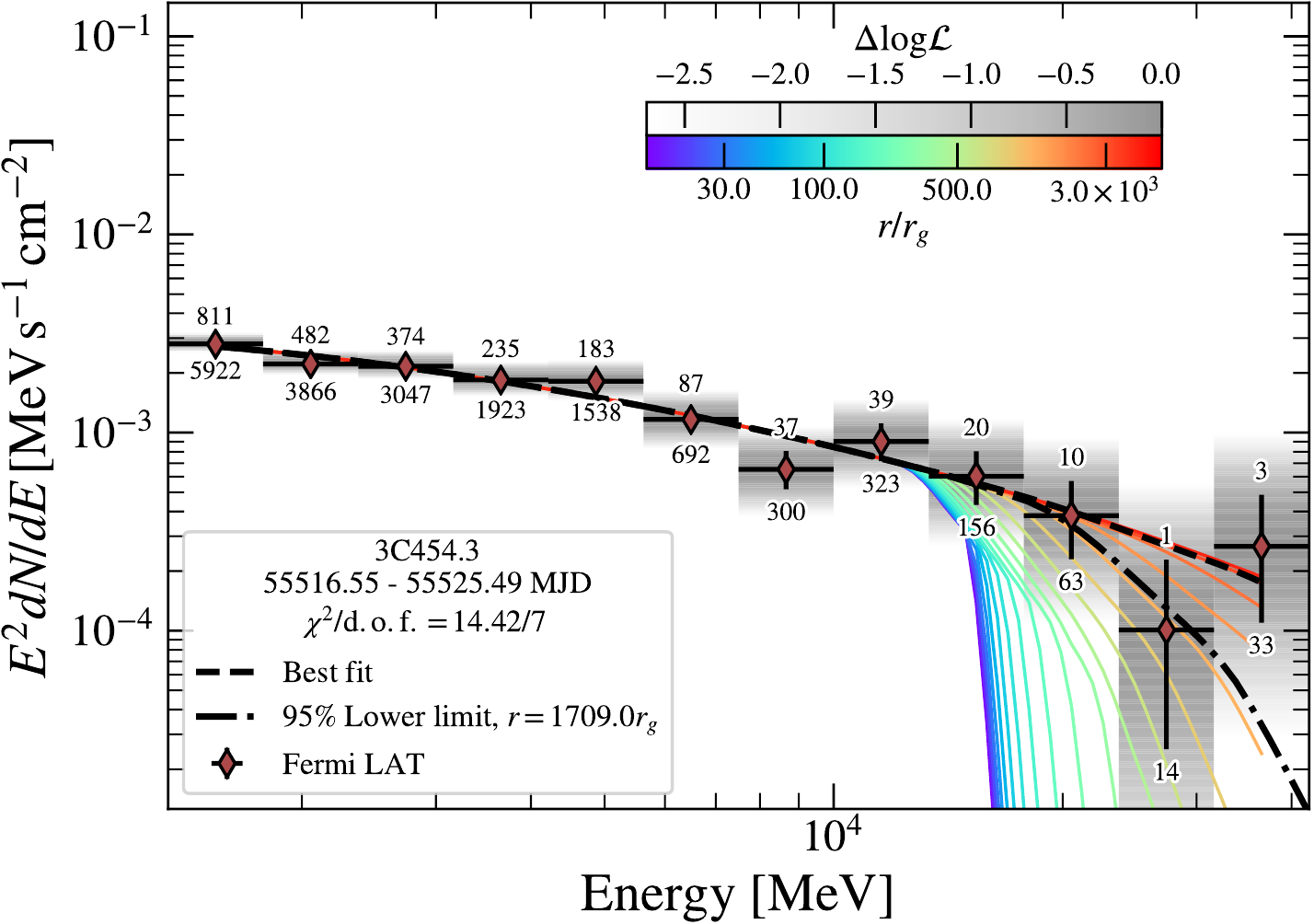}\\
    \includegraphics[width=0.32\linewidth]{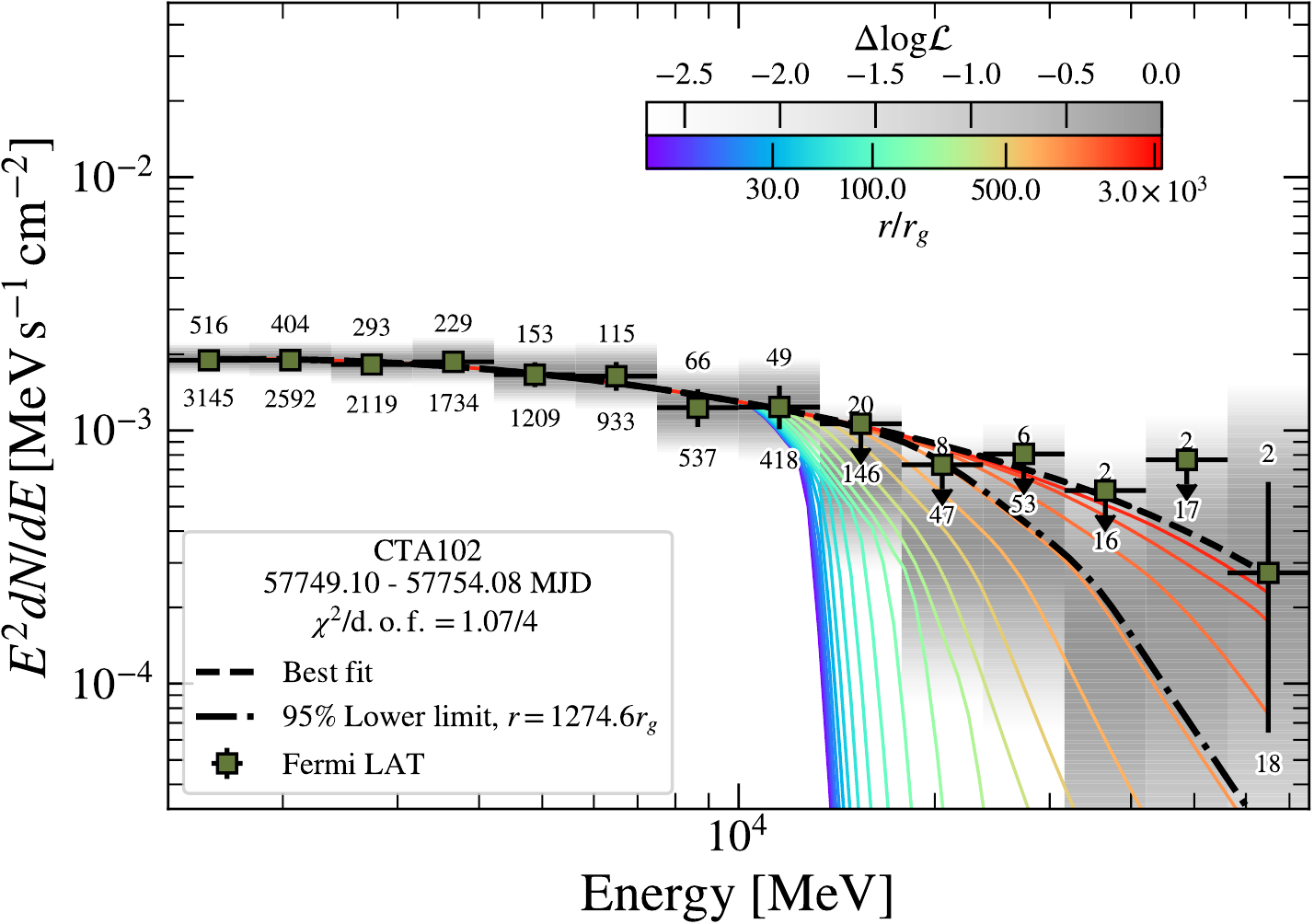} & 
    \includegraphics[width=0.32\linewidth]{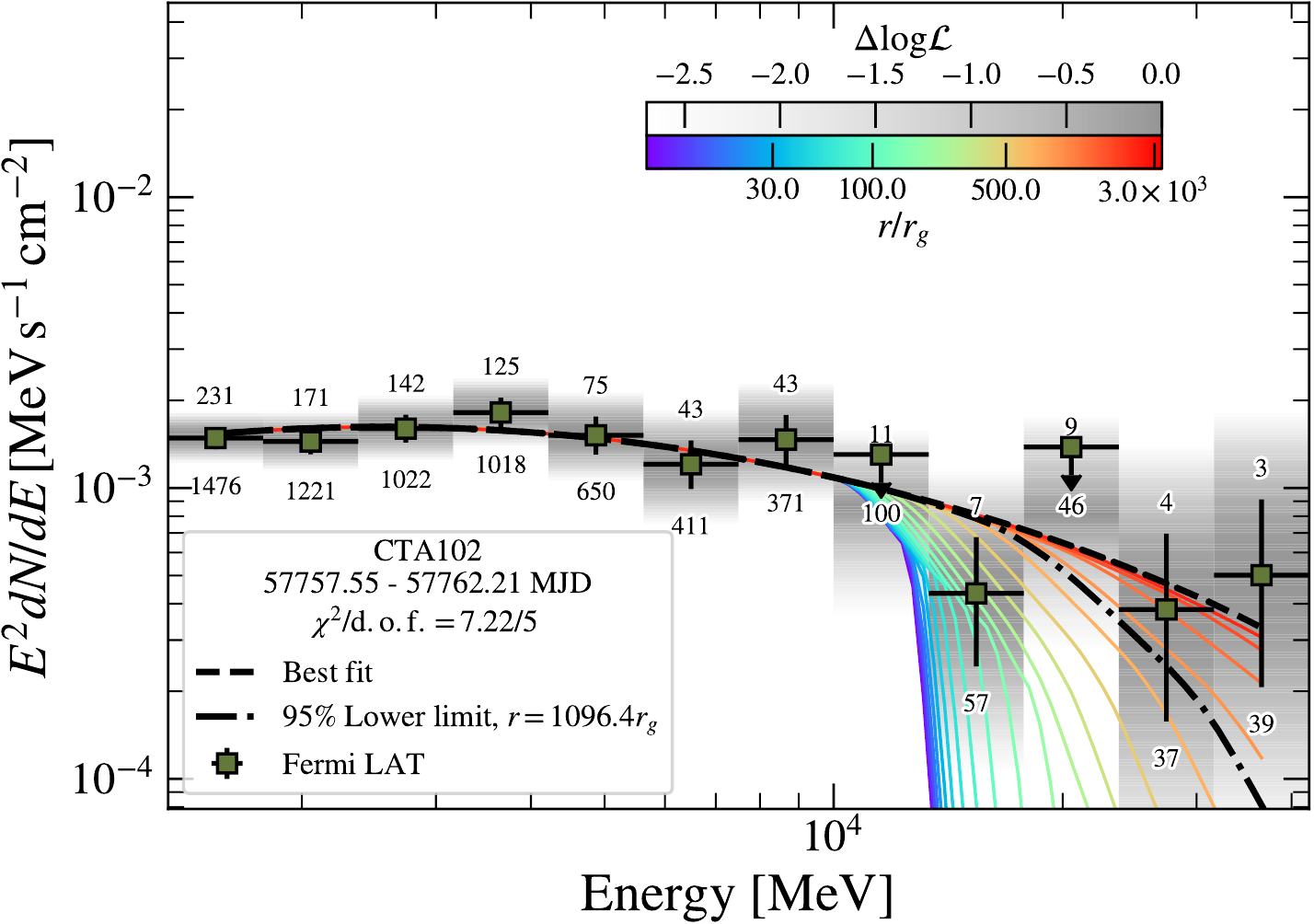} & 
    \includegraphics[width=0.32\linewidth]{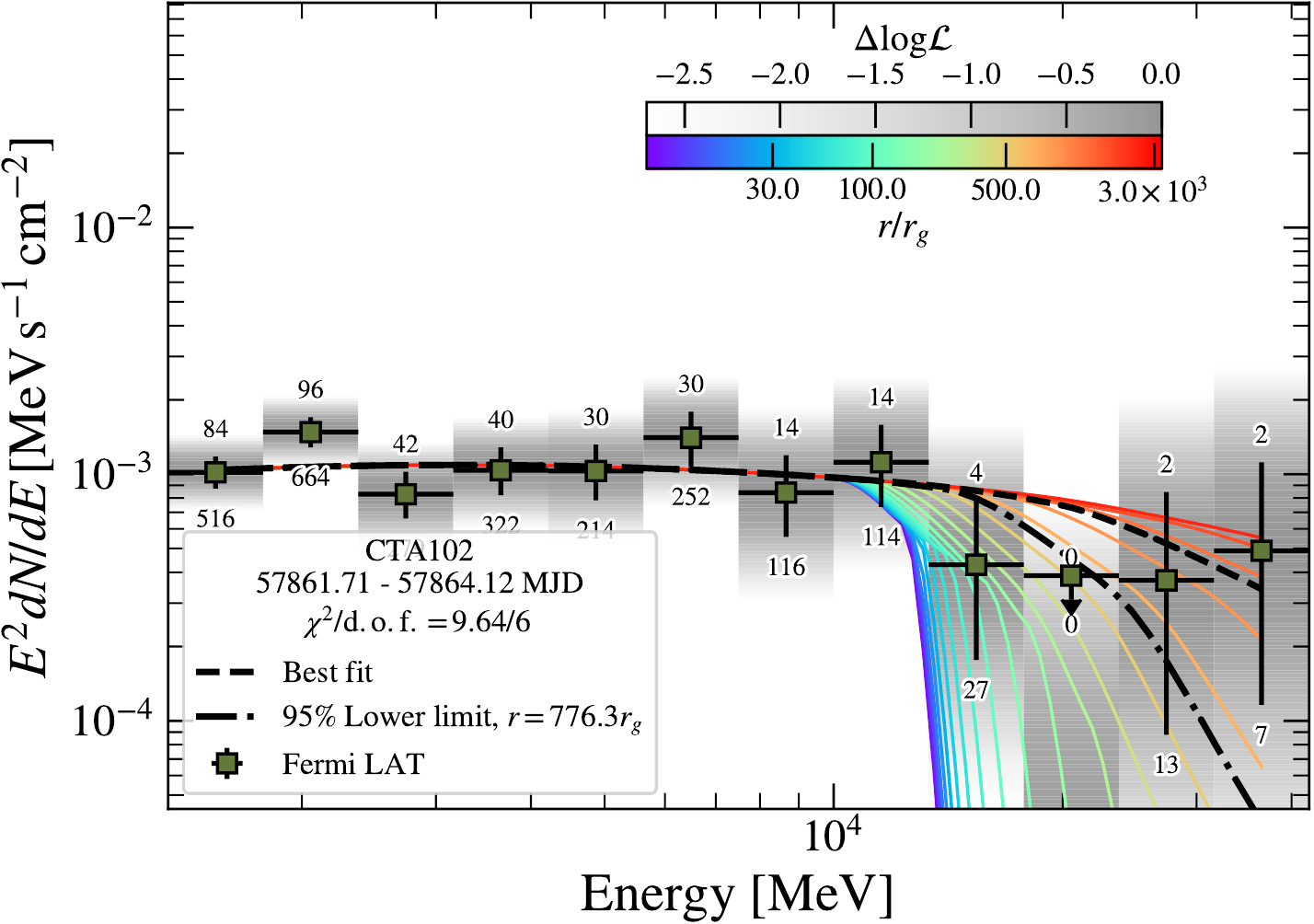}\\
    \includegraphics[width=0.32\linewidth]{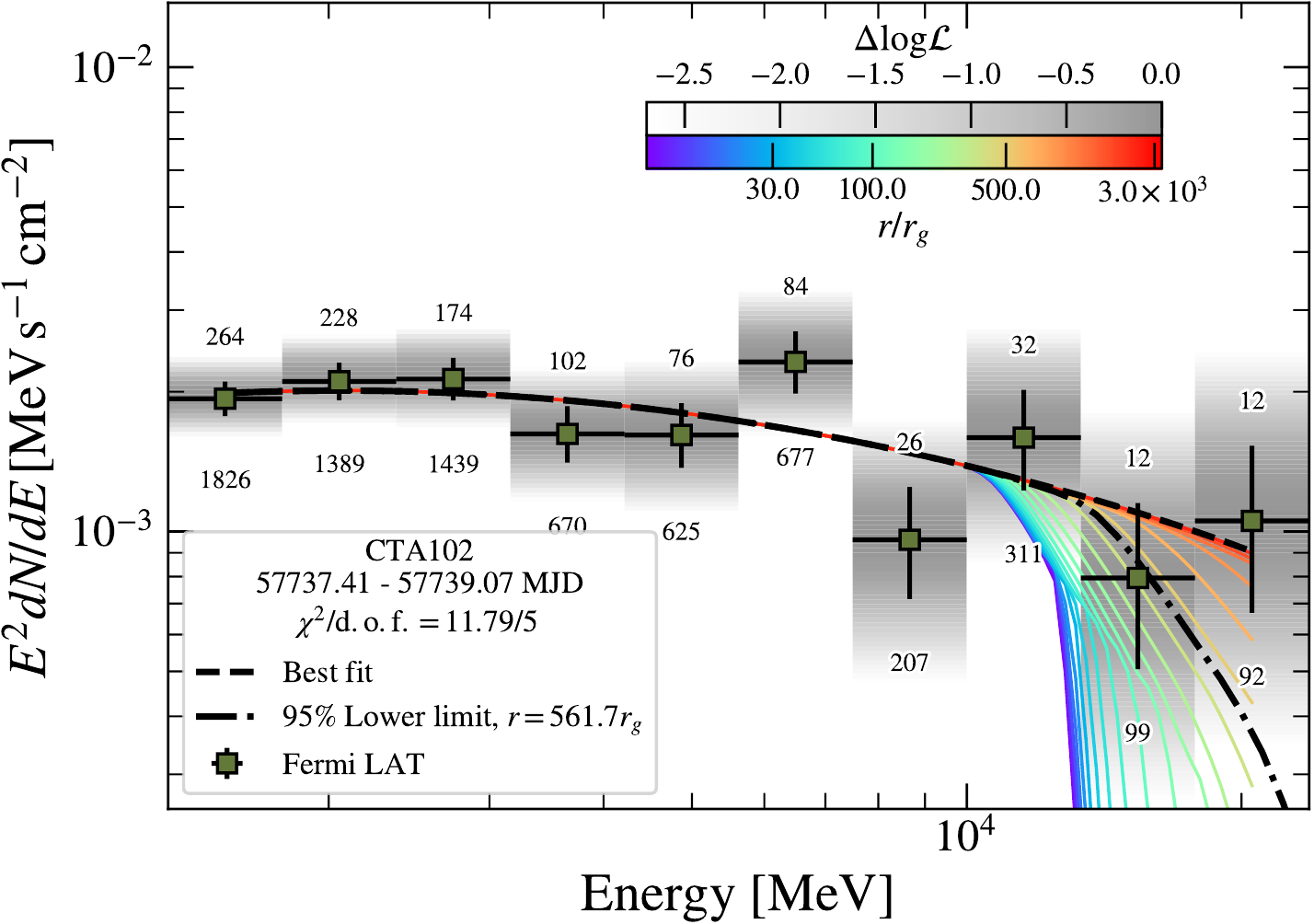}
    \end{tabular}

    \caption{Log-parabola fits above 1\,GeV to bright \gray flares detected at energies that correspond to an attenuation in the BLR of at least 20\,\% (for $r = 10^{-2}R_{\mathrm{Ly}\alpha}$). The attenuation due to the interactions with BLR photons (assuming the ring geometry) is shown as colored lines. The best fit (95\,\% lower limit on $r$) is shown as a black dashed (dashed-dotted) line. 
    The fit uses the bin-by-bin likelihood curves shown as gray bands. The numbers below and above the flux points show the $\mathrm{TS}$ values  with which each bin is detected and the number of \Grays associated with the source at a probability $>85\,\%$, respectively.}
    \label{fig:seds}
\end{figure*}

We can search for BLR absorption features by fitting the observed \gray spectra during the brightest flares with functions of the form 
\begin{eqnarray}
    f(E,\vec{\pi},r,z) &=& f_\mathrm{int}(E,\vec{\pi}) \times \nonumber \\ &{}& \exp\left[-\left(\tau_{\gamma\gamma}^\mathrm{BLR}(E,r) +\tau_{\gamma\gamma}^\mathrm{EBL}(E,z) \right)  \right],
\end{eqnarray} 
where $f_\mathrm{int}(E, \vec{\pi})$ describes the intrinsic spectrum at observed \gray energy $E$ emitted by the source, which depends on spectral 
source parameters, $\vec{\pi}$, such as, e.g.,  flux normalization, power-law index, and spectral curvature (see also Appendix~\ref{sec:avg-spec} for the definitions of the spectral models),
and $\tau_{\gamma\gamma}^\mathrm{BLR / EBL}$ is the optical depth due to interactions of \Grays with photons of the BLR and extragalactic background light (EBL), respectively. 
For the EBL optical depth, which depends on $E$ and the source redshift, $z$, we use the EBL model of \citet{2011MNRAS.410.2556D}.\footnote{
Since the FSRQ have curved spectra and are not analyzed beyond $\sim 50\,$GeV (see Figure~\ref{fig:seds}), the uncertainty introduced by choosing different EBL models is marginal.
}
The BLR optical depth is described by the stratified BLR model introduced by \citet{finke2016}, who models the BLR either as a collection of shells or rings perpendicular to the jet axis, in order to emulate a flattened BLR. 
Each shell or ring is assumed to have infinitesimal thickness and to emit a monochromatic UV or optical emission line. 
The radii $R_\mathrm{li}$ of the shells and rings as well as the line luminosities $L_\mathrm{li} = \xi_\mathrm{li}L_\mathrm{disk}$ are taken from templates of average spectra obtained in reverberation mapping campaigns and provide values relative to the radius and luminosity of the H$\beta$ line~\citep[see][for further details]{finke2016}.
With the H$\beta$ luminosities listed in Table~\ref{tab:src-select}, we fix the absolute luminosities (or conversely $\xi_{\mathrm{H}\beta}$) and radii of all lines included in the model.  
Together with the masses of the super-massive black holes, we can then calculate $\tau_{\gamma\gamma}^\mathrm{BLR}$ for both geometries as a function of $r$ and observed \gray energy $E$.
In the BLR model, the absorption is dominated by pair production with Ly$\alpha$ photons at rest-frame energy of $\epsilon_{\mathrm{Ly}\alpha}\sim10.2\,$eV emitted at radii between $\sim 8\times10^{16}$ and $2\times10^{17}$\,cm. 
In the ring geometry, the corresponding energy density, which is assumed to be isotropic in the stationary frame of the galaxy, becomes~\citep{finke2016}
    \begin{equation}
        u_\mathrm{BLR} \approx u_{\mathrm{Ly}\alpha}= \frac{\xi_{\mathrm{Ly}\alpha}L_\mathrm{disk}}{4\pi c(R_{\mathrm{Ly}\alpha}^2 + r^2)},
        \label{eq:u-blr}
    \end{equation}
and takes values $\sim 5\times10^{-2}\,\mathrm{erg}\,\mathrm{cm}^{-3}$ regardless of the source for $r = R_{\mathrm{Ly}\alpha}$.
These numbers can be compared against typical values for the BLR radius, $R_\mathrm{BLR} \sim 10^{17}\,\mathrm{cm}\, (L_\mathrm{disk} / 10^{45} \mathrm{erg}\,\mathrm{s}^{-1})^{1/2}$ \citep[e.g.][]{2007ApJ...659..997K,2009ApJ...697..160B} and energy density $u_\mathrm{BLR} \sim 10^{-2}\,\mathrm{erg}\,\mathrm{cm}^{-3} $ (again in the stationary galaxy frame) assuming  $L_\mathrm{BLR} = \xi_\mathrm{BLR} L_\mathrm{disk}$ with $\xi_\mathrm{BLR}\sim 0.1$. 
The chosen BLR model gives values broadly consistent with typical values within a factor of a few.

Typically, FSRQ spectra show intrinsic curvature, even below energies at which BLR absorption becomes important (see, e.g., the 3FGL). Therefore we chose a log-parabola for the intrinsic spectral function and also test a power law with super-exponential cut-off (see Eqs.~\ref{eq:avg-spec-lp} and \ref{eq:avg-spec-plexp} for the definition of the models). 
In the fit, we only include energy bins above 1\,GeV, as we expect the BLR cut-off at energies $\gtrsim 10\,$GeV. In this way, we avoid that the best fit is determined mainly by the high photon statistics at lower energies.
Additionally, we select narrow time intervals around the brightest flares (see Section~\ref{sec:zoom} and Figure~\ref{fig:gti}).
This is a compromise between sufficient photon statistics to probe energies above 10\,GeV and avoiding the mixing of different activity states with potentially different spectral states. 
From Figure~\ref{fig:specvar} we see that for the time bins with the highest fluxes spectral variability is only marginally present, which should render our results robust against potential variations of the intrinsic spectra. 
Also, in the fit we only include energy bins detected with $\mathrm{TS} > 0$ and skip flares where the absorption is below 80\,\% in the highest energy bin with $\mathrm{TS} > 0$ and for the smallest BLR distance tested ($r = 10^{-2}R_{\mathrm{Ly}\alpha})$. 
This excludes all flaring periods from 3C~273, for which we cannot obtain any limits from the \gray spectra.

We derive the best-fit values for the spectral parameters $\vec{\pi}$ and the distance $r$ with a likelihood maximization of the bin-by-bin likelihood curves, which we extract with \textsc{fermipy}\footnote{The bin-by-bin likelihoods are derived by fixing the spectral shape in each bin to a power law and mapping the likelihood as a function of the normalization. In the process, the spectral parameters of the neighboring point sources and diffuse backgrounds are fixed to their broadband best-fit values.} and that are shown as gray shaded bands in the panels of Figure~\ref{fig:seds}. 
The flux points in the figure coincide with the maximum likelihood. 
Also shown are the best-fit spectra and BLR attenuation for different values of $r$ (colored curves). 

\begin{figure}
    \centering
    \includegraphics[width = .9\linewidth]{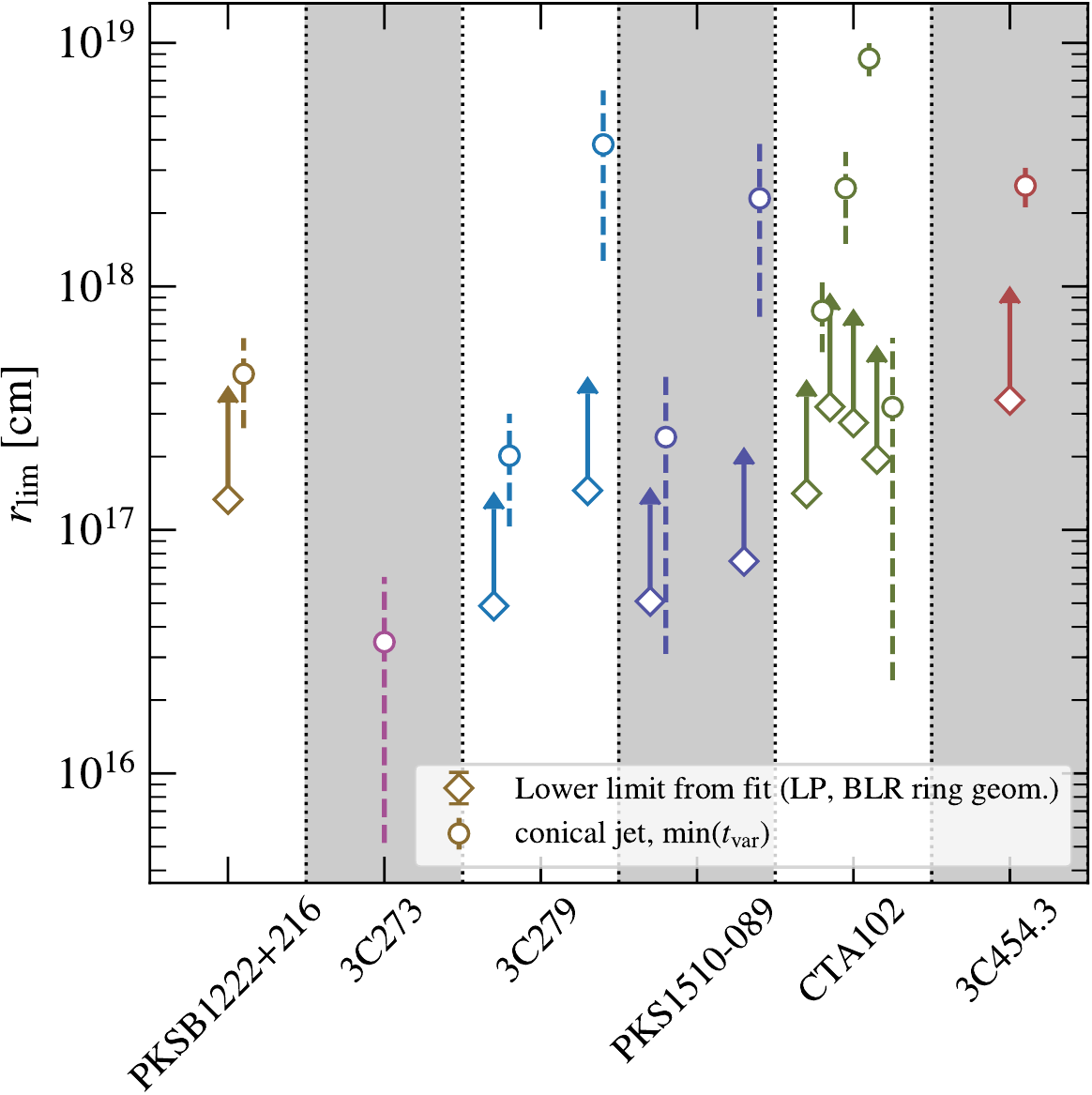}
    \caption{Lower limits on the distance $r$ of the \gray emitting region to the central black hole. Limits from fits to \gray spectra are shown as diamonds, and values derived from variability considerations are shown as circles. }
    \label{fig:blr_limits}
\end{figure}

For both tested BLR geometries, the best-fit value of $r$ is always close to or coincides with the maximum tested value, $r = 10R_{\mathrm{Ly}\alpha}$, and hence no significant absorption is found (dashed black lines in Figure~\ref{fig:seds}). Consequently,  we use \textsc{Minos} to derive the profile likelihood as a function of $r$ from which we determine the 95\,\% lower limit on $r$, $r_\mathrm{lim}$ (dashed-dotted black lines). 
The limit values are reported for each flare in Figure~\ref{fig:blr_limits} and  summarized in Table~\ref{tab:blrabs} for the ring BLR geometry and log-parabola spectrum.
Assuming instead a power law with superexponential cutoff yields consistent results. 
For the BLR shell geometry, the lower limits are a factor of $\sim2$-$3$ higher because this geometry predicts stronger absorption~\citep{finke2016}. The ring geometry is therefore the conservative choice. 

As can be seen from Table~\ref{tab:blrabs} and Figure~\ref{fig:blr_limits} the limits are of the order of $r_\mathrm{lim}\sim10^{17}\,$cm which translates to a distance close to or even beyond the Ly$\alpha$-emitting ring and, consequently, the BLR itself. In terms of gravitational radii, the emission regions are located at distances of at least $\sim10^3r_g$. 
Table~\ref{tab:blrabs} also reports the energy of the highest-energy photon (HEP) associated with the FSRQ with at least $99\,\%$ probability. For all but one source, this energy is larger than the energy where the optical depth due to absorption in the BLR exceeds $\tau_{\gamma\gamma}^\mathrm{BLR} > 1$ (assuming $r = r_\mathrm{lim}$).
Our limits generally agree with the results of \citet{2018MNRAS.477.4749C}, who could limit the maximum value of $\tau_{\gamma\gamma}^\mathrm{BLR}$ to be around $\sim 1$ for 3C~454.3 and PKS~B1222+216 and $\sim 0.2$ for CTA~102. 

The limits sensitively depend on the detection of the source at energies as high as possible. To demonstrate that the detections are not spurious, we also report the detection significance and the number of detected \Grays (associated with the source with a probability $>85\,\%$) for each energy bin below and above the flux points in Figure~\ref{fig:seds}, respectively. The highest-energy bins only contain a handful of source photons (one to four), which underlines the necessity to use the full Poisson likelihood information. 
Doing so, the energy bins are indeed detected with significances of $\sim \sqrt{\mathrm{TS}}$ $\gtrsim 4\,\sigma$.
The reason is that in the considered energy interval and short time spans (see  Table~\ref{tab:blrabs}) the number of expected background events is small. 

\begin{deluxetable*}{ccccccc|ccc}
\tablewidth{0pt}
\tablecaption{ \label{tab:blrabs}Results from BLR absorption fits to \gray spectra.}
\tablehead{\colhead{$t_0$} & \colhead{$\Delta t$}  & \colhead{$r_\mathrm{lim}$} & 
\colhead{$r_\mathrm{lim}$} & \colhead{$r_\mathrm{lim}$} &  \colhead{$E_\mathrm{HEP}$}  & \colhead{$E_{\tau_{\gamma\gamma} = 1}$} & \colhead{$t_\mathrm{cool,~BLR}$}  &
\colhead{$t_\mathrm{cool,~dt}$}  & \colhead{$\tau_\mathrm{decay}$}\\
{} [MJD] & [days] & $[10^{17}\mathrm{cm}]$ & $[R_{\mathrm{Ly}\alpha}]$ & $[r_g]$ & [GeV] & [GeV] &  [minutes] & [hr] & [hr]
}
\startdata
\hline
\multicolumn{10}{c}{PKS~B1222+216}\\
\hline
55,364.68 & 3.42 & 1.33 & 1.40 & 609 & 75.39 & 69.69 & 8.2 & 2.3 & $47.4\pm8.3$\\
\hline
\multicolumn{10}{c}{3C~279}\\
\hline
57,188.07 & 1.87 & 0.49 & 0.64 & 867 & 56.03 & 42.91 &  2.7 & 19.0 & $0.5\pm0.9$\\
58,133.34 & 5.32 & 1.45 & 1.91 & 2580 & 92.56 & 107.91& 9.0 & 19.0 & $8.2\pm6.3$\\
\hline
\multicolumn{10}{c}{PKS~1510-089}\\
\hline
57,114.16 & 1.42 & 0.51 & 0.66 & 1088 & 66.54 & 54.99 & 0.6 & 4.5 & $0.4\pm0.3$\\
57,243.84 & 4.53 & 0.74 & 0.97 & 1591 & 75.93 & 65.39 & 0.8 & 4.5 & $44.4\pm9.4$\\
\hline
\multicolumn{10}{c}{CTA~102}\\
\hline
57,737.41 & 1.67 & 1.41 & 0.86 & 562 & 36.25 & 21.23 & 1.0 & 1.5 & $0.3\pm0.5$\\
57,749.10 & 4.99 & 3.20 & 1.95 & 1275 & 73.80 & 37.94& 2.8 & 1.5 & $8.7\pm1.2$ \\
57,757.55 & 4.66 & 2.76 & 1.67 & 1096 & 39.19 & 32.38& 2.2 & 1.5 & $24.6\pm2.3$ \\
57,861.71 & 2.42 & 1.95 & 1.18 & 776 & 34.73 & 24.94& 1.4 & 1.5 & $1.2\pm0.7$ \\
\hline
\multicolumn{10}{c}{3C~454.3}\\
\hline
55,516.55 & 8.93 & 3.19 & 1.36 & 1598 & 41.19 & 28.73& 4.2 & 16.8 & $2.6\pm1.0$ \\
\enddata
{
\tablecomments{HEPs are given for source probabilities $> 0.99$. The decay times are given for the flare component with the highest peak flux as determined in the fit to the orbital light curves in Section~\ref{sec:results-local}.
For the cooling times, an observed \gray energy of $10^{8.5}\,\mathrm{eV} \approx 316\,$MeV is assumed. }
}
\end{deluxetable*}

We also compare the limits from fits to \gray spectra to considerations from variability arguments in Figure~\ref{fig:blr_limits}.
If the emission region $R'_\mathrm{blob}$ (the prime denotes the comoving frame) is causally connected during the flare, the shortest variability time $t_\mathrm{var}$ sets an upper limit on its size \citep[e.g.,][]{2008MNRAS.384L..19B},
\begin{equation}
    R'_\mathrm{blob} \leqslant \frac{ct_\mathrm{var}\delta_\mathrm{D}}{1+z},
    \label{eq:rblob}
\end{equation}
where $\delta_\mathrm{D} = \Gamma^{-1}_\mathrm{L}(1 - \beta\cos\theta_\mathrm{obs})^{-1}$ is the Doppler boost factor with the bulk Lorentz factor of the flow; $\Gamma_\mathrm{L}$, $\beta = \sqrt{1 - \Gamma_\mathrm{L}^{-2}}$ is the associated velocity; and $\theta_\mathrm{obs} $ is the angle between the line of sight and the jet axis.
\citet{2013A&A...558A.144C} found a correlation, 
$\theta_\mathrm{j} \sim \rho\Gamma^{-1}_\mathrm{L}$ with $\rho \sim 0.2$ from the data of the MOJAVE very large baseline interferometry (VLBI) blazar monitoring program.
Similar values are also found from Very Long Baseline Array (VLBA) monitoring observations \citep{2017ApJ...846...98J}.
Using this correlation and under the assumption that the plasma blob occupies half of the jet's cross section, we find 
$\theta_\mathrm{j} \sim 0.1\Gamma^{-1}_\mathrm{L}$ and obtain
an upper limit on the distance to the black hole, $r \sim 10R'_\mathrm{blob}\Gamma_\mathrm{L} \equiv r_\mathrm{j}$.
The values are plotted in Figure~\ref{fig:blr_limits} for the minimum of the rise and decay times of the brightest flares found in Figure~\ref{fig:gti}.
The average values for $\delta_\mathrm{D}$ and $\Gamma_\mathrm{L}$ obtained from VLBA  observations are used \citep[see also Table~\ref{tab:src-select}]{2017ApJ...846...98J},  and the total  uncertainty is obtained by summing the uncertainties on $\delta_\mathrm{D}$ and $\Gamma_\mathrm{L}$, and the fit uncertainty of $t_\mathrm{var}$ in quadrature.
It should be noted that the underlying assumption is that the \Grays are produced cospatially with the radio emission for which the Doppler factors are measured.  
In general, we find that $r_\mathrm{lim} \gtrsim r_\mathrm{j}$ indicating that the emission regions are at larger distances to the black hole than predicted from the conical jet scenario.  

In our BLR model, we use a simplified BLR geometry and do not include the hydrogen or He~II recombination continua or emission lines of the He~II Ly series~\citep[as done in, e.g.,][]{2010ApJ...717L.118P,2014ApJ...794....8S}. 
Comparing the optical depths of our model to the results of the sophisticated modeling of~\citet[see, in particular, their Figure~11]{2017MNRAS.464..152A}, who described the BLR as a collection of ionized gas clouds irradiated by the accretion disk, we find that our values of $\tau_{\gamma\gamma}^\mathrm{BLR}$ reach unity at energies a factor of $\sim1.5$ higher, but around 100\,GeV, the optical depth in the two models is similar. 
Also, the BLR model of \citet{finke2016} reproduces the trend observed in the sophisticated model that the absorption sets in at higher energies and is overall weaker for larger values of $r$. 
As we do not observe significant cutoffs in the spectra, we therefore conclude that the ring geometry adopted here provides a conservative limit on $r$. 

Furthermore, evidence exists that the the luminosity of BLR emission lines is variable in 3C~454.3, PKS~1510-089, and PKS~B1222+216~\citep{2013ApJ...763L..36L,2015ApJ...804....7I} and correlates with the \gray emission~\citep{2013ApJ...763L..36L,2015IAUS..313...43L}, which could indicate that \Grays are produced through IC scattering with BLR photons (see also Section~\ref{sec:tcool}).  
Additionally, \citet{2013ApJ...763L..36L} found that the BLR brightening coincides with the passage of a superluminal jet component through the radio core, which could indicate that BLR clouds are located at larger distances, $\gtrsim 1\,$pc, than assumed here.
A brighter BLR emission during \gray flares and BLR material located at larger distances would mean stronger \gray attenuation, which would shift our limits to even larger distances. 

\subsubsection{Jet Shielding by a Plasma Sheath}
\label{sec:plasma-sheath}

In view of the severity of these constraints, it is worth considering radical alternatives to the standard model of FSRQ \gray emission. The first possibility is that the inner jet is actually shielded from external soft photons. One way in which this can happen is if the broad line-emitting clouds derive from the accretion disk and are propelled to radii of $\sim0.1-1\,{\rm pc}$ by the centrifugal action of magnetic field lines attached to the accretion disk \citep{emmering:1992mac,1994ApJ...434..446K,bottorf:1997dyn}. The magnetic field channels the cool gas along outward trajectories with speeds of $\sim0.03c$ including some rotation.  Individual gas clouds can be confined transversely by magnetic pressure but will cool as they expand. 

Generic BLR models have filling factors of $\sim10^{-5}-10^{-4}$ and covering factors of $\sim0.1$. Now, suppose that some of these clouds derive from the inner disk and are attached to the toroidal field lines that are thought to collimate the jet. They will be
photoionized
and their thermal state will be a balance between photoionization heating plus expansion and radiative loss.
In order for a \gray of energy $E_\gamma$ to escape from the inner jet, we must have efficient shielding out to the $\gamma$-sphere  \citep{1995ApJ...441...79B} defined by the unshielded photons. If this outflow can remain cool enough, a column density $\Sigma\gtrsim\Sigma_{\rm shield}\sim5\times10^{-3}E_{X\,\rm keV}^{2.5}\,{\rm g\,cm}^{-2}$, for $0.014\lesssim E_{X\,\rm keV}\lesssim10$ suffices to shield the jet from photons of energy $E_{X\,\rm keV}$. Prominent line photons, notably Ly$\alpha$, should also be shielded. We can express $\Sigma_{\rm shield}$ in terms of $E_\gamma$ at the threshold for pair production, $\Sigma_{\rm shield}\sim2\times10^{-4}E_{\gamma\,\rm GeV}^{-2.5}\,{\rm g\,cm}^{-2}$, for $0.03\lesssim E_{\gamma\,{\rm GeV}}\lesssim20$ and a sufficiently large jet radius. 

Next, suppose that the cylindrical radius of the sheath at the $\gamma$-sphere is $10^{15}s_{\gamma{\rm sphere}\,15}\,{\rm cm}$, (typically $\sim0.1$ of the jet radius). The discharge associated with the shielding gas is then $\dot M_{\rm sheath}\sim10^{-5}s_{\gamma{\rm sphere}\,15}E_{\gamma\,\rm GeV}^{-2.5}\,{\rm M}_\odot\,{\rm yr}^{-1}$, which is quite modest for the energies of interest. An observer situated on the jet axis should be able to observe $\gamma$-rays from very close to the black hole at radii much smaller than that of the $\gamma$-sphere, where their observed variability timescales can be as short as minutes after correcting for relativistic time travel effects. Of course, there may be some opacity due to synchrotron photons emitted at smaller jet radii and beamed along the jet, but again, this need not be severe. Note that photons with energy below the Lyman continuum, including optical and infrared photons, should permeate the jet at all radii, so this model implies that there should not be rapid \gray variability with $E_{\gamma}\gtrsim30/(1+z)\,{\rm GeV}$ from FSRQs. The rapid variability seen  at TeV energy in some BLLs arises because
these sources lack a strong UVX continuum and the black hole masses are smaller.

\subsection{Considerations from radiative cooling}
\label{sec:tcool}
\subsubsection{Broad emission line radiation}
With the limits on $r$ it is possible to derive constraints on the energy density of external photon fields in the comoving frame, which could be responsible for the \gray emission due to IC scattering with relativistic electrons in the emission region. 
Because the cooling time depends on the energy density and, in turn, on $r$, a comparison between the predicted IC cooling times and the observed decay times can provide further information on where the \Grays are emitted.
We first focus on the BLR photon field, but the discussion also applies for IC scattering with photons of the dust torus, which we will discuss at the end of this section.

\begin{figure}
    \centering
    \includegraphics[width = .9\linewidth]{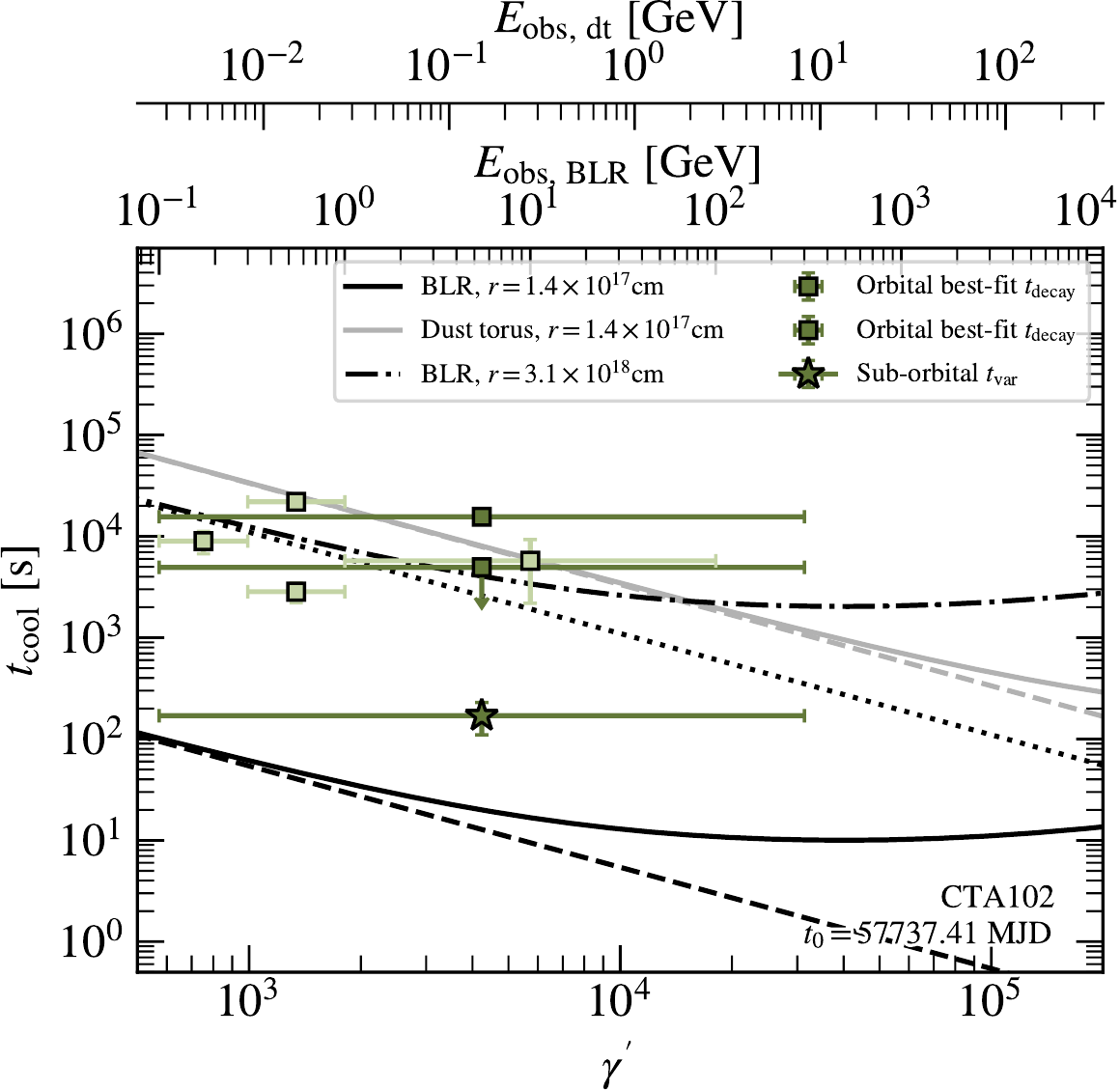}
    \caption{Cooling times for IC scattering with BLR photons and photons from the dust torus for one flaring episode of CTA~102. Also shown are observed decay times for the full energy range, energy-dependent light curves, and suborbital light curves.
    Note that the observed decay times are plotted with respect to IC scattering with BLR photons. For scattering with photons from the dust torus, the points need to be shifted to higher values of $\gamma'$ to match $E_\mathrm{obs,dt}$ (see second $x$-axis on the top of the figure).}
    \label{fig:tcool}
\end{figure}

In the galaxy frame, the energy density of the BLR in the ring geometry is approximately given by Equation~\ref{eq:u-blr}; hence, the photon number density is $n_\mathrm{BLR} \approx u_\mathrm{BLR} / \epsilon_{\mathrm{Ly}\alpha}\sim10^{9}\,\mathrm{cm}^{-3}$.
Assuming that the BLR photon field is isotropic and in the limit $\Gamma_\mathrm{L} \gg 1$, the energy density in the comoving frame becomes
$u'_\mathrm{BLR} = (4/3)\Gamma_\mathrm{L}^2 u_\mathrm{BLR}$~\citep{1994ApJS...90..945D,2002ApJ...575..667D}.
We calculate the energy loss of the electrons, $\dot{\gamma}_\mathrm{BLR}'$, due to IC scattering in the comoving frame numerically, in order to incorporate Klein-Nishina   effects following \citet{1970RvMP...42..237B}.
The observed cooling time is then given by
\begin{equation}
    t_\mathrm{cool,BLR} = \frac{1 + z}{\delta_\mathrm{D}} \frac{\gamma'}{\dot{\gamma}'_\mathrm{BLR}}.
    \label{eq:tcool}
\end{equation}
In the Thomson regime, this becomes
\begin{equation}
    t_\mathrm{cool,BLR} = \frac{1+z}{\delta_\mathrm{D}}\frac{3m_ec^2}{4c\sigma_\mathrm{T}u_\mathrm{BLR}'\gamma_\mathrm{BLR}'},
    \label{eq:tcool-thomson}
\end{equation}
where $m_e$ is the electron mass and $\sigma_\mathrm{T}$ is the Thomson cross section. 
In what follows, we approximate the electron Lorentz factor with~\citep[e.g.,][]{2009herb.book.....D,finke2016}
\begin{equation}
    \gamma'_\mathrm{BLR} = \frac{1}{ \delta_\mathrm{D}}\sqrt{\frac{E_\mathrm{obs,BLR}(1+z)}{2\epsilon_{\mathrm{Ly}\alpha}}},
    \label{eq:gammaprime}
\end{equation}
where $E_\mathrm{obs,BLR}$ is the observed \gray energy of IC-scattered BLR photons. 
From Eqs.~\ref{eq:u-blr}, \ref{eq:tcool-thomson}, and \ref{eq:gammaprime} it becomes clear that the cooling time scales as $t_\mathrm{cool, BLR}\propto r^2 \Gamma^{-2}$, i.e., the cooling becomes less efficient for large distances. 
Furthermore, if the \gray emission is produced very far away from the BLR, the external photons will appear as a point source illuminating the emission region from behind, so that $u'_\mathrm{BLR} = (1/4)\Gamma^{-2} u_\mathrm{BLR}$ \citep{1994ApJS...90..945D}, leading to an additional decrease of the cooling time. 
The BLR cooling time for one flare of CTA~102 is shown for $r=r_\mathrm{lim}$ and $r = 1\,$pc as a function of $\gamma'$ and $E_\mathrm{obs,BLR}$ in Figure~\ref{fig:tcool},  
assuming again the average values for $\delta_\mathrm{D}\sim31$ and $\Gamma_\mathrm{L}\sim22$. 
Klein-Nishina effects become important for $\gamma'_\mathrm{BLR} \gtrsim 2\times10^3$ or $E_\mathrm{obs,BLR}\gtrsim1\,$GeV and a clear departure from the Thomson regime in which 
$t_\mathrm{cool}\propto(E_\mathrm{obs,BLR})^{-1/2}$
becomes visible.
The decay times derived from the fit of the two bright flares of CTA~102 around MJD 57,738 are also shown in Figure~\ref{fig:tcool}
(see the first solid horizontal line in the panel in the fourth row and second column in Figure~\ref{fig:gti}), where one decay time is shown as an upper limit due to its large uncertainty. 
If the decay is indeed caused by IC scattering with BLR photons, a distance of $r\sim1\,$pc is still compatible with the observed decay times. 
Similar conclusions can be drawn for the other sources. 
We provide the cooling times at 300\,MeV and the shortest decay times in Table~\ref{tab:blrabs}.

Figure~\ref{fig:tcool} also shows the variability time $t_\mathrm{var}$ derived for the suborbital period in Section~\ref{sec:sub-gti}, which is, however, in the rising part of the flare (see Figure~\ref{fig:lc_minutes}). If equally short decay times were observed, this would suggest a distance $r \sim r_\mathrm{lim}$. 

As noted by \citet{2012ApJ...758L..15D}, the energy dependence of the cooling times could  further reveal the dominant photon field responsible for IC scattering: while Klein-Nishina effects become important already at 1\,GeV for scattering with BLR photons, the Thomson regime should be valid to higher energies for IC scattering with photons of the dusty torus (dt). 
Again following \citet{finke2016}, we assume that the torus also has a ring geometry and emits monochromatic photons with energy $\epsilon_\mathrm{dt} = 2.7 k_\mathrm{B} T_\mathrm{dt}$, with $k_\mathrm{B}$ the Boltzmann constant.
Generic values for the dust temperature $T_\mathrm{dt}$ are around 1000\,K, and the dust luminosity is taken to be $\xi_\mathrm{dt}L_\mathrm{disk}$ with the dust scattering fraction $\xi_\mathrm{dt} = 0.1$.
We adopt these values for all sources except 
 PKS~B1222+216, 3C~273, and CTA~102. 
 For PKS~B1222+216 and CTA~102  \citet{2011ApJ...732..116M} found $T_\mathrm{dt} = 1200\,$K and a dust luminosity of $7.9\times10^{45}\,\mathrm{erg}\,\mathrm{s}^{-1}$ and $7\times10^{45}\,\mathrm{erg}\,\mathrm{s}^{-1}$, respectively. 
\citet{2005ApJ...625L..75H} observed silicate emission in 3C~273 from which they deduced a silicate temperature of 140\,K and luminosity of $\sim 10^{45}\,\mathrm{erg}\,\mathrm{s}^{-1}$ but were unable to derive a temperature of the dust \citep[see also the discussion in][]{2011ApJ...732..116M}. \citet{2008A&A...486..411S} instead found a dust temperature of 1200\,K.
We adopt the latter value and again set $\xi_\mathrm{dt} = 0.1$.
The sublimation radius of the dust torus, $R_\mathrm{dt} = 3.5\times10^{18}\,\mathrm{cm}(L_\mathrm{disk}/10^{45}\,\mathrm{erg}\,\mathrm{s}^{-1})^{1/2}(T_\mathrm{dt}/10^3\,\mathrm{K})^{-2.6}$ is used as the ring radius. 
Making the appropriate substitutions in Equations~\ref{eq:u-blr} and \ref{eq:tcool}-\ref{eq:gammaprime}, we plot the cooling time $t_\mathrm{cool,dt}$ for $r = r_\mathrm{lim}$ as a grey line in Figure~\ref{fig:tcool} (note the additional $x$-axis, since $E_\mathrm{obs, BLR} \neq E_\mathrm{obs,dt}$). 
Indeed, Klein-Nishina effects only become relevant at $E_\mathrm{obs,dt} \gtrsim 10^2\,$GeV or $\gamma' \gtrsim 10^5$. The cooling times at $E_\mathrm{obs,dt} = 316\,$MeV are also provided in Table~\ref{tab:blrabs}.

\begin{figure*}
    \centering
    \includegraphics[width = .9 \linewidth]{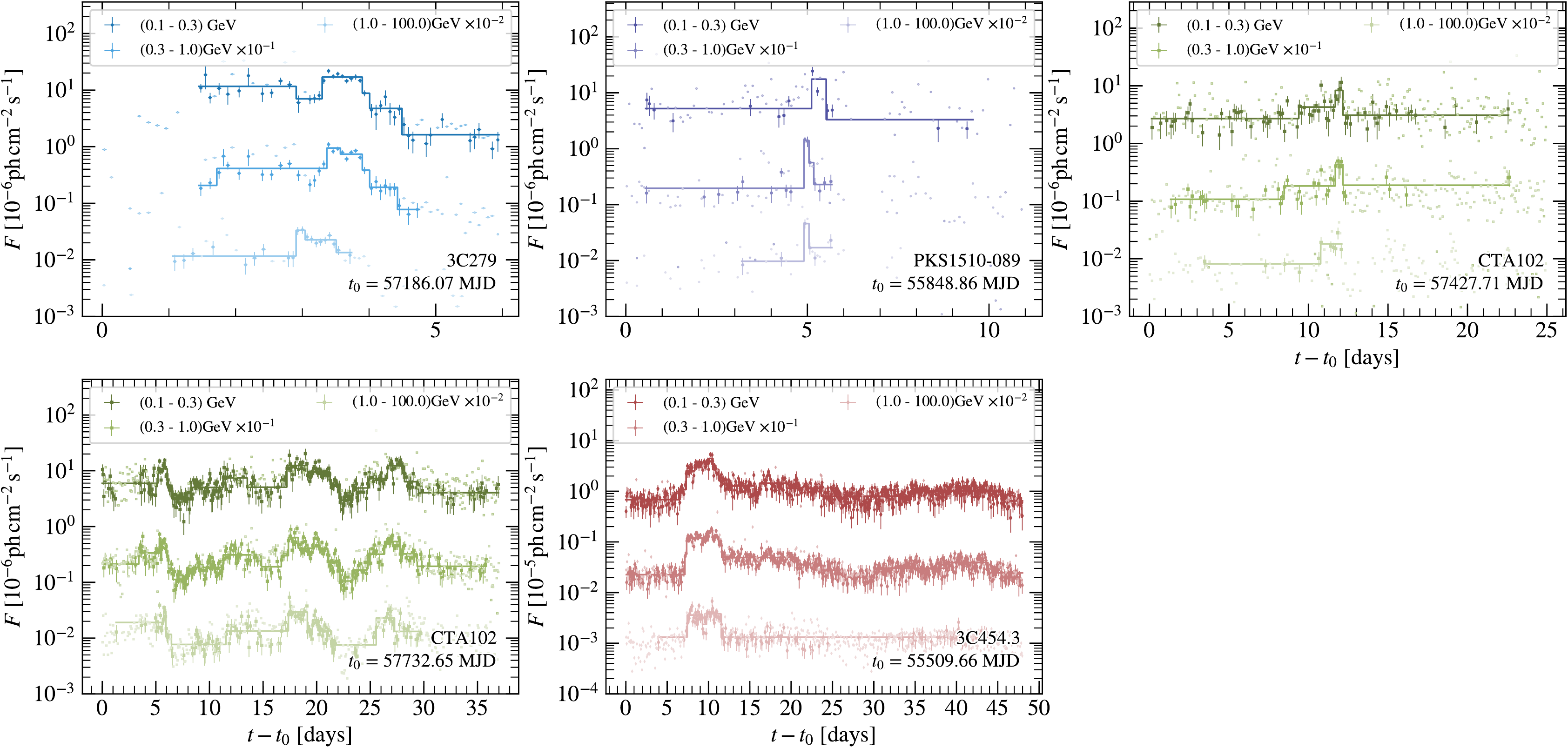}
    \caption{\fermiLAT orbital light curves for three energy bins. The same time binning is used as in Figure~\ref{fig:gti}. The thick colored lines indicate the BB representation. Only light curves are shown for which at least two BBs are identified for each energy bin. The fluxes of the light curves of the energy bins 0.3-1\,GeV and 1-100\,GeV are shifted by $10^{-1}$ and $10^{-2}$, respectively, for better visibility. }
    \label{fig:lcebins}
\end{figure*}

To further investigate the energy dependence of the decay times, we split the energy range of our analysis into three energy bins, 
 from 100\,MeV-300\,MeV, 300\,MeV-1\,GeV, and 1\,GeV-100\,GeV and recompute the orbital light curves.
The energy bins are chosen as a compromise between the number of bins and sufficient photon statistics in each bin. 
The light curves for which at least two BBs are identified in each energy bin are shown in Figure~\ref{fig:lcebins}.
We repeat the fits of the exponential profiles to the energy-dependent light curves but allow only one flare profile per HOP group. 
The resulting decay times are also plotted in Figure~\ref{fig:tcool}. 
Only for the 300\,MeV-1\,GeV energy bin is the double peak of the flare resolved, and in general, the fit qualities are rather poor with $\chi^2$ per degree of freedom between 1.67 and 2.26.
The steep $\chi^2$ curves also explain the rather small error bars on $\tau_\mathrm{decay}$. 
From the fit values, we cannot draw a conclusion as to whether the decay times evolve with $(E_\mathrm{obs})^{-1/2}$ as expected in the Thomson regime.
Due to the lack of high photon statistics at energies beyond 10\,GeV, where the differences in cooling times between the dust torus and BLR become more pronounced, we are not able to use the method suggested by \citet{2012ApJ...758L..15D} to determine the photon field dominating the IC scattering. 
This conclusion also holds for the other sources. 

\begin{figure*}
    \centering
    \includegraphics[width = .9 \linewidth]{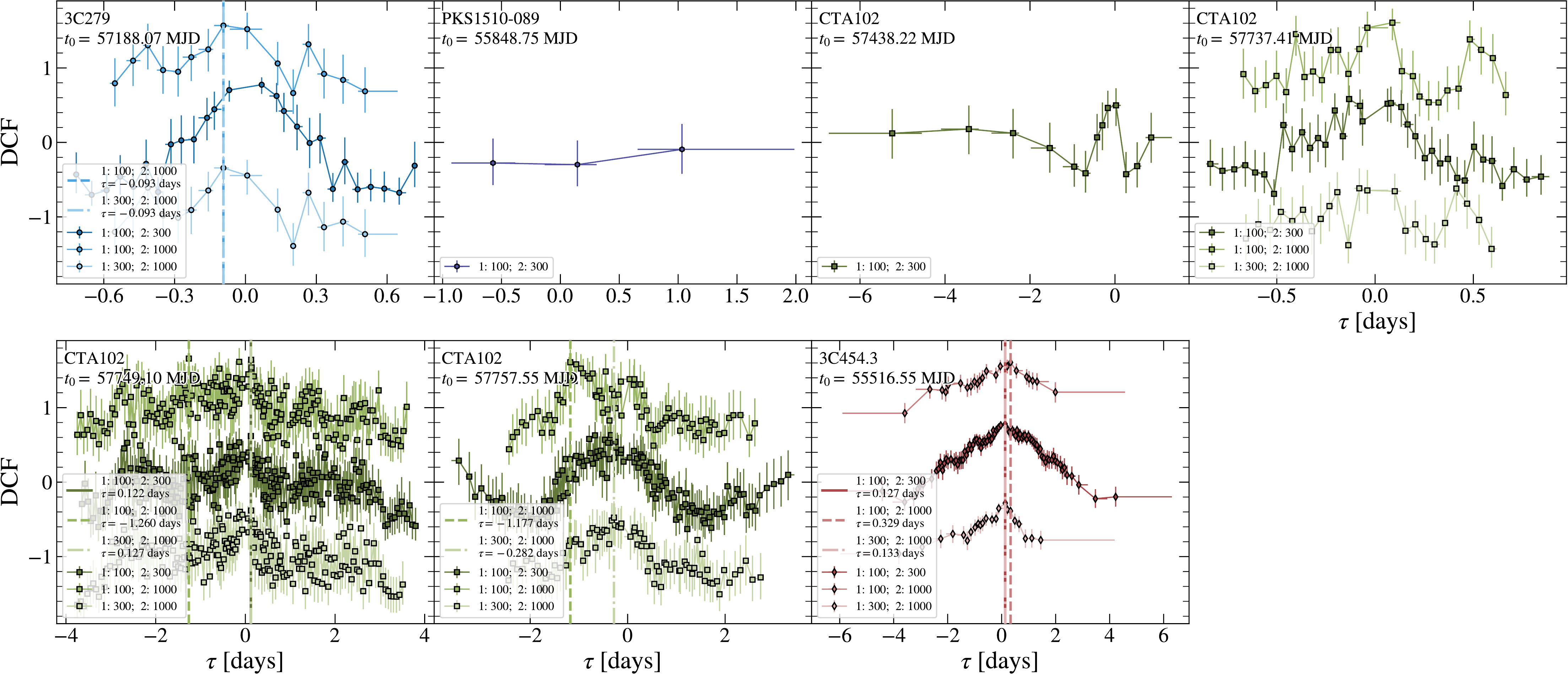}
    \caption{Results for the DCF analysis for the light curves in Figure~\ref{fig:lcebins}. In order to detect time lags for single flares, the three flares of CTA~102 starting at MJD 57,733 are separated using the HOP groups. For time lags $\tau < 0$, the high-energy light curve leads the low-energy one. The DCFs between the energy bins (0.1-300\,MeV,1-100\,GeV) and (0.3-0.3\,GeV,1-100\,GeV) are shifted by $\pm 1$ for better visibility. 
    Vertical lines mark the maximum of the DCFs if $\mathrm{max}(\mathrm{DCF}) > 2 \sqrt{\mathrm{Var}(\mathrm{DCF})}$.}
    \label{fig:zdcf}
\end{figure*}

Interestingly, the BBs for the energy-dependent light curves of the flares of 3C~279, PKS~1510-089, and the last flare of CTA~102 seem to show time lags between the energy bands, with the high-energy emission leading the low-energy \Grays.
For the 3C~279 flare around MJD 57,188, \citet{2015ApJ...808L..48P} could not find any time lags between the energy bins 0.1-1\,GeV and above 1\,GeV using the $Z$-transformed discrete correlation function \citep[DCF;][]{1997ASSL..218..163A,2013arXiv1302.1508A}.
Using the same methodology, we show the DCFs for our energy-dependent light curves with at least two BBs per energy bin in Figure~\ref{fig:zdcf}.
We mark the time lags $\tau$ with horizontal lines at the maximum DCF values if $\mathrm{max}(\mathrm{DCF}) > 2 \sqrt{\mathrm{Var}(\mathrm{DCF})}$.
In contrast to \citet{2015ApJ...808L..48P}, we find evidence that the emission above 1\,GeV leads the emission at lower energies with $\sim 0.1$\,days. 
However, from the fits to the light curves, the decay time at higher energies is actually longer ($0.45\pm0.16$\,days above 1\,GeV versus $0.22 \pm 0.03$ between 0.1 and 0.3\,GeV). 
Therefore, the lag might not be associated with cooling but rather with a changing particle injection. 
From the spectral variation (Figure~\ref{fig:specvar}) it seems that the time bin before the peak of the flare has a harder spectrum; however, the uncertainties are too large to draw firm conclusions.
For the CTA~102 flare around MJD 57,758 we also find that the high-energy emission is leading, whereas for the flare at MJD 57,749 the picture is reversed. 
For 3C~454.3 the DCF also indicates that the low-energy emission is leading the high-energy emission, again suggesting that these lags are connected to the injection of particles rather than radiative cooling.

\subsubsection{Synchrotron \GRays}
\label{sec:gammasync}

A second alternative to the standard model is that the \gray emission mechanism is electron synchrotron radiation \citep{TheFermi-LAT:2016dss}, not IC scattering, as usually supposed \citep[e.g.,][]{Madejski:2016oqg}.
Electron synchrotron radiation is mostly dismissed because there is a $\sim70\,{\rm MeV}$ radiation reaction limit on the \gray energy in the comoving frame \citep[e.g.,][]{1975ctf..book.....L,blandford:2017mag}. 
However, if there is sufficient plasma entrainment beyond the outer light cylinder of the black hole magnetosphere, the dominant, positively charged particles in the jet will be protons even after allowing for some additional pair production. 
Large electric field components along the magnetic field may be created through a dynamical untangling of large magnetic flux ropes at relativistic speed---magnetoluminescence---and, when the plasma density is low, will lead to a conversion of electromagnetic energy to relativistic particles and \Grays across much larger volumes than can be processed by magnetic reconnection. 

Under these circumstances, half of the electromagnetic energy that is dissipated should go into the protons, which can be accelerated to much higher energy than the electrons. For an electromagnetic jet of power $L_{\rm jet}=10^{45}L_{\rm jet\,45}\,{\rm erg\,s}^{-1}$ bulk Lorentz factor $\Gamma_{\rm L}=10\Gamma_{\rm L1}$ and width $s=10^{15}s_{15}\,{\rm cm}\sim r/\Gamma_{\rm L}$, the comoving magnetic field strength is $B'\sim30L_{\rm jet\,45}^{1/2}s_{15}^{-1}\Gamma_{\rm L1}^{-1}\,{\rm G}$, 
the comoving accelerating electric field strength, which we designate as $\cal E'$, could be as large as ${\cal E}'_{\rm max}\sim1L_{\rm jet\,45}^{1/2}s_{15}^{-1}\Gamma_{\rm L1}^{-1}\,{\rm MV\,m}^{-1}$
and the total potential difference across the jet could be as large as $V_{\rm max}\sim100\,L_{\rm jet\,45}^{1/2}\,{\rm EV}$. 

Proton acceleration is likely to be limited by the Bethe-Heitler process, where a photon of energy $E''_\gamma>1\,{\rm MeV}$ in the proton rest frame creates an electron-positron pair with a cross section that rises slowly from $\sigma_{\rm BH}\sim1\,{\rm mb}$ at $E_\gamma''\sim5\,{\rm MeV}$ to $\sim 10\,{\rm mb}$ at $E_\gamma''\sim1\,{\rm GeV}$ \citep[e.g.,][]{2009herb.book.....D}. Pions will be created at higher energy when $E''_{\gamma}\gtrsim150\,{\rm MeV}$ and could be responsible for very high-energy neutrino emission but need not concern us here. 

If we focus on the $\sim3\,{\rm min}$ flare in 3C~279 with 
$r_g\sim5.6\times10^{13}\,{\rm cm}$, 
$L_{\rm jet\,45}\sim1$, and assume that $\Gamma_{\rm L}\sim10$, then the constraint of Equation~\ref{eq:rblob} suggests that the size of the emitting region associated with the flare is $R'_{\rm blob}\sim10^{14}\,{\rm cm}$. There is a second constraint in that the electromagnetic energy contained within the blob should be large enough to account for the amplitude of the flare. This suggests that $r\sim10^{16}\,{\rm cm}$, $s_{15}\sim1$ and a fraction $\sim0.01$ of the jet area is involved with this flare. 

Next, suppose that the inner jet is effectively shielded blueward of the Lyman continuum, and so the highest-energy external photons in the jet originating from the accretion disk, with energy $E_{\rm UV}\sim10\,{\rm eV}$, will have a number density $n_{\rm UV}\sim10^{12}\,{\rm cm}^{-3}$ (assuming a distance where the photon density is the logarithmic mean between $10^9\,\mathrm{cm}^{-3}$ at $R_\mathrm{Ly\alpha}$ and the photon density of a blackbody with $T = 3\times10^4\,$K, which gives $n\sim10^{15}\,\mathrm{cm}^{-3}$) and energy density $\sim10\,{\rm erg\,cm}^{-3}$, roughly a tenth of the magnetic energy density. These photons will have energies $E_\gamma''\sim\Gamma_{\rm L}\gamma_p'E_{\rm UV}\sim100\,{\rm MeV}$, where $\gamma_p'\sim10^6$ is the proton Lorentz factor, in the comoving frame. Pairs will then be created at a rate $R'\sim\Gamma_{\rm L}n_{UV}\sigma_{\rm BH}\sim10^{-11}\,{\rm m}^{-1}$ in the comoving frame, and the associated pairs will have energies $E_e'\sim \gamma_P'E_\gamma''\sim100\,{\rm TeV}$. The proton energy loss rate rate in the comoving frame is then $E_e'R'\sim1\,{\rm keV}\,{\rm m}^{-1}\propto\Gamma_L^2\gamma_P'^2$. 
The electric field $\cal E$ needed to balance this loss is only $\sim10^{-3}$ of ${\cal E}_{\rm max}'$. 
The proton acceleration/radiation lengths are then $\sim10^{14}\,{\rm cm}\sim R_{\rm blob}'$ and the protons can be maintained at $\sim\,{\rm PeV}$ energies for the duration of the flare.

The pairs will rapidly cool by synchrotron emission (IC scattering is strongly Klein-Nishina suppressed), radiating \Grays with comoving energy $\sim (E_e'/m_ec^2)^2 B' \sim 1\,{\rm GeV}$ and active galactic nucleus (AGN) frame energies boosted by a factor $\sim\Gamma_{\rm L}$ to energies $E_\gamma\lesssim10\,{\rm GeV}$. 
These \Grays are just below the threshold energy for pair production ($E_\gamma\sim25\,{\rm GeV}$) and should be visible when the line of sight lies within the jet and its absorbing sheath.
The \Grays emitted below the jet $\gamma$-sphere will create more pairs that will emit lower-energy \gray photons, which should escape unimpeded. The overall process is electromagnetic and should be very efficient, unlike with photopion production, where there will be neutrino and neutron losses. 

Electrons will also be directly and rapidly accelerated by the electric field to energies of $\sim300\,{\rm GeV}$ until they are limited through radiating synchrotron \Grays of energy $\sim0.5\,{\rm MeV}$ in the AGN reference frame. These should escape unimpeded with comparable power to the GeV \Grays and could be detectable. (IC scattering is also Klein-Nishina suppressed but could be significant.) In order to dissipate the energy at relativistic speed, the current density must be $\sim3\,\mu{\rm A\,m}^{-2}$, and the associated proton pressure would have to be $\sim100\,{\rm dyne\,cm}^{-2}$, comparable with the magnetic pressure at the height of the flare.

The  case of 3C~279 is extreme and may require specialized, not generic, conditions, including especially the necessary efficacy of the shielding at photon energies above the Lyman continuum. However, even in this case, it seems that the surprisingly rapid variation observed can be explained by making simple, though not mandatory, assumptions. Modeling the larger sample of variable FSRQs described here introduces many more possibilities. In particular, the presumption that the emission originates in a single ``zone,'', while appropriate for an extreme flare, is surely quite wrong when modeling a more slowly varying \gray spectrum. Most of the emission is likely to originate over a range of larger jet radii with lower radiation density. The details will be largely dictated by the interaction of the jet with the surrounding outflow and the dynamics of the jet electromagnetic field.

A fuller account of the processes involved will be presented elsewhere.

\subsection{Results from Radio-\GRay correlation analysis}
\label{sec:gammaradio}

Cross-correlating \gray light curves with radio light curves provides an alternative method to locate the \gray emitting region~\citep[e.g.,][]{2014MNRAS.441.1899F}.
Under the assumption that the flares are produced in a common compact emission  region moving down the jet~\citep[e.g.,][]{2014MNRAS.445..428M},
the distance $d_{\gamma,r\nu}$ between the 
the \gray sphere, where the \gray opacity due to, e.g., absorption in the BLR, becomes less than unity~\citep{1995ApJ...441...79B}, and the 
radio core, where synchrotron self-absorption becomes negligible~\citep[][]{1981ApJ...243..700K}, 
can be estimated from the time lag between the light curves, 
\begin{equation}
    d_{\gamma,r\nu} = \frac{\Gamma\delta_\mathrm{D}\beta c\tau_{\mathrm{peak},\gamma,r\nu}}{1 + z},
    \label{eq:dgamma-r}
\end{equation}
where $\tau_{\mathrm{peak},\gamma,r\nu}$ is the time lag corresponding to a peak in the cross-correlation function between the \gray and radio light curves obtained at frequency $\nu$.
Under the assumption that the radio emission lags the \Grays, the distance of the \gray emission region to the central black hole is thus $d_\gamma = d_{\mathrm{core},r\nu} - d_{\gamma,r\nu}$, where $d_{\mathrm{core},r\nu}$ is the position of the radio core at frequency $\nu$.
The core position itself is frequency-dependent~\citep[the core shift effect; see, e.g.,][]{1998A&A...330...79L},
\begin{equation}
    d_{\mathrm{core},r\nu} = \frac{\Omega_{r\nu}}{\nu^{1/k_r}\sin\theta_\mathrm{obs}},
     \label{eq:core-shift1}
\end{equation}
where $k_r$ depends on the electron energy spectrum and the magnetic field in the emitting region~\citep{1981ApJ...243..700K} and
\begin{equation}
    \Omega_{r\nu} = 4.85\times10^{-9} \frac{\Delta r_\mathrm{mas} d_\mathrm{L}}{1 + z}\frac{\nu^{1/k_r}\nu_0^{1/k_r}}{\nu_0^{1/k_r}-\nu^{1/k_r}},
    \label{eq:core-shift2}
\end{equation}
where $d_\mathrm{L}$ is the luminosity distance and $\Delta r_\mathrm{mas}$ is the offset between the radio cores in milliarcseconds at frequencies $\nu$ and $\nu_0$. 
The offest is related to the time lag between two radio light curves $\tau_{\mathrm{peak},r\nu,r\nu_0}$ through 
$\Delta r_\mathrm{mas} = \mu \tau_{\mathrm{peak},r\nu r\nu_0}$, where $\mu$ is the jet proper motion. 

The proper motion and the core position at 15\,GHz have been determined from
the MOJAVE VLBI blazar monitoring program  \citep[][]{2012A&A...545A.113P,2016AJ....152...12L}.
The following distances $d_\mathrm{core,~15GHz}$ were determined  under the assumption that $k_r = 1$: for PKS~B1222+216 $d_\mathrm{core,~15GHz}= 23.41\,$pc; for 3C~279 $d_\mathrm{core,~15GHz}<7.88$\,pc; for PKS~1510-089 $d_\mathrm{core,~15GHz} = 17.71\,$pc; for 
CTA~102 $d_\mathrm{core,~15GHz} =46.7\,$pc; and for 3C~454.3 $d_\mathrm{core,~15GHz} = 20.36\,$pc.
Dedicated analyses have also been carried out and found for 3C~454.3 $k_r = 0.6$-$0.8$ and $d_\mathrm{core,~15GHz} \sim 38\,$pc, and, since $d_{\mathrm{core},r\nu}\propto\nu^{-1/k_r}$,  $d_\mathrm{core,~43GHz} \sim 9\,$pc~\citep{2014MNRAS.437.3396K}. 
For 3C~273 \citet{2013ARep...57...34V} found $k_r = 1.4$ and $d_{\mathrm{core},r\nu} = 134\nu^{-1/1.4}$ using radio observations at frequencies between 4.8 and 362 GHz.
Lastly, \citet{2015A&A...576A..43F} conducted VLBA observations of CTA~102 ranging from 5 to 86\,GHz and found $k_r = 1.0$ as a best-fit value and $d_{\mathrm{core,~86GHz}}\sim7\,$pc.
Provided that we can estimate $\tau_{\mathrm{peak,15GHz},r\nu}$, it is possible with the above results to estimate the core position at an arbitrary radio frequency using Eqs.~\ref{eq:core-shift1} and \ref{eq:core-shift2}.
In order to arrive at an estimate for $d_\gamma$ the only remaining task is to perform a cross-correlation study between \gray and radio light curves.

\begin{figure*}
    \centering
    \includegraphics[width = .8\linewidth]{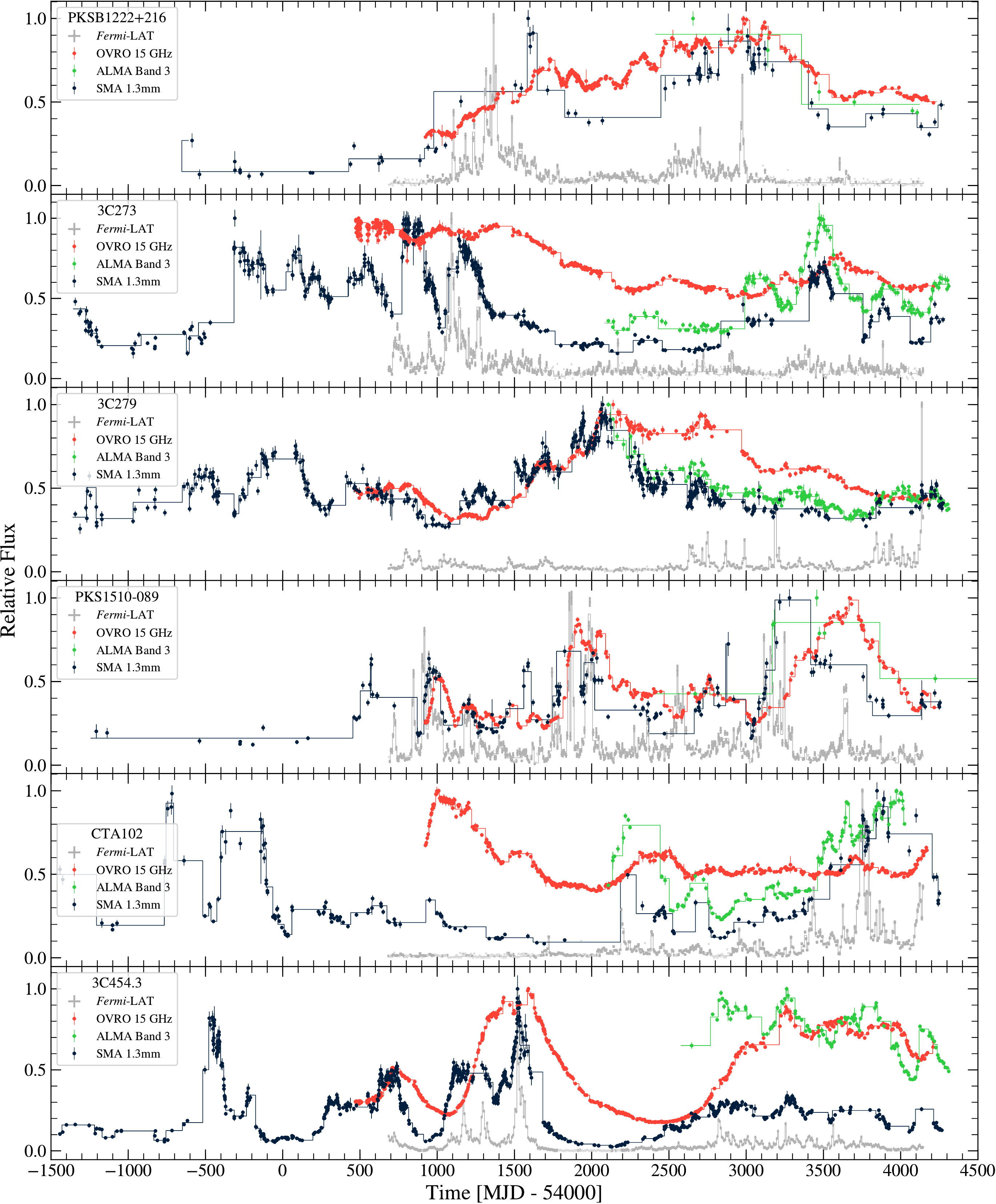}
    \caption{Radio and \gray light curves normalized to the respective maximum flux.}
    \label{fig:lc-radio}
\end{figure*}

We search for time lags between the \fermiLAT light curves and radio light curves obtained with the Owens Valley Radio Observatory (OVRO) at 15\,GHz, the Atacama Large submillimeter/millimeter Array (ALMA) between 84 and 116\,GHz (Band 3, 3.6-2.6\,mm), and the Submillimeter Array (SMA) at 230\,GHz (1.3\,mm). 
All of the studied FSRQs are included in an ongoing blazar monitoring program at OVRO~\citep{2011ApJS..194...29R} and SMA~\citep{2007ASPC..375..234G}, and also serve as calibrators for SMA and ALMA~\citep{2018MNRAS.478.1512B} at millimeter wavelengths.\footnote{Data from the observatories are available at \url{http://www.astro.caltech.edu/ovroblazars}, \url{https://almascience.eso.org/alma-data/calibrator-catalogue}, and \url{http://sma1.sma.hawaii.edu/callist/callist.html}.}
We show all radio and \gray light curves in Figure~\ref{fig:lc-radio}.
It is evident that the OVRO light curves show variations on longer time scales and with less flicker noise behavior. 
For 3C~454.3 there appears to be a correlation between the radio and \gray flux, at least for the giant flare in 2010 (around MJD 55,500). 
Due to scarce and uneven sampling, we do not use the ALMA and SMA light curves of PKS~B1222+216 and PKS~1510-089.
We also do not include the SMA light curve of CTA~102 in the following analysis for the same reason.

To quantify the correlations, we again closely follow the methodology laid out by \citet{2014MNRAS.445..437M}. 
For two light curves with fluxes $a_i$ and $b_j$ 
measured at times $t_{ai}$ and $t_{bj}$ we compute the local cross-correlation function (LCCF)
\begin{equation}
\mathrm{LCCF}(\tau) = \frac{1}{M}\frac{\sum(a_i - \bar{a}_\tau)(b_j - \bar{b}_\tau)}{\sigma_{a\tau}\sigma_{b\tau}},
\end{equation}
where the sum runs over the $M$ pairs for which $\tau \leqslant t_{ai} - t_{bj} < \tau + \Delta t$ for some chosen time step $\Delta t$, and $\bar{a}_\tau$ and $\bar{b}_\tau$ and $\sigma_{a\tau}$ and $\sigma_{b\tau}$ are the flux averages and standard deviations over the $M$ pairs, respectively~\citep{1999PASP..111.1347W}. 
The LCCF is bound between $-1$ and $1$ and has much larger efficiency in recovering linear correlations between light curves compared to the DCF \citep{2014MNRAS.445..437M}.
For the binning of the time lags $\tau$ we choose half the maximum of the median of the time separations between consecutive data points in the two light curves.
The minimum and maximum values of $\tau$ are chosen to be $\pm0.5$ times the length of the shortest light curve~\citep{2014MNRAS.445..428M}.

We determine the significance of a peak in the LCCF by cross-correlating pairs of simulated light curves. 
For the \gray light curves, the simulation proceeds in the same way as described in Section~\ref{sec:results-global}, where we use the best-fit values $\hat{\beta}$ for the assumed PSDs.
For the radio light curves, we proceed in a similar way. 
First, we determine the best-fit PSDs similar to the \gray light curves. 
In order to achieve a good fit between the observed and simulated PSDs, we change the methodology used for the \gray light curves in the following ways.
Instead of matching the flux probability distribution of the light curves, as suggested by \citet{2013MNRAS.433..907E}, we use variance matching~\citep{2014MNRAS.445..437M}.\footnote{For radio light curves, the best-fit slopes are much closer to $\beta \sim 2$, so that Parseval's theorem applies.}
Furthermore, we do not apply uncertainties to the simulated light curves. 
Doing so generally leads to a strong flattening of the periodograms at high frequencies when they become dominated by white noise introduced by the uncertainties. 
This is not observed in the periodogram derived from observations. 
The reason might be a correlation between uncertainties and flux, which is not taken into account by the adopted simulation scheme and can lead to an overestimation of the simulated uncertainties.
Lastly, the radio light curves are unevenly sampled and can show large observational gaps. 
Simply applying the interpolation scheme used for the the \gray light curves would mean that most data points that enter the calculation of the periodogram are actually interpolated. 
To mitigate this problem, we split the light curves where they show large gaps. 
We found that a split at gaps that are 20 (4.5) times larger than the median separation between consecutive measurements for the OVRO and SMA (ALMA) light curves provides a good compromise between minimizing the number of splits and too few data points within a light-curve segment. 
Furthermore, for the interpolated light curves, we use a time step equal to the 80\,\% quantile of the observed separation (the median would correspond to the 50\% quantile). 
In this way, we lose sensitivity to the highest frequencies but end up with interpolated light curves with roughly the same number of data points as the observed ones. 
We do not average the interpolated flux points as in the \gray case, since radio observations are usually short in duration and report flux densities instead of integrated fluxes. 
The periodograms of the individual light-curve segments are finally log-averaged following \citet{1993MNRAS.261..612P}.

We report the best-fit slopes of the assumed power-law PSDs $\hat{\beta}$, their confidence interval, and the $p_\beta$-value of the fits in Table~\ref{tab:lccf}.
The confidence interval and $p_\beta$-value are determined in the same way as described in Section~\ref{sec:results-global}.
All fits show a high fit quality. 
For the SMA and OVRO light curves for 3C~454.3, as well as for the 3C~273 light curve obtained with ALMA, we are only able to provide upper bounds on $\beta$.
In general, we confirm the trend that the OVRO light curves show a softer PSD $\hat{\beta} \gtrsim 2$ for all sources. 
Moving to higher frequencies with ALMA and SMA, the PSD hardens and becomes more flicker noise-like. 
Comparing our results for the OVRO light curves to previous analyses of \citet{2014MNRAS.445..428M}, who used 4~yr of data, we find them to be consistent within the uncertainties. 

Having determined the best-fit PSDs, we use them to create artificial light curves in the same way as for fitting the periodogram itself.
We then calculate the LCCF between 5000 pairs of uncorrelated simulated light curves and derive confidence bands on the LCCF. We use the confidence bands to determine the $p_\tau$-value, which gives the probability of finding an LCCF value at a given $\tau$ greater or equal to the observed value under the assumption that the light curves are uncorrelated. 

For observed LCCFs where we find time lags with a significance $1 - p_\tau > 0.95$, we estimate the uncertainty on the peak time using flux randomization and random subsample selection (drawing 1000 samples) following \citet{1998PASP..110..660P}, as suggested by \citet{2014MNRAS.445..437M}.

\begin{deluxetable*}{l|cc|ccc}
\tablewidth{0pt}
\tablecaption{ \label{tab:lccf}Results from PSD analysis of radio light curves, as well as \gray and radio LCCF results.}
\tablehead{\colhead{Source} & \colhead{$\hat{\beta}$} & \colhead{$p_\beta$} & \colhead{$\tau_\mathrm{peak}$ [days]} & \colhead{$p_\tau$} & \colhead{$d_{\gamma, r}$ [pc]}}
\startdata
\hline
\multicolumn{6}{c}{OVRO}\\
\hline
PKS~B1222+216 & $1.92^{+0.39}_{-0.59}$ & 0.59 & --- & --- & ---\\
3C~273 & $2.38^{+0.30}_{-0.97}$ & 0.94 & $-416.5^{+217.0}_{-140.0}$ & 0.0068 & $10.96~[5.2,14.6]~\pm4.4$\\
3C~279 & $2.29^{+0.32}_{-0.94}$ & 0.71 & --- & --- & ---\\
PKS~1510-089 & $1.89^{+0.45}_{-0.84}$ & 0.34 & --- & --- & ---\\
CTA~102 & $2.23^{+0.26}_{-0.92}$ & 0.84 & --- & --- & ---\\
3C~454.3 & $2.20^{+0.36}_{-2.20}$ & 0.40 & $-101.5^{+49.0}_{-112.0}$ & 0.0156 & $15.39~[8.0,32.4]~\pm2.8$\\
\hline
\multicolumn{6}{c}{ALMA Band 3}\\
\hline
3C~273 & $2.12^{+0.40}_{-2.12}$ & 0.73 & --- & --- & ---\\
3C~279 & $1.82^{+0.38}_{-0.45}$ & 0.89 & --- & --- & ---\\
CTA~102 & $1.94^{+0.42}_{-1.33}$ & 0.45 & $-216.0^{+209.0}_{-11.0}$ & 0.0092 & $58.85~[1.9,61.8]~\pm7.3$\\
3C~454.3 & $1.73^{+0.36}_{-0.30}$ & 0.25 & $-27.0^{+30.0}_{-30.0}$ & 0.0164 & $4.09~[-0.5,8.6]~\pm0.7$\\
\hline
\multicolumn{6}{c}{SMA 1.3\,mm}\\
\hline
3C~273 & $1.48^{+0.40}_{-0.33}$ & 0.17 & $-122.5^{+84.0}_{-7.0}$ & 0.0088 & $3.22~[1.0,3.4]~\pm1.3$\\
3C~279 & $1.61^{+0.16}_{-0.28}$ & 0.97 & --- & --- & ---\\
3C~454.3 & $1.64^{+0.31}_{-1.64}$ & 0.21 & $10.5^{+21.0}_{-28.0}$ & 0.0002 & $-1.59~[-4.8,2.7]~\pm0.3$\\
\enddata
{
\tablecomments{For the LCCF analysis, we only report time lags $< 0 $ and with a significance $p_\tau < 0.05$. The range of $d_{\gamma, r}$ values reported in square brackets is due to the uncertainty on $\tau_\mathrm{peak}$, the remaining uncertainties are the propagated errors on $\Gamma$ and $\delta_\mathrm{D}$.}
}
\end{deluxetable*}

\begin{figure*}
    \centering
    \includegraphics[width = .32\linewidth]{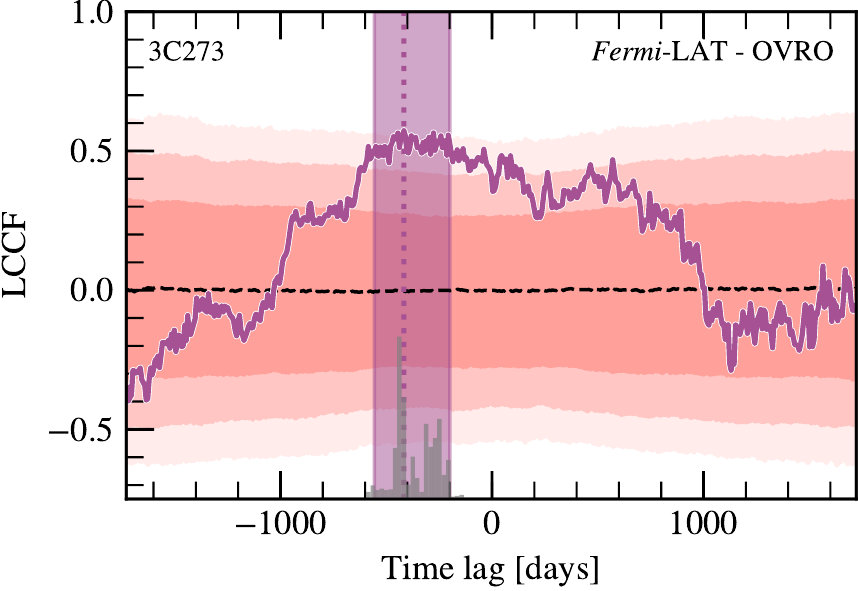}
    \includegraphics[width = .32\linewidth]{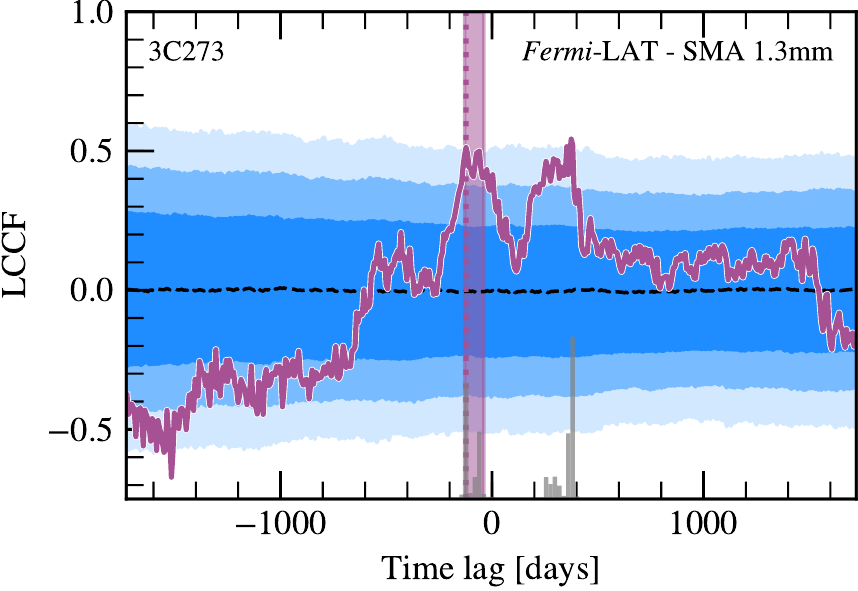}
    \includegraphics[width = .32\linewidth]{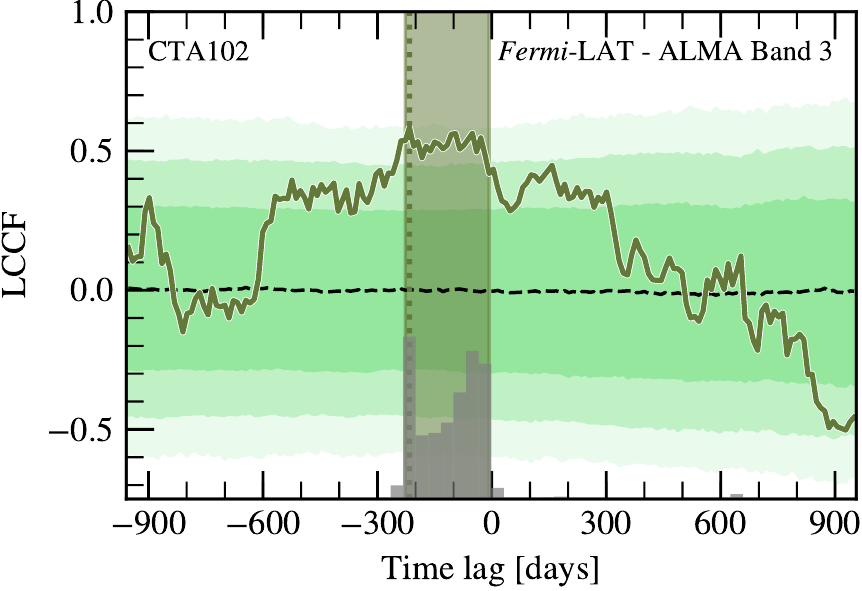}
    \includegraphics[width = .32\linewidth]{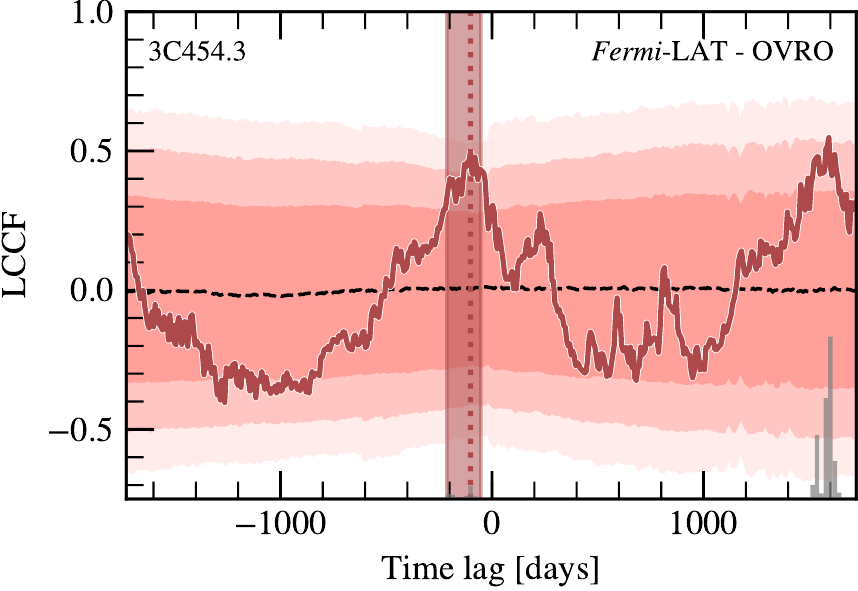}
    \includegraphics[width = .32\linewidth]{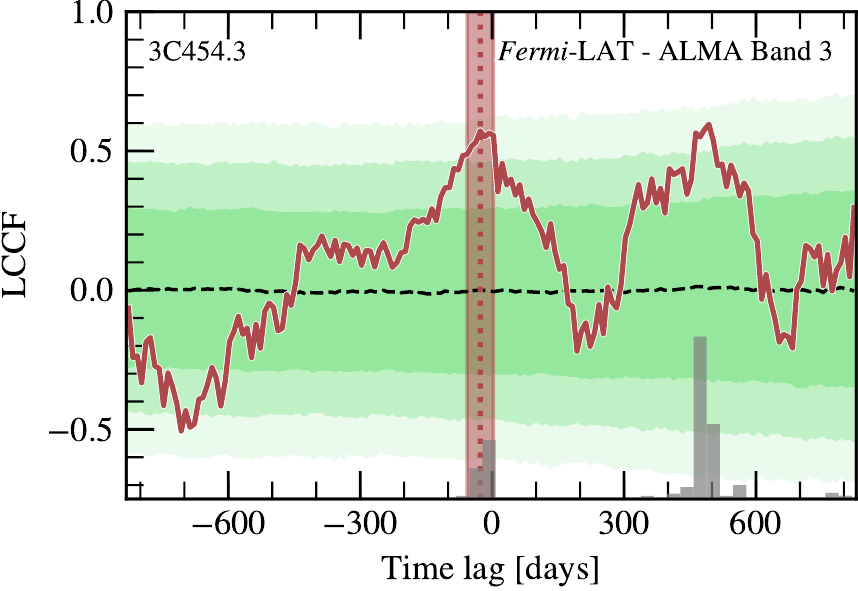}
    \includegraphics[width = .32\linewidth]{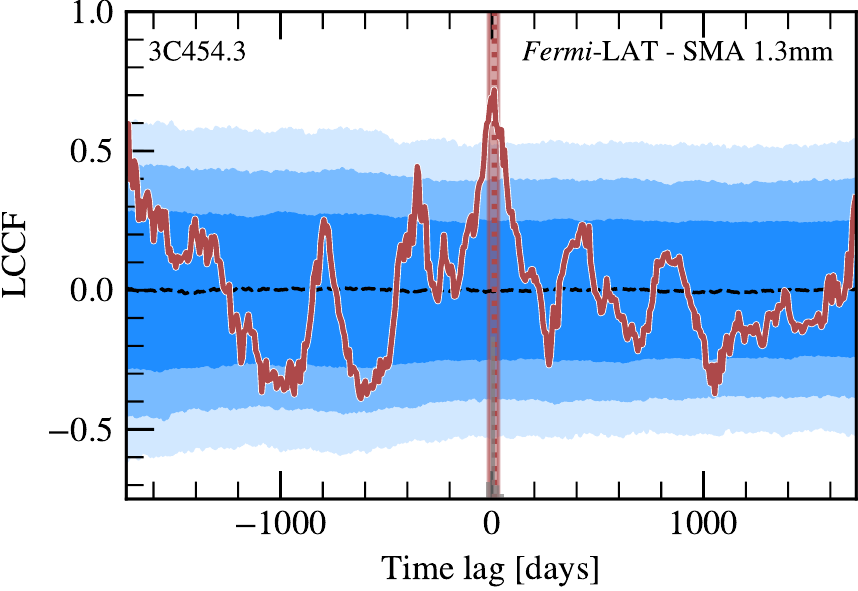}
    \caption{The LCCFs between \gray and radio light curves for the cases where a significant time lag, $p_\tau < 0.05$, is found. The vertical dotted line and shaded region show the time lag peaks (for $\tau < 0$) and their uncertainty. 
    The colored regions denote the 68\,\%, 95\,\%, and 99\,\% envelopes (from dark to light) derived from simulated uncorrelated light-curve pairs. 
    The gray histograms show the number of $\tau_\mathrm{peak}$ values obtained from flux randomization and random subsample selection to estimate the uncertainty on $\tau_\mathrm{peak}$.}
    \label{fig:lccf}
\end{figure*}

We show the observed LCCF between \gray and radio light curves in Figure~\ref{fig:lccf} for the cases where we find a significant peak in the LCCF. This is the case for 3C~273 (OVRO and SMA), CTA~102 (ALMA), and 3C~454.3 (all radio and millimeter light curves correlate with the \gray light curve).
The correlation between the SMA and \gray light curve of 3C~454.3 is the most significant, with $p_\tau = 2\times10^{-4}$ ($3.72\sigma$). 
The peak and the uncertainties are marked by a dotted line and shaded region in Figure~\ref{fig:lccf}.
They are also summarized, together with the $p_\tau$ values, in Table~\ref{tab:lccf}.
In the table, we only consider peaks with $\tau < 0$ (the \gray light curve leads the radio light curve); however, 
for 3C~273 and 3C~454.3 peaks with $\tau > 0$ are also visible. 
For 3C~454.3 these peaks occur at large values of $\tau$ that might be due to the several small flares observed at \gray energies. Since the overlap between the light curves is smaller for larger values of $|\tau|$, we deem these peaks less credible. 
For 3C~273, the situation is less clear. 
The SMA light curve shows two prominent flares around the 
\gray flare, which gives rise to the two peaks in the LCCF. 
The low state of the source in recent years and the less dense sampling of the SMA light curve render it difficult to draw firm conclusions. As we see below, even the peak at $\tau < 0$ does not lead to constraints on the position of the \gray emitting region.

Using Equaiton~\ref{eq:dgamma-r} and the peaks identified in the LCCF, we can now derive the distance between the \gray and radio emitting regions. 
For 3C~454.3 the time lags between the \gray and millimeter light curves of ALMA and SMA are consistent with zero; hence, $d_{\gamma,r}$ is consistent with zero as well.
For the correlation with the OVRO light curve, a longer time lag of $\tau_{\mathrm{peak},\gamma,15\mathrm{GHz}} = -102$~days is found, placing $d_{\gamma,r}$ between $\sim 5$ and 35\,pc.
This time lag is consistent with the recent DCF analysis carried out by \citet{2018MNRAS.480.5517L}, who found a time lag of $(115\pm6)$\,days at $2.5\,\sigma$ significance using 8\,yr of OVRO data and the \fermiLAT weekly monitored light curve. 

For CTA~102, we only find a significant lag between the ALMA and \gray light curves. 
The time lag translates into $d_{\gamma,r} \sim [-5;69]\,$pc, taking uncertainties on $\tau_{\mathrm{peak},\gamma,100\mathrm{GHz}}$ as well as $\Gamma$ and $\delta_\mathrm{D}$ into account.
Hence, the distance is also consistent with zero. 

We also find a significant correlation between the \gray light curve of 3C~273 and both light curves of SMA and OVRO. 
For the 15\,GHz OVRO light curve, $d_{\gamma,r}$ is found between 0.8 and 19\,pc, whereas for 230\,GHz, the distance falls between $-0.7$ and 4.7\,pc and is consistent with zero. 

In order to determine the core positions at the $\sim100$\,GHz (ALMA) and $230\,$GHz (SMA) cores, we carry out a cross-correlation between the radio light curves to derive $\tau_{\mathrm{peak},15\mathrm{GHz},r\nu}$.
The results are reported in Table~\ref{tab:lccf-radio}.
Using Equations~\ref{eq:core-shift1} and \ref{eq:core-shift2}, we arrive at the new core position, which we also show in Table~\ref{tab:lccf-radio}, where we use the average values of the measured jet proper motion~\citep[see Table 4 in][]{2016AJ....152...12L}, the observation angle $\theta_\mathrm{obs}$ reported in \citet{2017ApJ...846...98J},
and the values of $k_r$ found in the dedicated analyses discussed above. 
For 3C~273 and 3C~454.3, we find that our core positions are consistent with values calculated from the $d_\mathrm{core}\propto\nu^{-1/k_r}$ relation obtained by \citet{2013ARep...57...34V} and \citet{2014MNRAS.437.3396K}, respectively. 
This is a nontrivial result, since we have combined our time lags with jet proper motions from MOJAVE and jet properties from VLBA observations. 

Combining the core positions and $d_{\gamma,r}$ we find that for 3C~454.3 the \gray emitting region is consistent with the position of SMA millimeter core at $\sim 0.8^{+0.4}_{-0.5}$\,pc, also in agreement with the LCCF from ALMA. The OVRO result is only marginally in agreement with this result, suggesting instead that $d_\gamma \gtrsim 3\,$pc.
Given the larger uncertainty on the time lag and lower significance of the correlation, we deem the results at millimeter wavelengths as more robust. 
They also agree with the findings of \citet{2014MNRAS.441.1899F}.

For CTA~102 we do not find a significant correlation with the OVRO light curve; therefore, we use the core position at 86\,GHz, $d_{\mathrm{core},86\,\mathrm{GHz}} \sim 7\,$pc~\citep{2015A&A...576A..43F}.
With the results on $d_{\gamma,r}$, we also find that for CTA~102 the distance of the \gray emitting region is consistent with the millimeter core.

For 3C~273, we find that the derived core shift between 230 and 15\,GHz of $\sim 4\,$pc is not consistent with the $\nu^{-1/1.4}$ relation obtained by \citet[][]{2013ARep...57...34V}. 
Nevertheless, within the uncertainties, $d_{\gamma,r}$ is also consistent with zero, and hence \gray and millimeter emission could be produced cospatially.

\begin{deluxetable}{lccc}
\tablewidth{0pt}
\tablecaption{ \label{tab:lccf-radio}Results for time lags and core positions from a radio/radio LCCF analysis.}
\tablehead{\colhead{Source} & \colhead{$\tau_{\mathrm{peak},r\nu_1,r\nu_2}$ [days]} & \colhead{$p_{\tau}$}  & \colhead{$d_{\mathrm{core},r\nu_1}$ [pc]}  }
\startdata
\hline
\multicolumn{4}{c}{ALMA Band 3 \& OVRO}\\
\hline
3C~273 & $-161^{+72}_{-36}$ & 0.0222 & $1.5~[0.8,1.8]~\pm0.6$ \\
3C~279 & $-622^{+154}_{-168}$ & 0.0036 & $8.1~[6.1,10.3]~\pm2.6$ \\
CTA~102 & --- & --- & --- \\
3C~454.3 & $-667^{+100}_{-30}$ & 0.0782 & $4.6~[3.9,4.8]~\pm2.6$ \\
\hline
\multicolumn{4}{c}{SMA 1.3mm \& OVRO}\\
\hline
3C~273 & $-427^{+293}_{-87}$ & 0.0256 & $4.0~[1.2,4.8]~\pm1.5$ \\
3C~279 & $-165^{+12}_{-125}$ & 0.0152 & $2.2~[2.0,3.8]~\pm0.7$ \\
3C~454.3 & $-110^{+14}_{-0}$ & 0.0924 & $0.8~[0.7,0.8]~\pm0.4$ \\
\hline
\multicolumn{4}{c}{SMA 1.3\,mm and ALMA Band 3}\\
\hline
3C~273 & $1^{+9}_{-36}$ & 0.0000 & --- \\
3C~279 & $-188^{+175}_{-119}$ & 0.0002 & --- \\
3C~454.3 & $3^{+10}_{-20}$ & 0.0004 & --- \\
\enddata
{
\tablecomments{For the LCCF analysis, we only report time lags and with a significance $p_\tau < 0.1$. The range of core positions in square brackets is due to the uncertainty on $\tau_\mathrm{peak}$, whereas the remaining uncertainties are the propagated errors on $\theta_\mathrm{obs}$ and the jet proper motion $\mu$.}
}
\end{deluxetable}

We conclude this section with a word of caution. 
One should note that a peak in the LCCF does not necessarily measure the lag in the intrinsic processes that generate the ``injections'' that produce flares.
There will be an offset that depends on the relative shapes of the flare light curves at the different wavelenghts. 
To estimate the size of this effect, we compute the autocorrelation functions (ACFs) for the the \gray and  radio/millimeter light curves for the sources for which we find a significant lag. 
For each source, the central peaks of the ACFs in radio/millimeter and \Grays have comparable widths (within a factor of 2); the only exception is for 3C~273, for which the central peak in the ACF of the OVRO light curve is much broader than the ACF of the \emph{Fermi} light curve. 
From this, we conclude that it is possible that the intrinsic flare shapes at the different wavelengths are similar. 
However, one has to keep the caveat in mind that different flare shapes can produce similar ACFs, as demonstrated in Fig. 14 of \citet{1981ApJS...45....1S}.
A more quantitative exploration of this issue will be presented elsewhere.

\section{Summary and Conclusion}
\label{sec:conclusion}

We have carried out a comprehensive temporal and spectral analysis of  \gray data of six FSRQs that have exhibited the brightest \gray flares within 9.5~yr of \fermiLAT observations.
In order to identify flaring episodes in an objective way, we have introduced a novel combination of BBs \citep[][]{2013ApJ...764..167S} and a hill-climbing algorithm inspired by the HOP group finding algorithm~\citep{1998ApJ...498..137E}, see Section~\ref{sec:zoom}.
We have derived daily, orbital, and suborbital light curves for the brightest \gray flares identified in this way.

\subsection{Global Light-curve Properties}
From the weekly binned full 9.5~yr light curves, we are able to determine global temporal properties (Section~\ref{sec:results-global}) such as the flux distributions and PSDs. 
The former are well described with BPLs or lognormal distributions, while the latter, derived following \citet{2014MNRAS.445..437M}, indicate that the light curves show power-law-type PSDs with indices $\beta\sim 1$ indicating flicker noise.
With the exception of 3C~279, our slopes are also compatible within 1,$\sigma$--2\,$\sigma$ with the slopes found in the optical $R$ band for the FSRQs considered by \citet[][3C~273, 3C~279, PKS~1510-089, 3C~454.3]{2012ApJ...749..191C}. 
This could indicate that the emission in the \gray band and at optical wavelengths is produced by the same underlying electron population through external Compton scattering and synchrotron emission, respectively \citep{2014ApJ...791...21F,2015ApJ...809...85F}.  
We also developed a novel objective algorithm to determine the flux level of the QB (see Section~\ref{sec:qb}).
The determination of the quiescent flux can have important implications for emission models for large-scale blazar jet components~\citep{2014ApJ...780L..27M}.
A thorough analysis of the quiescent state of the studied sources will be provided elsewhere.

\subsection{Local Light-curve Properties and Variability on Timescales of Minutes}
We use the light curves with one flux bin per orbit to derive local temporal flare properties by fitting the light curves with exponential  profiles (Section~\ref{sec:results-local}).
The obtained rise and decay times show that rapid variability at time scales of the order of the horizon-crossing timescale, which range from $\sim 0.5$ to $\sim 2$\,hr for the considered black hole masses, are a common feature of all sources.
In general, we find a large variety of flares showing both FRED behavior and the opposite,
similar to the results found by \citet[][see their Figure 10]{2019MNRAS.482..743R}.
No clear trend between flare asymmetry and other flare parameters, such as peak flux, integrated flux, or flare duration, is found. 
No apparent evolution of these quantities with time is observed either. 
This variety of flare profiles could be explained in the scheme of magnetic reconnection with different orientations of the reconnection layers leading to a variety of Doppler factors of the injected plasmoids~\citep[e.g.,][]{2016MNRAS.462.3325P,2018MNRAS.tmp.2522C}. 
With our novel approach of identifying flares, it will be possible in the future to build large statistical samples of flares (selected not only by their peak flux but also, e.g., by their integrated flux) whose properties could be compared in more detail to predictions of the reconnection scenario.
Small and fast plasmoids injected close to the line of sight could also explain minute-scale variability~\citep{2016MNRAS.462.3325P} for which we find evidence at the 2\,$\sigma$ significance level (posttrial) in suborbital light curves of at least two sources, 3C~279 and CTA~102 (Table~\ref{tab:minute}). 
For 3C~454.3 and PKS~1510-089, the BBs also indicate variability on such short timescales; however, a fit with a constant flux to the light curves of these orbits cannot be rejected beyond the $2\,\sigma$ (posttrial) level.
Other possible explanations of such short variability include an energy-dependent kinetic beaming of particles during reconnection~\citep{2012ApJ...754L..33C}, radiative cooling of a plasma accelerated by recollimation shocks~\citep{Bodo:2017qqn}, or synchrotron radiation by electrons accelerated to energies beyond $\gamma' \gtrsim 10^6$~\citep{TheFermi-LAT:2016dss}, a scenario motivated by the \gray flares of the Crab nebula~\citep{2011Sci...331..739A} and also discussed for the flare of PKS~B1222+216~\citep{2012MNRAS.425.2519N}.
We have proposed two alternative scenarios in which the \Grays are shielded from low-energy photons by a plasma sheath (Section \ref{sec:plasma-sheath}) and \Grays are produced by synchrotron emission of electron-positron pairs created by the interaction of protons with low-energy photons (Section \ref{sec:gammasync}). Our discussion is of a qualitative nature only, and a full treatment will be presented elsewhere. 

\subsection{Location of the $\gamma$-Ray Emitting Region}
We have also investigated the location of the \gray emitting region  through three approaches: searches for a spectral cutoff, a comparison of decay times with predictions of radiative cooling times, and a correlation between \gray and radio light curves. 
Using the BLR model of \citet{finke2016}, we find no significant spectral cutoff due to the interaction of \Grays with BLR photons, which places the \gray emission region outside or on the edge of the BLR, $r \gtrsim R_{\mathrm{Ly}\alpha}$ or $\gtrsim 10^3r_g$ (see Table~\ref{tab:blrabs}). 
These lower limits are conservative in the sense that more sophisticated models of the BLR~\citep[e.g.,][]{2017MNRAS.464..152A}, which include continuum emission and a more realistic geometry, predict even larger optical depths than the model used here.
We also do not account for a possible brightening of the BLR emission during \gray flares, as found by \citet{2013ApJ...763L..36L}.
The observed spectra over different time periods are provided for completeness in Appendix~\ref{sec:avg-spec}.

The observed decay times of the brightest flares are compatible with the radiative cooling times predicted from IC scattering of electrons with BLR photons for distances of the \gray emitting regions up to $r\sim 1\,$pc. 
The IC scattering with photons of the dust torus yields cooling times on the order of hours for values of $r$ up to the sublimation radius of $\sim 3\,$pc. 
This is only compatible with a subset of the analyzed flares; see Table~\ref{tab:blrabs}.
In order to reconcile cooling times with minute-scale variability, the emission region would have to be close to the obtained lower limits. 
At the same time, the value of $r$ needs to be compatible with the amount of observed \gray emission. The IC emission scales as $\delta_\mathrm{D}^3 / x^2$, where $x^2 = R_\mathrm{li}^2 + r^2$ \citep[see, e.g., Equation (87) in][]{finke2016}. 
A future comparison of the distance derived from IC emission predicted from multi\-wavelength modeling and cooling times could provide further insight into this issue.

In principle, the energy dependence of the observed decay times can be used to distinguish cooling with BLR and dust torus photons, as proposed by~\citet{2012ApJ...758L..15D}. 
At high electron energies, cooling with BLR photons occurs in the Klein-Nishina regime, whereas cooling with dust torus photons occurs in the Thompson regime (see Figure~\ref{fig:tcool}).
However, the fit of the exponential flare profiles to light curves in different energy bands yields inconclusive results, since the photon statistics are not sufficient to distinguish between the two scenarios.

We find correlations significant beyond the 2.1\,$\sigma$ level between the \gray light curves and radio light curves for 3C~273 (correlations found with OVRO and SMA light curves), CTA~102 (with ALMA), and 3C~454.3 (with OVRA, ALMA, and SMA). 
The time lags between the \gray and millimeter light curves of ALMA and SMA are consistent with zero, which could indicate a cospatial production. 
This is consistent with the picture of superluminal knots passing through a standing shock associated with the radio core, as argued for the flares in PKS~1510-089 and 3C~454.3 \citep{2010ApJ...710L.126M,2012ApJ...758...72W,2013MNRAS.428.2418O}. 
This would entail values of $r$ on the order of parsecs. 
One has to keep in mind, however, that the inferred  uncertainties on the time lags are large. 
Taken together with the uncertainties on the Doppler boost and bulk Lorentz factor, 
smaller values of $r$ cannot be ruled out by our analysis. 

Our three approaches to constrain the location of the \gray emitting region are all consistent with rather large distances from the central black hole, $r \sim 1\,$pc. 
If the distances are even larger, the radiative cooling with BLR photons becomes inefficient, and cooling through IC scattering with dust torus photons does not reproduce the observed flare decay times. 
Such large distances are at odds with the evidence with minute-scale variability, which we, however, only observe at $\sim2\,\sigma$ posttrial significance in the rising parts of the flares. 
Densely sampled light curves at other wavelenghts could provide further insight if this short variability is indeed present and possibly connected to the injection of relativistic particles.
As noted by~\citet{2018MNRAS.477.4749C}, the fact that we do not find significant absorption provides more promising prospects to detect these sources with future observations with the Cerenkov Telescope Array (CTA). 
The improved point-source sensitivity of CTA, together with its energy range between 20\,GeV and 300\,TeV~\citep{2017arXiv170907997C}, could lead to the detection of spectral absorption features during FSRQ flares due to the interaction of \Grays with infrared photons from the dust torus that should become important at $~\sim$TeV energies~\citep[see, e.g., Figure 14 in][]{finke2016}.
The CTA observations could further effectively probe the shortest-variability time scales at very high \gray energies.

\begin{acknowledgments}
We would like to thank the anonymous referee for providing helpful comments on the manuscript.
M.M.  is  a  Feodor-Lynen Fellow and acknowledges the support of the Alexander von Humboldt  Foundation.
We also thank Christoph Wendel for carefully reading the manuscript.
The \textit{Fermi} LAT Collaboration acknowledges generous ongoing support
from a number of agencies and institutes that have supported both the
development and the operation of the LAT as well as scientific data analysis.
These include the National Aeronautics and Space Administration and the
Department of Energy in the United States, the Commissariat \`a l'Energie Atomique
and the Centre National de la Recherche Scientifique / Institut National de Physique
Nucl\'eaire et de Physique des Particules in France, the Agenzia Spaziale Italiana
and the Istituto Nazionale di Fisica Nucleare in Italy, the Ministry of Education,
Culture, Sports, Science and Technology (MEXT), High Energy Accelerator Research
Organization (KEK) and Japan Aerospace Exploration Agency (JAXA) in Japan, and
the K.~A.~Wallenberg Foundation, the Swedish Research Council and the
Swedish National Space Board in Sweden.
Additional support for science analysis during the operations phase is gratefully
acknowledged from the Istituto Nazionale di Astrofisica in Italy and the Centre
National d'\'Etudes Spatiales in France. This work performed in part under DOE
Contract DE-AC02-76SF00515.

The Submillimeter Array is a joint project between the Smithsonian Astrophysical Observatory and the Academia Sinica Institute of Astronomy and Astrophysics and is funded by the Smithsonian Institution and the Academia Sinica.

This research has made use of data from the OVRO 40-m monitoring program~\citep{2011ApJS..194...29R} which is supported in part by NASA grants NNX08AW31G, NNX11A043G, and NNX14AQ89G and NSF grants AST-0808050 and AST-1109911.
This paper also makes use of the following ALMA data: ADS/JAO.ALMA\#2011.0.00001.CAL. ALMA is a partnership of ESO (representing its member states), NSF (USA) and NINS (Japan), together with NRC (Canada), MOST and ASIAA (Taiwan), and KASI (Republic of Korea), in cooperation with the Republic of Chile. The Joint ALMA Observatory is operated by ESO, AUI/NRAO and NAOJ. The National Radio Astronomy Observatory is a facility of the National Science Foundation operated under cooperative agreement by Associated Universities, Inc.
\end{acknowledgments}

\begin{appendix}

\section{Average Spectra for Different Time Intervals}
\label{sec:avg-spec}
In Table~\ref{tab:avg-spec} we present the average observed best-fit spectra for the analyzed FSRQs for different time ranges. 
The same spectral shapes as in the 3FGL are assumed:  either a log-parabola or a power law with superexponential cutoff, which are given by
\begin{eqnarray}
    \frac{\mathrm{d}N_\mathrm{LP}}{\mathrm{d}E} = N_0 \left(\frac{E}{E_0}\right)^{-(\Gamma + \kappa\ln(E / E_0))}, \label{eq:avg-spec-lp}\\
    \frac{\mathrm{d}N_\mathrm{PLsupExp}}{\mathrm{d}E} = N_0 \left(\frac{E}{E_0}\right)^{-\Gamma}\exp\left[-\left(\frac{E}{E_\mathrm{cut}} \right)^{\Gamma_2}\right]\label{eq:avg-spec-plexp}.
\end{eqnarray}

\startlongtable
\begin{deluxetable}{ccccccccc}
\tablewidth{0pt}
\tablecaption{ \label{tab:avg-spec}Average spectral parameters for the time ranges $[t_0, t_0 + \Delta t]$ over which the light curves are derived.}
\tablehead{\colhead{$t_0$} & \colhead{$\Delta t$}  & \colhead{$F(E \geqslant0.1\,\mathrm{GeV})$} & \colhead{$N_0$} & \colhead{$\Gamma$} & \colhead{$\kappa$} & \colhead{$E_\mathrm{cut}$} & \colhead{$\Gamma_2$} & \colhead{$E_0$} \\
\colhead{MJD} & \colhead{days} & \colhead{$[10^{-6}\,\mathrm{cm}^{-2}\,\mathrm{s}^{-1}]$} & \colhead{$[10^{-9}\,\mathrm{MeV}^{-1}\,\mathrm{cm}^{-2}\,\mathrm{s}^{-1}]$} & & & \colhead{[GeV]} & & \colhead{[GeV]}} 
\startdata
\hline
\multicolumn{9}{c}{\textit{Daily Light Curves}}\\
\multicolumn{9}{c}{PKS~B1222+216}\\
 \hline
55,088.65  & 35.00 & $0.866\pm0.040$ &  $0.948\pm0.037$ & $2.063 \pm 0.053$ & $0.042 \pm 0.021$ & $\ldots$ & $\ldots$ & 0.31\\
55,249.65  & 259.00 & $1.557\pm0.018$ &  $1.707\pm0.018$ & $2.117 \pm 0.014$ & $0.064 \pm 0.006$ & $\ldots$ & $\ldots$ & 0.31\\
56,915.65  & 98.00 & $0.815\pm0.026$ &  $0.849\pm0.024$ & $2.298 \pm 0.038$ & $0.074 \pm 0.021$ & $\ldots$ & $\ldots$ & 0.31\\
\hline
\multicolumn{9}{c}{3C~273}\\
 \hline
55,004.65  & 189.00 & $1.196\pm0.022$ &  $2.038\pm0.034$ & $2.409 \pm 0.027$ & $0.116 \pm 0.016$ & $\ldots$ & $\ldots$ & 0.25\\
55,242.65  & 56.00 & $1.070\pm0.043$ &  $1.834\pm0.064$ & $2.360 \pm 0.058$ & $0.107 \pm 0.031$ & $\ldots$ & $\ldots$ & 0.25\\
\hline
\multicolumn{9}{c}{3C~279}\\
 \hline
56,733.65  & 70.00 & $1.325\pm0.034$ &  $1.148\pm0.027$ & $2.158 \pm 0.028$ & $0.076 \pm 0.015$ & $\ldots$ & $\ldots$ & 0.34\\
57,174.65  & 42.00 & $3.078\pm0.054$ &  $2.783\pm0.046$ & $2.082 \pm 0.020$ & $0.095 \pm 0.011$ & $\ldots$ & $\ldots$ & 0.34\\
57,797.65  & 91.00 & $1.396\pm0.034$ &  $1.243\pm0.027$ & $2.117 \pm 0.024$ & $0.092 \pm 0.013$ & $\ldots$ & $\ldots$ & 0.34\\
58,084.65  & 70.00 & $2.964\pm0.078$ &  $2.816\pm0.053$ & $1.969 \pm 0.026$ & $0.111 \pm 0.011$ & $\ldots$ & $\ldots$ & 0.34\\
\hline
\multicolumn{9}{c}{PKS~1510-089}\\
 \hline
54,892.65  & 42.00 & $2.541\pm0.055$ &  $1.256\pm0.027$ & $2.205 \pm 0.021$ & $0.094 \pm 0.014$ & $\ldots$ & $\ldots$ & 0.45\\
55,830.65  & 63.00 & $2.369\pm0.050$ &  $1.145\pm0.023$ & $2.207 \pm 0.019$ & $0.074 \pm 0.012$ & $\ldots$ & $\ldots$ & 0.45\\
55,928.65  & 126.00 & $2.001\pm0.048$ &  $0.915\pm0.016$ & $2.311 \pm 0.020$ & $0.090 \pm 0.012$ & $\ldots$ & $\ldots$ & 0.45\\
57,090.65  & 49.00 & $2.311\pm0.074$ &  $1.135\pm0.025$ & $2.195 \pm 0.025$ & $0.081 \pm 0.013$ & $\ldots$ & $\ldots$ & 0.45\\
57,230.65  & 28.00 & $2.027\pm0.077$ &  $1.119\pm0.035$ & $1.972 \pm 0.035$ & $0.067 \pm 0.016$ & $\ldots$ & $\ldots$ & 0.45\\
\hline
\multicolumn{9}{c}{CTA~102}\\
 \hline
57,391.65  & 112.00 & $2.031\pm0.033$ &  $2.843\pm0.739$ & $1.936 \pm 0.104$ & $\ldots$ & $3.742 \pm 3.771$ & $0.536 \pm 0.172$ & 0.31\\
57,650.65  & 238.00 & $4.241\pm0.029$ &  $5.065\pm0.210$ & $1.931 \pm 0.026$ & $\ldots$ & $8.638 \pm 1.767$ & $0.704 \pm 0.080$ & 0.31\\
57,972.65  & 182.00 & $1.925\pm0.028$ &  $8.507\pm11.070$ & $1.682 \pm 0.255$ & $\ldots$ & $0.107 \pm 0.350$ & $0.321 \pm 0.121$ & 0.31\\
\hline
\multicolumn{9}{c}{3C~454.3}\\
 \hline
55,123.65  & 84.00 & $5.165\pm0.064$ &  $11.102\pm3.312$ & $1.678 \pm 0.070$ & $\ldots$ & $0.210 \pm 0.137$ & $0.388 \pm 0.038$ & 0.41\\
55,263.65  & 56.00 & $6.786\pm0.090$ &  $8.840\pm4.906$ & $1.938 \pm 0.134$ & $\ldots$ & $0.543 \pm 0.789$ & $0.404 \pm 0.101$ & 0.41\\
55,459.65  & 189.00 & $8.390\pm0.053$ &  $125.464\pm99.276$ & $1.467 \pm 0.102$ & $\ldots$ & $0.002 \pm 0.004$ & $0.231 \pm 0.025$ & 0.41\\
\hline
\\
\multicolumn{9}{c}{\textit{Orbital Light Curves}}\\
\multicolumn{9}{c}{PKS~B1222+216}\\
 \hline
55,359.65  & 10.00 & $4.670\pm0.138$ &  $5.396\pm0.144$ & $1.940 \pm 0.035$ & $0.082 \pm 0.015$ & $\ldots$ & $\ldots$ & 0.31\\
56,966.65  & 17.00 & $2.364\pm0.090$ &  $2.504\pm0.098$ & $2.260 \pm 0.046$ & $0.079 \pm 0.028$ & $\ldots$ & $\ldots$ & 0.31\\
\hline
\multicolumn{9}{c}{3C~273}\\
 \hline
55,091.65  & 38.00 & $2.225\pm0.070$ &  $3.814\pm0.110$ & $2.363 \pm 0.045$ & $0.108 \pm 0.025$ & $\ldots$ & $\ldots$ & 0.25\\
\hline
\multicolumn{9}{c}{3C~279}\\
 \hline
56,748.65  & 8.00 & $3.549\pm0.120$ &  $3.331\pm0.107$ & $2.076 \pm 0.038$ & $0.144 \pm 0.024$ & $\ldots$ & $\ldots$ & 0.34\\
57,185.65  & 6.00 & $10.632\pm0.198$ &  $10.215\pm0.189$ & $1.956 \pm 0.022$ & $0.121 \pm 0.013$ & $\ldots$ & $\ldots$ & 0.34\\
58,129.65  & 14.00 & $9.092\pm0.163$ &  $8.421\pm0.138$ & $2.035 \pm 0.020$ & $0.105 \pm 0.011$ & $\ldots$ & $\ldots$ & 0.34\\
\hline
\multicolumn{9}{c}{PKS~1510-089}\\
 \hline
54,908.65  & 14.00 & $3.890\pm0.102$ &  $1.937\pm0.052$ & $2.188 \pm 0.025$ & $0.090 \pm 0.017$ & $\ldots$ & $\ldots$ & 0.45\\
55,848.65  & 11.00 & $5.014\pm0.196$ &  $2.472\pm0.087$ & $2.133 \pm 0.036$ & $0.047 \pm 0.019$ & $\ldots$ & $\ldots$ & 0.45\\
55,865.65  & 10.00 & $5.898\pm0.195$ &  $2.930\pm0.090$ & $2.188 \pm 0.031$ & $0.087 \pm 0.020$ & $\ldots$ & $\ldots$ & 0.45\\
55,971.65  & 28.00 & $4.240\pm0.083$ &  $2.012\pm0.042$ & $2.273 \pm 0.019$ & $0.100 \pm 0.014$ & $\ldots$ & $\ldots$ & 0.45\\
57,112.65  & 6.00 & $4.901\pm0.162$ &  $2.533\pm0.079$ & $2.105 \pm 0.031$ & $0.075 \pm 0.018$ & $\ldots$ & $\ldots$ & 0.45\\
57,242.65  & 10.00 & $4.043\pm0.166$ &  $2.377\pm0.080$ & $1.859 \pm 0.039$ & $0.077 \pm 0.017$ & $\ldots$ & $\ldots$ & 0.45\\
\hline
\multicolumn{9}{c}{CTA~102}\\
 \hline
57,427.65  & 25.00 & $3.410\pm0.080$ &  $4.707\pm1.235$ & $1.785 \pm 0.134$ & $\ldots$ & $3.538 \pm 3.425$ & $0.644 \pm 0.241$ & 0.31\\
57,732.65  & 37.00 & $9.203\pm0.079$ &  $11.852\pm0.911$ & $1.783 \pm 0.042$ & $\ldots$ & $6.111 \pm 2.083$ & $0.643 \pm 0.095$ & 0.31\\
57,789.65  & 13.00 & $6.703\pm0.171$ &  $13.779\pm7.056$ & $1.606 \pm 0.156$ & $\ldots$ & $1.045 \pm 1.696$ & $0.428 \pm 0.133$ & 0.31\\
57,860.65  & 5.00 & $4.988\pm0.211$ &  $5.437\pm0.233$ & $1.873 \pm 0.064$ & $\ldots$ & $17.370 \pm 5.938$ & $1.275 \pm 0.623$ & 0.31\\
58,127.65  & 22.00 & $3.996\pm0.080$ &  $5.015\pm0.100$ & $1.965 \pm nan$ & $\ldots$ & $3.354 \pm nan$ & $0.817 \pm nan$ & 0.31\\
\hline
\multicolumn{9}{c}{3C~454.3}\\
 \hline
55,159.65  & 29.00 & $8.889\pm0.158$ &  $10.164\pm4.488$ & $1.843 \pm 0.137$ & $\ldots$ & $0.914 \pm 1.025$ & $0.487 \pm 0.119$ & 0.41\\
55,286.65  & 21.00 & $11.062\pm0.135$ &  $9.695\pm4.868$ & $2.030 \pm 0.151$ & $\ldots$ & $1.659 \pm 2.383$ & $0.506 \pm 0.196$ & 0.41\\
55,509.65  & 48.00 & $17.918\pm7.110$ &  $402.758\pm155.384$ & $1.288 \pm 0.048$ & $\ldots$ & $0.002 \pm 0.001$ & $0.237 \pm 0.014$ & 0.41\\
55,561.65  & 10.00 & $11.888\pm0.248$ &  $145.234\pm551.525$ & $1.569 \pm 0.446$ & $\ldots$ & $0.002 \pm 0.017$ & $0.211 \pm 0.117$ & 0.41\\
\hline
\\
\multicolumn{9}{c}{\textit{Suborbital Light Curves}}\\
\multicolumn{9}{c}{PKS~B1222+216}\\
 \hline
55,364.68  & 3.42 & $7.907\pm0.281$ &  $9.309\pm0.308$ & $1.853 \pm 0.043$ & $0.092 \pm 0.019$ & $\ldots$ & $\ldots$ & 0.31\\
56,972.49  & 2.75 & $4.083\pm0.257$ &  $4.096\pm0.278$ & $2.328 \pm 0.078$ & $0.034 \pm 0.045$ & $\ldots$ & $\ldots$ & 0.31\\
\hline
\multicolumn{9}{c}{3C~273}\\
 \hline
55,094.74  & 15.29 & $3.231\pm0.128$ &  $5.657\pm0.203$ & $2.273 \pm 0.058$ & $0.116 \pm 0.031$ & $\ldots$ & $\ldots$ & 0.25\\
\hline
\multicolumn{9}{c}{3C~279}\\
 \hline
57,188.07  & 1.87 & $22.058\pm0.405$ &  $21.570\pm0.436$ & $1.923 \pm 0.023$ & $0.132 \pm 0.014$ & $\ldots$ & $\ldots$ & 0.34\\
56,749.87  & 6.87 & $4.695\pm0.162$ &  $4.301\pm0.143$ & $2.116 \pm 0.039$ & $0.130 \pm 0.024$ & $\ldots$ & $\ldots$ & 0.34\\
58,133.34  & 5.32 & $15.058\pm0.343$ &  $14.438\pm0.301$ & $1.988 \pm 0.026$ & $0.131 \pm 0.015$ & $\ldots$ & $\ldots$ & 0.34\\
\hline
\multicolumn{9}{c}{PKS~1510-089}\\
 \hline
55,850.60  & 4.34 & $6.758\pm0.268$ &  $3.490\pm0.134$ & $2.090 \pm 0.036$ & $0.066 \pm 0.021$ & $\ldots$ & $\ldots$ & 0.45\\
54,914.79  & 2.69 & $6.795\pm0.291$ &  $3.659\pm0.155$ & $2.062 \pm 0.042$ & $0.090 \pm 0.025$ & $\ldots$ & $\ldots$ & 0.45\\
55,867.64  & 2.16 & $10.018\pm0.553$ &  $5.203\pm0.260$ & $2.105 \pm 0.051$ & $0.080 \pm 0.029$ & $\ldots$ & $\ldots$ & 0.45\\
55,872.41  & 1.37 & $10.087\pm0.541$ &  $5.692\pm0.322$ & $2.026 \pm 0.053$ & $0.116 \pm 0.035$ & $\ldots$ & $\ldots$ & 0.45\\
57,114.16  & 1.42 & $6.200\pm0.342$ &  $3.193\pm0.180$ & $2.156 \pm 0.053$ & $0.103 \pm 0.036$ & $\ldots$ & $\ldots$ & 0.45\\
57,243.84  & 4.53 & $6.042\pm0.282$ &  $3.882\pm0.151$ & $1.696 \pm 0.047$ & $0.107 \pm 0.020$ & $\ldots$ & $\ldots$ & 0.45\\
\hline
\multicolumn{9}{c}{CTA~102}\\
 \hline
57,737.41  & 1.67 & $14.560\pm0.372$ &  $16.714\pm0.939$ & $1.762 \pm 0.065$ & $\ldots$ & $10.473 \pm 3.823$ & $0.932 \pm 0.273$ & 0.31\\
57,749.10  & 4.99 & $15.144\pm0.277$ &  $20.082\pm2.754$ & $1.698 \pm 0.079$ & $\ldots$ & $5.071 \pm 2.997$ & $0.643 \pm 0.151$ & 0.31\\
57,757.55  & 4.66 & $13.567\pm0.350$ &  $30.027\pm52.650$ & $1.603 \pm 0.416$ & $\ldots$ & $0.902 \pm 5.436$ & $0.363 \pm 0.353$ & 0.31\\
57,861.71  & 2.42 & $8.206\pm0.321$ &  $8.939\pm0.370$ & $1.847 \pm 0.054$ & $\ldots$ & $17.767 \pm 5.612$ & $1.327 \pm 0.586$ & 0.31\\
57,435.39  & 8.22 & $5.008\pm0.157$ &  $6.068\pm0.841$ & $1.821 \pm 0.109$ & $\ldots$ & $5.733 \pm 3.476$ & $0.857 \pm 0.343$ & 0.31\\
57,446.36  & 0.89 & $3.658\pm0.471$ &  $5.723\pm5.589$ & $1.477 \pm 0.612$ & $\ldots$ & $1.735 \pm 4.707$ & $0.761 \pm 0.823$ & 0.31\\
57,449.98  & 0.64 & $4.005\pm0.611$ &  $81.278\pm272.361$ & $0.576 \pm 0.827$ & $\ldots$ & $0.019 \pm 0.088$ & $0.361 \pm 0.187$ & 0.31\\
58,127.40  & 1.82 & $5.614\pm0.362$ &  $42.578\pm48.687$ & $1.445 \pm 0.275$ & $\ldots$ & $0.042 \pm 0.099$ & $0.324 \pm 0.096$ & 0.31\\
\hline
\multicolumn{9}{c}{3C~454.3}\\
 \hline
55,516.55  & 8.93 & $37.920\pm0.468$ &  $114.329\pm2.884$ & $1.574 \pm 0.028$ & $\ldots$ & $0.100 \pm 0.000$ & $0.333 \pm 0.008$ & 0.41\\
55,166.29  & 5.55 & $14.132\pm0.387$ &  $10.652\pm4.021$ & $1.965 \pm 0.172$ & $\ldots$ & $2.757 \pm 2.926$ & $0.706 \pm 0.334$ & 0.41\\
\hline
\enddata
{
\tablecomments{The light curves for the respective time intervals are shown in Figs.~\ref{fig:daily}, \ref{fig:gti}, and, for the cases where at least one orbital bin is detected with $\mathrm{TS}\geqslant150$, in Figure~\ref{fig:lc_minutes}.
If the uncertainties are $nan$ (not a number), the parameters have been fixed during the fit to achieve convergence.}
}
\end{deluxetable}



\end{appendix}

\bibliography{mainbib}



\end{document}